\documentstyle [12pt,a4wide]{report}
\def\diffsigma{{\rm Diff}\Sigma}

\def\ldiffsigma{{\rm LDiff}\Sigma}

\def\gabd{\gamma_{\alpha\beta}}
\def\gabu{\gamma^{\alpha\beta}}
\def\ham{{{\cal H}^{gr}}_{\perp}}
\def\hom{{{\cal H}^{gr}}_{\perp ,i}}
\def\mom{{{\cal H}^{gr}}_i }
\def\momj{{{\cal H}^{gr}}_j }
\def\hamphi{{\cal H}_{\perp}^{\phi}}
\def\momphi{{\cal H}_i^{\phi} }
\def\momjphi{{\cal H}_j^{\phi} }
\def\hamtot{{\cal H}_{\perp}^T}
\def\momtot{{\cal H}_i^T }

\def\xxp{(x\leftrightarrow x')}
\def\xxpp{(ix\leftrightarrow jx')}
\def\ltilde{\lambda}
\def\mtilde{\mu}
\def\rtilde{R}

\def\f{\phi}
\def\p{\pi}

\def\fdot{\dot{\phi}}

\def\%{\hskip -0.25in}

\begin{document}  
\thispagestyle{empty}
\setcounter{page}{1}
\begin{titlepage}

\begin{flushright}
{\Large February 2000}
\end{flushright}
\vskip 7cm
\begin{center}
{\Large Ph. D. Thesis}
\vskip 1cm
{\LARGE{{  Classical Histories In Hamiltonian Systems }}}   
\vskip 6cm
{{\Large Ioannis Kouletsis}}
\end{center}

\pagebreak

\end{titlepage}
\renewcommand{\theequation}{\arabic{equation}}
\let\ssection=\section
\renewcommand{\section}{\setcounter{equation}{0}\ssection}

\begin{center}
\% {\LARGE Contents}
\end{center}

\vskip -1.5cm
\begin{center}
{\Large Part I: Introduction} 
\end{center}

\begin{tabbing}
{\large Chapter I-1. Overview Of The Thesis} \\[-0.5cm]
$\bullet$ The Aim Of The Thesis{\hskip 10.28cm} \=
{\hskip 0.05cm} \= 6. \\[-0.5cm] 
$\bullet$ The Context Of The Thesis  \> \> 7. \\[-0.5cm] 
$\bullet$ The Structure Of The Thesis \> 1 \> 0. \\[-0.0cm] 
{\large Chapter I-2. Hamiltonian General Relativity}  \\[-0.5cm] 
$\bullet$ The Required Assumptions For Spacetime{\hskip 7cm} \= 12.
{\hskip 0.05cm} \=  \\[-0.5cm] 
$\bullet$ The Lapse Function And The Shift Vector  \> 14. \\[-0.5cm] 
$\bullet$ The Canonical Form Of General Relativity \> 15. \\[-0.0cm] 
{\large Chapter I-3. The Problem Of Time} \\[-0.5cm] 
$\bullet$ Time In Quantum Theory And In General Relativity \> 16. \\[-0.5cm]  
$\bullet$ The Approaches To Quantum Gravity \> 17. \\[-0.0cm] 
{\large Chapter I-4. The Internal Time Approach}  \\[-0.5cm] 
$\bullet$ Parametrized Particle Dynamics \> 18. \\[-0.5cm] 
$\bullet$ The Internal Time Formalism \> 20. \\[-0.5cm] 
$\bullet$ Problems With The Internal Time Formalism \> 21. \\[-0.0cm] 
{\large Chapter I-5. The Gaussian Time Formulation} \\[-0.5cm] 
$\bullet$ Coordinate Conditions  \> 23. \\[-0.5cm] 
$\bullet$ The Reference Fluid    \> 23. \\[-0.0cm] 
\end{tabbing}

\vskip -1.5cm
\begin{center}
{\Large Part II: A New Lie Algebra For Vacuum General Relativity}
\end{center}

\begin{tabbing}
{\large Chapter II-1. The Discovery Of A Genuine Lie Algebra} \\[-0.5cm] 
$\bullet$ Incoherent Dust  {\hskip 11.57cm} \= 26. \\[-0.5cm] 
$\bullet$ Self-Commuting Combinations \> 27. \\[-0.5cm] 
$\bullet$ The Weight-$\omega$ Equation  \> 28. \\[-0.5cm] 
$\bullet$ A Possibility  \> 29. \\[-0.0cm] 
{\large Chapter II-2. Treating The Algebra Algebraically} \\[-0.5cm] 
$\bullet$ Preliminary Remarks About The $\omega$-Equation \> 31. \\[-0.5cm] 
$\bullet$ The $\omega$-Ansatz And The Linear Equation \>
32. \\[-0.0cm]  
{\large Chapter II-3. Actions Leading To The Algebra} \\[-0.5cm] 
$\bullet$ The Scalar Field Action \> 34. \\[-0.5cm] 
$\bullet$ The Hamiltonian Analysis Of The Coupled System \> 35. \\[-0.5cm] 
$\bullet$ The Two Equations For $M$ And $\pi$ \> 36. \\[-0.5cm] 
$\bullet$ Solving The Two Equations For $M$ And $\pi$. The General
Case \> 37. \\[-0.5cm]  
$\bullet$ Solving The Two Equations For $M$ And $\pi$. The Special
Cases \> 39. \\[-0.0cm]  
{\large Chapter II-4. The Inverse Procedure} \\[-0.5cm] 
$\bullet$ Actions For Given Solutions \> 42. \\[-0.5cm] 
$\bullet$ An Application \> 45. \\[-0.0cm] 
{\large Chapter II-5. The Algebra In Vacuum Gravity} \\[-0.5cm] 
$\bullet$ The Time Evolution Generated By The Solutions \> 46. \\[-0.5cm] 
$\bullet$ Finding Vacuum Solutions Of The Algebra \> 48. \\[-0.0cm] 
\end{tabbing}

\vskip -1.5cm
\begin{center}
{\Large Part III: Introducing Classical Histories}
\end{center}

\begin{tabbing}
{\large Chapter III-1. Motivation} \\[-0.5cm] 
$\bullet$ Description {\hskip 12.39cm} \= 51. \\[-0.5cm]  
$\bullet$ The Dirac Algebra And The Principle Of Path Independence \>
53. \\[-0.5cm]   
$\bullet$ The Problem Of Deriving A Theory From Just The
Canonical Algebra \> 54. \\[-0.5cm] 
$\bullet$ Selecting The Physical Representations Of The Dirac Algebra
\> 56. \\[-0.5cm]  
$\bullet$ The Full List Of The Selection Postulates 
\> 58. \\[-0.5cm]  
$\bullet$ The Need For A Detailed Understanding Of The Selection
Postulates  \> 59. \\[-0.0cm]  
{\large Chapter III-2. The History Phase Space} \\[-0.5cm] 
$\bullet$ The Unconstrained Hamiltonian \> 62. \\[-0.5cm] 
$\bullet$ Incorporating The Fixed Functions \> 64. \\[-0.5cm] 
$\bullet$ The Constrained Hamiltonian \> 66. \\[-0.0cm] 
{\large Chapter III-3. The Evolution Postulate} \\[-0.5cm] 
$\bullet$ The Inverse Procedure And The Evolution Postulate \> 68. \\[-0.5cm]
$\bullet$ The Evolution Postulate In An Equivalent Form \> 70. \\[-0.0cm] 
\end{tabbing}

\vskip -1.5cm
\begin{center}
{\Large Part IV: Canonical Theories Derived From first Principles}
\end{center}

\begin{tabbing}
{\large Chapter IV-1. The Indirect Method Applied To Gravity} \\[-0.5cm] 
$\bullet$ The New Set Of Postulates
{\hskip 9.58cm} \= 74. \\[-0.5cm] 
$\bullet$ Recovery Of The Re-shuffling And Ultra-locality Postulates \> 75. \\[-0.5cm] 
$\bullet$ The Two Jacobi Identities \> 77. \\[-0.5cm] 
$\bullet$ Recovery Of The Super-momentum Constraint \> 80. \\[-0.5cm] 
$\bullet$ Recovery Of The Weak Representation Postulate \> 81. \\[-0.5cm] 
$\bullet$ The Principle Of Path Independence \> 83. \\[-0.5cm] 
$\bullet$ Recovery Of The Super-Hamiltonian Constraint \> 84. \\[-0.5cm]
$\bullet$ The Representations Of The Weak Principle \> 87. \\[-0.0cm]  
{\large Chapter IV-2. The Direct Method Applied To Field Theory} \\[-0.5cm] 
$\bullet$ Preliminaries \> 89. \\[-0.5cm] 
$\bullet$ The Unknown Legendre Transformation \> 91. \\[-0.5cm] 
$\bullet$ The Unknown Lagrangian \> 94. \\[-0.0cm] 
\end{tabbing}

\vskip -1.5cm
\begin{center}
{\Large Part V: The Geometric Interpretation Of The New Algebra}
\end{center}

\begin{tabbing}
{\large Chapter V-1. The Generators Of Deformations} \\[-0.5cm] 
$\bullet$ Preliminaries {\hskip 12.121cm} \= 97. \\[-0.5cm]  
$\bullet$ The Representation For The Deformation Generators \> 98. \\[-0.0cm]  
{\large Chapter V-2. The Canonical Generators} \\[-0.5cm] 
$\bullet$ The Decomposition Of The Spacetime Theory \> {\hskip -0.35cm} 101. \\[-0.5cm] 
$\bullet$ An Application  \> {\hskip -0.35cm} 102. \\[-0.0cm] 
\end{tabbing}

\vskip -1.5cm
\begin{center}
{\Large Part VI: Discussion, Acknowledgments And References}
\end{center}

\begin{tabbing}
{\large Chapter VI-1. The Results, Their Relation, Their Extent And Limitations} \\[-0.5cm] 
$\bullet$ The Genuine Algebra {\hskip 10.454cm} \= 105. \\[-0.5cm]    
$\bullet$ The Use Of Classical Histories In The Thesis \>
106. \\[-0.5cm] 
$\bullet$ Self-Commuting Combinations In The Vacuum Theory \>
107. \\[-0.5cm]   
$\bullet$ Histories And Explicit Spacetime Invariance \> 108. \\[-0.5cm]  
$\bullet$ The Consistent Histories Approach  \> 109. \\[-0.0cm] 
{\large Acknowledgments} \> 110. \\[-0.0cm] 
{\large Appendix A. } {\hskip 11.70cm}  111. \\[-0.0cm] 
{\large Appendix B. } {\hskip 11.71cm} 112. \\[-0.0cm] 
{\large Appendix C. } {\hskip 11.70cm}  116. \\[-0.0cm] 
{\large Bibliography. } {\hskip 11.455cm}  118. \\[-0.0cm] 
\end{tabbing}

\pagebreak

\begin{center}
{\LARGE Part I:}
\vskip 9cm
{\LARGE Introduction}
\end{center}

\pagebreak

\% {\LARGE Chapter I-1. Overview Of The Thesis.} 
\vskip 1cm
\hskip -0.25in
{\Large $\bullet$} The Aim Of The Thesis: The Hamiltonian formulation of general relativity was developed originally as a preliminary step to a quantum theory of gravity.
Time is separated from the remaining spacetime coordinates and is treated as a privileged parameter, in accordance with the standard algorithm of quantization. The incompatibility between the treatment of time in the classical and in the quantum theory results in the so-called problem of time in canonical quantum gravity[1,2]. Several attempts have been made to devise alternative algorithms of quantization which may accommodate the covariance of the classical theory from the outset. 

\% One of the most prominent of these attempts is based on the notion of continuous histories[3,4,5,6] in the context of the consistent histories approach to quantum theory[7,8,9,10]. A {\it history} is defined as a sequence of {\it time-ordered} propositions about the properties of the physical system. It is precisely this intrinsic temporality of histories that may provide a solution to the problem of time in quantum gravity. Such a solution is the ultimate aim of the histories program. The analogue of continuous histories in a classical theory is the main theme of this thesis. 

\% By the term {\it classical histories} it is implied that the canonical fields, as well as the symplectic structure of the theory, depend on the full foliation of spacetime into hyper-surfaces, rather than on the embedding at a single instant of time. For example, the history Poisson bracket between a scalar field and its conjugate momentum is defined by $\{ \phi(x,t), \pi(x',t') \} := \delta (x,x') \delta (t,t')$. The aim of the thesis is to illustrate that, even at the classical level, the advantages of a history theory over the standard canonical approach are significant, especially when it comes to discussing spacetime issues. This fact strongly suggests a history canonical approach when considering quantum gravity.

\% {\Large $\bullet$} The Context Of The Thesis: The need for using classical histories came from an attempt to deal with an apparently unrelated issue; namely, the lack of a genuine Lie algebra in canonical general relativity. More precisely, the Dirac algebra[11] is the algebra according to which the Hamiltonian and momentum constraints of canonical relativity close under the Poisson bracket operations. However, the induced metric on the three-dimensional hyper-surface appears explicitly in the Poisson bracket between the Hamiltonian constraints defined at two spatially distinct points. This means that the Dirac algebra is not a genuine Lie algebra; a fact that creates many serious difficulties in canonical quantum gravity, especially in group-oriented approaches[1,2,12].

\% The coupling of gravity to dust helped Brown and Kucha\v{r}[13] toparameterized discover simple quadratic combinations of the gravitational Hamiltonian and momentum constraints whose Poisson brackets vanish strongly. If these combinations replace the Hamiltonian constraint to form an equivalent system of constraints for {\it vacuum} general relativity, a genuine Lie algebra is created. It is natural to ask whether the coupling of gravity to other sources yields alternative combinations of the gravitational constraints whose Poisson brackets also vanish strongly. Kucha\v{r} and Romano[14] illustrated how this can be done by coupling gravity to a massless scalar field. Brown and Marolf[15] produced other combinations by coupling gravity to fluids, and Markopoulou[16] found an equation satisfied by all such combinations by treating the problem algebraically.

\% The physical relevance of this equation is not clear. However, an insight into its origin can be gained if it is shown to be related to phenomenologically physical systems like the ones discussed by Kucha\v{r} {\it et al}. Indeed, it is shown (here, and in ref. 17) that all such combinations can be derived from a generalized action functional that is coupled to gravity. The action functional depends on a scalar field and is  by two arbitrary functions of a Lagrange multiplier. After the elimination of the multiplier from the action, the resulting theory is interpreted as a theory of gravity coupled to scalar fields with nonlinear self-interactions. 

\% The momentum conjugate to the scalar field can be solved solely in terms of the gravitational Hamiltonian and momentum generators. This leads to scalar densitized combinations of the gravitational generators that are parametrized by an arbitrary function of one variable. In the case of {\it vacuum} gravity, these combinations provide exactly the general solution of the equation constructed algebraically by Markopoulou. 

\% The coincidence of the purely algebraic result with the result arising from the action principle is indeed remarkable. It suggests that the role of couplings in general relativity ought to be investigated further. However, returning to the case of vacuum gravity, an important problem arises concerning the usefulness of these self-commuting combinations. Namely, it is shown (here) that most of the combinations are ill-defined on the constraint surface of vacuum gravity, while the remaining (well-defined) ones lead to a trivial dynamical evolution if used to replace the Hamiltonian constraint. 

\% The question arises whether alternative gravitational constraints for vacuum gravity can be constructed. These should not only satisfy the new Lie algebra but also generate the appropriate dynamical evolution. In order to investigate this possibility, a further insight into the origin of the new Lie algebra is required. Indeed, significant progress is achieved by comparing the new algebra with the Dirac one (here, and in ref. 18). The geometric interpretation of the latter is known and a method for finding its physically relevant representations is also available. 

\% The underlying geometry of the Dirac algebra was first recognized by Teitelboim[19], while an algorithm for constructing its representations was developed by Hojman, Kucha\v{r} and Teitelboim in their derivation of geometrodynamics from first principles[20]. The attempt to apply their algorithm unambiguously to the case of the new algebra results in a re-examination of the standard canonical formalism and in the introduction of classical histories. This is precisely how the notion of classical histories arises in the thesis.

\% If the phase space is defined over the space of classical histories, it is shown (here, and in ref. 18) that the precise relation between the individual postulates used by Kucha\v{r} {\it et al} is clarified. The original set of postulates can then be replaced by an {\it evolution} postulate which is related directly to the spacetime picturfavore. To be precise, the original aim of Kucha\v{r} {\it et al} was to use postulates that depend exclusively on the {hyper-surface}. The assumption of a surrounding spacetime appears only implicitly in their arguments, and this is the reason why the connection of the postulates is not clear.

\% It is only after choosing to make this assumption explicit that the precise relationship of the postulates is revealed. If this choice is not taken, nothing can be gained or lost through the introduction of histories. What is remarkable, however, is that in the standard formalism there is {\it no} second choice, and the postulates of Kucha\v{r} {\it et al} are then the best one can hope for. This happens precisely because of the use of equal-time Poisson brackets in the standard approach. It is only in the history formalism that the assumption of a surrounding spacetime can be made explicit, thus leading to the understanding of the postulates.

\% This observation, alone, illustrates the genuine advantages of the canonical history approach over its standard counterpart and speaks in its  when approaching quantum gravity. In addition, the use of a history phase space results in certain corrections of the results of Kucha\v{r} {\it et al}. These lead to the discovery of additional representations of the {evolution} postulate for general relativity. The meaning of these new representations is only partially clarified in the thesis. 

\% Finally, by applying the history formalism to the original problem of the new algebra, its geometric interpretation is found. The representations of the algebra that generate the correct dynamical evolution are also constructed. This is achieved by decomposing the Lagrangian with respect to an appropriate choice of foliation. The Hamiltonian and momentum constraints that arise through this decomposition are then the required representations of the algebra. Since the choice of foliation does not affect the physical content of the theory, it follows that these representations generate the correct dynamical evolution. 
This procedure works for an arbitrary algebra and for any field theory that is parametrized. For general relativity a more elaborate scheme needs to be devised; this project is under preparation[21].

\% {\Large $\bullet$} The Structure Of The Thesis: A brief review of the standard canonical formulation of general relativity is presented in the remaining part of the introduction. The problem of time as well as the main approaches to quantization are discussed. Particular emphasis is placed on the internal-time approach and on the related Gaussian-time formulation, the knowledge of which is needed for the chapters that follow. Part II begins with the presentation of the genuine Lie algebra discovered by Brown and Kucha\v{r} in the context of the Gaussian-time approach. Certain representations of this algebra are combinations of the super-Hamiltonian and super-momentum constraints of canonical general relativity. The connection of these combinations with a generalized scalar field action functional that is coupled to gravity is presented. 

\% In the case of pure gravity, these representations are ill-defined and an alternative interpretation is needed. This need leads to the use of classical histories which are discussed in two parts. In part III the phase space over the space of histories is defined while an algorithm for interpreting the
algebra is explained. It modifies and completes the algorithm devised by Kucha\v{r} {\it et al}[20] for deriving representations of the canonical generators from first principles. In the new algorithm, the principles of Kucha\v{r} {\it et al} are replaced by the {evolution} postulate. 

\% In part IV, the evolution postulate is used to find the most general canonical representations of (i) vacuum general relativity and (ii) a scalar field theory on a given metric background. These two theories act as examples of a constrained and an unconstrained system, respectively. In the former case, new canonical representations arise. 
In part V, the geometric interpretation of the new Lie algebra is presented. The revised algorithm is used in finding representations of the algebra for a parametrized scalar field theory. The current results, their extent and their limitations are discussed in part VI. Some future directions, both in the classical and in the quantum domain, are pointed out.

\vskip 13cm

\% {\LARGE Chapter I-2. Hamiltonian General Relativity.} 
\vskip 1cm
\hskip -0.25in
{\Large $\bullet$} The theory of general relativity was written in 
Hamiltonian form initially by Dirac[11], Arnowitt, Deser and Misner[22]. A
feature of the original formulation was the selection of a specific coordinate
system on the spacetime manifold but, later, a global
geometric approach was developed by Kucha\v{r}[23]. The following summary
is based on an article by Isham[1].   

\% {\Large $\bullet$} The Required Assumptions For Spacetime: The
starting point is a 3-dimensional manifold $\Sigma$ that 
represents physical space. It is assumed to be compact.
Without this assumption the resulting theory has to be augmented by
surface terms. The topology of the spacetime manifold
${\cal M}$ is assumed to be such that it can be foliated by an
one-parameter family of embeddings, 
\begin{equation}
{\cal X}_t : \Sigma \rightarrow {\cal M} , 
\end{equation}
$t \in R$, of $\Sigma$ in ${\cal M}$. This implies that
the spacetime manifold is limited topologically to be diffeomorphic to
$\Sigma \times R$, since the map 
\begin{equation}
{\cal X} : \Sigma \times R \rightarrow {\cal M},
\end{equation}
defined by
$(x,t) : \rightarrow {\cal X}(x,t) := {\cal X}_t(x)$, $x \in \Sigma$,  
is a diffeomorphism of $\Sigma \times R$ with ${\cal M}$. 

\% For each $x \in \Sigma$ the map 
\begin{equation}
{\cal X}_x : R \rightarrow {\cal M},
\end{equation} 
defined by $t \rightarrow {\cal X}(x,t)$, 
is a curve in ${\cal M}$ and therefore has an one-parameter family of
tangent vectors in ${\cal M}$. This is known as the deformation vector
field of the foliation and is defined by 
\begin{equation}
{\dot{\cal X}}(x,t) := {\dot{{\cal X}_x}}(t). 
\end{equation}
For a particular choice of foliation there is a unique deformation
vector field.  

\% In general, if ${\cal X}_t : \Sigma \rightarrow {\cal M}$ is an
embedding, the normal vector field $n$ to the embedding is
defined by the relation 
\begin{equation}
n_{\alpha}{{\cal X}^{\alpha}}_{,i} = 0.
\label{eq:norm1}
\end{equation}
The indices $\alpha = 0,1,2,3$ and $i = 1,2,3$ correspond to the
coordinate systems in ${\cal M}$ and in $\Sigma$ respectively. In the
Hamiltonian formulation of general relativity the 
embedding is required additionally to be space-like with respect to the
Lorentzian metric on ${\cal M}$. The defining relation
for the normal vector field must then be supplemented by the normalization
condition 
\begin{equation}
{\gamma}^{\alpha \beta} n_{\alpha} n_{\beta} = - 1.
\label{eq:norm2}
\end{equation}
The last relation implies that $n$ is a time-like vector field in
$({\cal M},\gamma)$ when the signature of the Lorentzian metric is
$(-1,1,1,1)$.  

\hskip -0.25in
{\Large $\bullet$} {The Lapse Function And The Shift Vector:} For each value of
the time parameter $t$, the deformation
vector can be decomposed into two components, one of which lies along
the hyper-surface ${\cal X}_t(\Sigma)$ and the other of which is
parallel to $n_t$,
\begin{equation}
\dot{{\cal X}^{\alpha}}
= N \gamma^{\alpha \beta} n_{\beta} +
N^i {{\cal X}^{\alpha}}_{,i}.
\label{eq:lsdecom}
\end{equation}
The functions $N(x,t)$ and $N^i(x,t)$ are known as the lapse function
and the shift  
vector respectively. Their geometric interpretation can be deduced from this
relation as follows. The lapse function specifies the proper time
separation between 
the hyper-surfaces ${\cal X}_t(\Sigma)$ and ${\cal X}_{t + \delta
t}(\Sigma)$ measured in the direction normal to the first
hyper-surface. The shift vector determines how, for
each $x \in \Sigma$, the point ${\cal X}_{t + \delta t}(x)$ in ${\cal
M}$ is displaced with respect to the intersection of the hyper-surface
${\cal X}_{t + \delta t}(\Sigma)$ with the normal geodesic drawn from
the point ${\cal X}_t(x)$.

\% In order to define the canonical theory, the spacetime metric
must be ``pulled-back'' from ${\cal M}$ to $\Sigma \times R$. In local
coordinates, the induced spatial metric is written as
\begin{equation}
g_{ij} :={\gamma}_{\alpha \beta} {{\cal X}^{\alpha}}_{,i} {{\cal
X}^{\beta}}_{,j}.
\label{eq:spatmetr}
\end{equation}
It is a positive definite tensor of signature (1,1,1) since the
embedding is space-like with respect to the Lorentzian structure. The
spatial metric contains six of 
the ten degrees of freedom of the original theory. The normal vector
field, the lapse function and the shift vector are the only other
quantities that depend on the spacetime metric, so
are all candidates for the remaining degrees of freedom.  
However, if the theory is decomposed with respect to the basis
$(n^{\alpha}, {{\cal X}^{\alpha}}_i)$, $n^{\alpha}$ cannot appear in
the pulled-back Lagrangian. The reason is that the right sides of equations
(\ref{eq:norm1})-(\ref{eq:norm2}) consist of pure numbers. The lapse  
function and the shift vector may therefore be identified with the remaining degrees of
freedom and treated as canonical coordinates.

\hskip -0.25in
{\Large $\bullet$} {The Canonical Form Of General Relativity:} The canonical
theory is obtained by decomposing the Hilbert-Einstein Lagrangian
with respect to the spatial metric, the shift vector and the
lapse function. A Legendre transformation is performed to
replace any time derivative of these variables with their conjugate
momenta. The lapse function and the shift
vector become non-dynamical Lagrange multipliers. They enforce on the
canonical variables the super-Hamiltonian and super-momentum constraints[22],
\begin{eqnarray}
{\cal S}&=&\int d^3xdt [p^{ij} \dot{g_{ij}} - N \ham - N^i \mom],
\label{eq:action}\\
\ham&=&{1\over2}g^{-{1\over2}}\biggl(g_{ik}g_{jl}+g_{il}g_{jk}-g_{ij}g_{kl}\biggr)p^{ij}p^{kl}-g^{1\over2}R \simeq0, \label{eq:ham} \\ 
\mom&=&-2D_jp^j_i \simeq 0.
\label{eq:mom}
\end{eqnarray}

\% The constraints (\ref{eq:ham})-(\ref{eq:mom}) satisfy the Dirac algebra[11] 
\begin{eqnarray}
\{\ham(x), \ham(x')\}&=&g^{ij}(x) \mom(x) \delta_{,j}(x,x')-\xxp,\label{eq:D1}
\\
\{\ham(x), \mom(x')\}&=&\hom(x) \delta (x,x')+
\ham(x) \delta_{,i}(x,x'),\label{eq:D2}
\\
\{\mom(x),\momj (x')\}&=&\momj(x)\delta_{,i}(x,x')-\xxpp. \label{eq:D3}
\end {eqnarray}
They are first-class since the right hand side of equations
(\ref{eq:D1})-(\ref{eq:D3}) vanishes on the constraint surface
(\ref{eq:ham})-(\ref{eq:mom}). The presence of the
spatial metric $g^{ij}$ in equation (\ref{eq:D1}) implies that the
Dirac algebra is not a genuine Lie algebra. This becomes a
problem in most attempts for a quantum theory of gravity[1,2,12]. 
\vskip 9cm
\hskip -0.25in
{\LARGE Chapter I-3. The Problem Of Time.}
\vskip 1cm
\hskip -0.25in
{\Large $\bullet$} {Time In Quantum Theory And In General Relativity:}
The formulation of quantum theory is 
grounded on the idea of a measurement made at a particular instant of
time. The time parameter is external to the system and 
belongs to the ``classical world'' according to the conventional 
Copenhagen interpretation. The fact that time is not a physical
observable is expressed mathematically by the lack of a time
operator in the quantum theory[1]. This special property of time applies 
for non-relativistic quantum theory and for relativistic particle
dynamics as well as for quantum field theory. 

\% However, this view of time cannot be maintained when the theory
of general relativity is taken into account. The reason is that 
the equations of general relativity transform covariantly under
changes of spacetime coordinates and physical results remain 
invariant under such changes. The invariance of the
theory under the action of the group of active point transformations,
Diff$({\cal M})$, implies that no intrinsic physical significance can
be assigned to the individual mathematical spacetime points. This
amounts to the lack of physical observables in the vacuum
theory[1]. It also suggests that fundamental features
of quantum field theory cannot be maintained, like the notion of a
space-like separation, the interpretation of the so-called
micro-causality condition and even the canonical commutation
relations themselves[1].  

\hskip -0.25in
{\Large $\bullet$} {The Approaches To Quantum Gravity:} Most approaches to
dealing with the contradictory roles assigned to time by the quantum
and the classical theory identify time in terms
of the internal structure of the system. They can be divided mainly into
three categories[1]: 

\% (i) An internal time is identified as a
functional of the canonical variables and the constraints are solved,
before the system is quantized. The
resolved constraints are linear in the momentum conjugate to the
internal time, so a linear Schr\"{o}dinger equation is produced with
respect to this choice of time.

\% (ii) The constraints are imposed at the
quantum level as restrictions on the allowed state vectors, and time is
identified only after this step. The resulting functional
differential equation is known as the Wheeler-DeWitt equation. It is
quadratic in the functional derivatives. The notion of time must be
recovered from the solutions of this equation, and the final
probabilistic interpretation of the theory must be made after this
identification of time. 

\% (iii) The timeless nature of general relativity is preserved. It is
assumed that a quantum theory of gravity can be constructed without any
mention to the concept of time. The latter is
considered to have a status that is purely phenomenological. 
\vskip 9cm
\hskip -0.25in
{\LARGE Chapter I-4. The Internal-Time Approach.}
\vskip 1cm
\hskip -0.25in
{\Large $\bullet$} The internal-time approach belongs to the first
type of quantization schemes. Time is assumed to be  
hidden among the canonical variables and must be identified and
separated before the theory is quantized. The procedure 
corresponds to the so-called ``de-parametrization'' of general relativity. 
It is motivated by the parallel that can be drawn between general
relativity and parametrized particle dynamics[24]. 

\hskip -0.25in
{\Large $\bullet$} {Parametrized Particle Dynamics:} The canonical action
describing the motion of a single 
non-relativistic particle of mass m in the potential field ${\cal F}(x^i, t)$
has the form[24]
\begin{equation}
S[x^i, p_i] = \int dt \Biggl({p_i}{{dx^i}\over dt}- H(x^i, p_i, t)\Biggr).
\label{eq:field}
\end{equation}
The precise expression for the Hamiltonian $H(x^i, p_i, t)$ is 
\begin{equation}
H(x^i, p_i, t)= {1\over {2m}}p_ip_i + {\cal F}(x^i, t).
\label{eq:'81}
\end{equation}

\% If the path of the particle is parametrized by an
arbitrary label time $\tau$, and the absolute Newtonian time $t$ is
adjoined to the configuration variables $x^i$, 
\begin{equation}
x^{\alpha} = (t, x^i),~~~~~~~x^i =
x^i(\tau),~~~~~~~p_i=p_i(\tau),~~~~~~~t=t(\tau), 
\end{equation}
the action (\ref{eq:field}) becomes
\begin{equation}
S[x^{\alpha}(\tau), p_i(\tau)] = \int d\tau
\Biggl({p_i(\tau)}{{dx^i(\tau)}\over d\tau}- H[x^{\alpha}(\tau), 
p_i(\tau)]  {dt(\tau)\over d\tau}\Biggr).
\label{eq:param}
\end{equation}

\% Expression (\ref{eq:param}) is numerically equal to the original
expression (\ref{eq:field}) so its variation with respect
to the canonical variables $x^i$ and $p_i$ gives the correct equations of
motion. In addition, the variation with respect to the Newtonian time
$t$ yields
\begin{equation}
{dH(\tau)\over d\tau} = {\partial
H(t)\over {\partial
t}}{dt(\tau)\over d\tau},  
\end{equation} 
which is valid by virtue of the original
Hamilton  equations for $x^i$ and $p_i$. The parametrized action
(\ref{eq:param}) is therefore consistent. 

\% The Hamiltonian process requires the definition
of the momentum $p_0$ conjugate to the Newtonian time $t$,
\begin{equation}
p_0 := - H[x^{\alpha}, p_i] .
\label{eq:gjr}
\end{equation}
The canonical action then becomes
\begin{equation}
S[x^{\alpha}(\tau), p_{\alpha}(\tau)] = \int d\tau
\Biggl({p_{\alpha}(\tau)}{{dx^{\alpha}(\tau)}\over d\tau}- {\cal
N}{\cal H}[x^{\alpha}(\tau), 
p_{\alpha}(\tau)]\Biggr) .
\label{eq:param2}
\end{equation}
Notice that a new quantity $p_{\alpha}$ has been introduced, defined by 
\begin{equation}
p_{\alpha} = (p_0, p_i),~~~~~~~p_0=p_0(\tau) .
\end{equation}

\% The geometric meaning of ${\cal N}$ is recovered by varying this
action with respect to $p_0$, 
\begin{equation}
{\cal N} = {dt(\tau)\over d\tau}.
\label{eq:mult}
\end{equation}
The variation with respect to the non-dynamical
Lagrange multiplier ${\cal N}$ enforces on the canonical data the constraint
\begin{equation}
{\cal H}[x^{\alpha}(\tau), p_{\alpha}(\tau)] = p_0(\tau) +
H[x^{\alpha}(\tau), p_i(\tau)] = 0.
\label{eq:ehmdote}
\end{equation}
Finally, the variation with respect to the remaining
variables leads to valid equations by virtue of the original theory. 

\hskip -0.25in
{\Large $\bullet$} {The Internal Time Formalism:}
The parallel that can be drawn between this procedure and
canonical geometrodynamics is the following. The parametrized Hamiltonian in
(\ref{eq:param2}) is constrained to vanish, the Newtonian time behaves
as an one-dimensional embedding, and the Lagrange
multiplier enforcing the constraint is the analogue of
the lapse function. The observation is that the constraint ${\cal H}=0$ arises
in the parametrized model because the Newtonian time $t$ and its
conjugate momentum $p_0$ have been adjoined to the true dynamical
variables $x^i$, $p_i$. It could be the case that the action for
geometrodynamics is already in parametrized form and can even be
de-parametrized; that is, reduced to its basic true degrees of freedom.

\% The conjecture is that there exists a canonical transformation[25],
\begin{equation}
\Biggl(g_{ij}(x), p^{ij}(x)\Biggr)~~~to~~~\Biggl(X^A(x), P_A(x),
\phi^q(x), \pi_q(x)\Biggr),
\end{equation}
that separates the four embedding variables $X^A(x)$ that specify the
hyper-surface from the two true gravitational
degrees of freedom $\phi^q(x)$ ($A=0, 1, 2, 3, \; q= 1, 2$).   
The constraint equations (\ref{eq:ham})-(\ref{eq:mom}) are then 
replaced by the equivalent set 
\begin{equation}
H_A(x) = P_A(x) + h_A(x; X, \phi, \pi] = 0.
\label{eq:ehmdote2}
\end{equation}
The modified constraints (\ref{eq:ehmdote2}) correspond to equation
(\ref{eq:ehmdote}) valid for the parametrized particle.
Drawing the analogy even further, the quantity $h_A(x; X, \phi, \pi]$
is interpreted as the energy density and the energy flux carried
by the variables $\phi^q(x)$ and $\pi_q(x)$ through the hyper-surface $X^A(x)$.
In the form (\ref{eq:ehmdote2}), the constraints are imposed on
the physical states according to the Dirac quantization algorithm. 
Because of their linearity they lead to a first-order
Schr\"{o}dinger equation. If this separation between embedding
variables and true degrees of freedom can be achieved, a
considerable progress towards a quantum theory of gravity will have been made.

\hskip -0.25in
{\Large $\bullet$} {Problems With The Internal Time Formalism:}
However, besides the standard technical issues associated
with a Schr\"{o}dinger-type equation[1], the internal time approach
faces problems that render it almost unattainable:  

\% 1. Global problem. Calculations performed in simple
models have shown that the new system of constraints
(\ref{eq:ehmdote2}) cannot be made globally equivalent to the original system
(\ref{eq:ham})-(\ref{eq:mom}). It is reasonable to surmise that this
problem becomes worse in the case of full geometrodynamics.  

\% 2. Multiple choice problem. If a natural
choice of internal time does not exist, the ensuing
quantum theories must either be equivalent or, at least, related in
a specific way. Simple examples show that this is not the case.

\% 3. Spacetime problem. What is implied in the internal-time
program is that the time coordinate should be 
constructed purely of the canonical data. Such a time must be
independent of the particular foliation relative to which the
canonical formalism has been defined, so it can only be a
spacetime scalar. Assuming only locality, it has been 
shown that scalar internal-time functionals do not exist[2].

\vskip 9cm
\hskip -0.25in
{\LARGE Chapter I-5. The Gaussian Time Formulation.}
\vskip 1cm
\hskip -0.25in
{\Large $\bullet$} The probable failure of the
internal-time approach suggests that the analogy drawn between
geometrodynamics and parametrized dynamics is a misleading
one. Nevertheless, this analogy motivated an alternative approach to
quantum gravity, where the framework for dealing with the 
conceptual aspects of time arises naturally.   

\% {\Large $\bullet$} {Coordinate Conditions:} Isham and Kucha\v{r}[26] discussed
the issue of representing spacetime diffeomorphisms in canonical
general relativity. They needed to construct 
a homomorphic mapping of spacetime vector fields into the Poisson
bracket algebra of the geometrodynamical phase space. Contrary to the
internal time scheme, they did not consider canonical gravity as being
already parametrized but, instead, they parametrized it once more. 
The geometrodynamical phase space is extended by the
space of embeddings of the spatial manifold $\Sigma$ in the spacetime
${\cal M}$ and, therefore, the required homomorphic mapping is constructed.

\% When extending the phase space, the need for consistency 
led them to restrict the spacetime metrics by Gaussian coordinate
conditions with respect to an auxiliary foliation structure. The
constraints (\ref{eq:ham})-(\ref{eq:mom}) are suspended 
temporarily and are re-introduced later, after the embedding canonical
variables have been adjoined to the gravitational ones. This is
achieved through varying the auxiliary structure.

\% The procedure of breaking the invariance of general relativity, and
restoring it again by parametrization, leads to the modification of the
constraints by terms which are linear in the momenta conjugate to the
Gaussian coordinates. As in the case of an internal time approach, the
equation obtained by imposing the new constraints as a
restriction on the physical states is a Schr\"{o}dinger
equation. The ensuing theory can be viewed as vacuum quantum gravity
but it can also be criticized as lacking physical interpretation[2]. 

\hskip -0.25in
{\Large $\bullet$} {The Reference fluid:}
Addressing this issue, Kucha\v{r} and Torre[27] gave a phenomenological
interpretation for the Gaussian conditions by 
taking them into account through a different technical
procedure. The Gaussian coordinates are adjoined to the
Hilbert-Einstein action by Lagrange multipliers and the total
action is varied. The additional variables introduce a 
source term into Einstein's field equations and are interpreted
as a reference fluid. 
The canonical analysis of the fluid results in a set of 
constraints similar to that in [26], and a Shcr\"{o}dinger
equation is obtained through the Dirac quantization scheme. 

\% The main advantages of the Gaussian-time approach over the 
internal-time one are the following: 

\% (i) The fluid variables are spacetime scalars by construction,
so there is no spacetime problem.

\% (ii) The introduction of the reference system is associated
with a privileged time so there is no multiple choice problem, either.

\% Unfortunately, although these gains are considerable, the Gaussian
reference fluid suffers from a basic problem.
Its energy-momentum tensor does not satisfy the energy conditions of
general relativity, so the fluid can only be given an interpretation
that is phenomenological.
 
\pagebreak

\begin{center}
{\LARGE Part II: }
\vskip 9cm
{\LARGE A New Lie Algebra For Vacuum General Relativity}
\end{center}
\pagebreak 

\% {\LARGE Chapter II-1. The Discovery Of A Genuine Lie Algebra.}
\vskip 1cm 
\hskip -0.25in
{\Large $\bullet$} {Incoherent dust:}
In their search for a realistic medium for the reference fluid, Brown and
Kucha\v{r}[13] avoided any mention to coordinate conditions. Instead, they 
constructed a physical Lagrangian that describes a globally hyperbolic
spacetime filled with incoherent dust,
\begin{equation}
{\cal S}=\int d^4X (-{1 \over 2})|\gamma|^{1 \over 2}M \bigg(
\gamma^{\alpha \beta} U_{\alpha} U_{\beta} + 1\bigg).
\label{eq:BKaction}
\end{equation}
The four-velocity $U_{\alpha}$ of the dust is defined by its
decomposition in the co-basis ${Z^{K}}_{, \alpha}$,
\begin{equation}
U_{\alpha} = -T_{, \alpha} + W_i {Z^i}_{, \alpha}.
\label{eq:fourvel}
\end{equation}
The scalars $Z^K = (T,Z^i)$ are assumed to be four independent
functions of the spacetime coordinates. The values of the variables
$Z^i$ correspond to the co-moving coordinates of the dust particles, and the
value of the variable $T$ corresponds to the proper time measured along the
particle flow lines. The three spatial components $W_i$ of the
four-velocity in the dust frame $\{ Z^i \}$ and the multiplier M are all
state variables, whose physical interpretation follows from the
ensuing equations of motion[13].

\% The co-moving coordinates of
the dust particles and the proper time along the dust world-lines
are treated as canonical coordinates, so a privileged dynamical
reference frame and time foliation are introduced into
spacetime. Disregarding certain problems concerning the factor
ordering[1,2], the Dirac quantization of the coupled system provides an
improved phenomenological approach to the problem of time
in quantum gravity. The work of Brown and Kucha\v{r} is the starting
point of this thesis. 

\% {\Large $\bullet$} Self-Commuting Combinations: While
studying the canonical decomposition of the dust action, the authors
of [13] came across   
a weight-two scalar combination of the gravitational constraints,  
\begin{equation}
G(x):=\ham ^2(x)-g^{ij}\mom(x) \momj(x).
\label{eq:G}
\end {equation}
The Poisson brackets of $G(x)$ with itself vanish strongly. If $G(x)$
replaces the usual Hamiltonian constraint to form an equivalent 
set of constraints for {\it vacuum} general relativity,   
\begin{equation}
G(x)=0=\mom(x),
\end{equation}
a genuine Lie algebra is created. 

\% The new algebra takes the form  
\begin{eqnarray}
\{G(x),G(x')\}&=&0 , \,\label{eq:GG} \\
\{G(x), \mom (x')\}&=&G_{,i}(x)\delta (x,x')+2G(x)\delta_{,i}(x,x') , \,
\label{eq:GH}\\
\{\mom (x), \momj (x')\}&=&\momj(x)\delta_{,i}(x,x')-\xxpp . \label{eq:HH} 
\end {eqnarray} 
It corresponds to the semi-direct product of the Abelian algebra 
generated by $G(x)$, equation (\ref{eq:GG}), with the algebra of 
spatial diffeomorphisms 
$\ldiffsigma$ generated by $\mom(x)$, equation (\ref{eq:HH}). 
The Poisson bracket (\ref{eq:GH}) reflects the transformation
of $G(x)$, under $\diffsigma$, as a scalar density of weight two. 

\% A similar result was obtained by Kucha\v{r} and Romano[14]. They
coupled gravity to a massless scalar field
and extracted another weight-two scalar combination of the
gravitational constraints,    
\begin{equation}
\Lambda_{\pm}(x):=g^{1\over2}(x)\biggl(-\ham
(x)\pm\sqrt{G(x)}\biggr). \label{eq:L} 
\end {equation} 
These results are significant because in the presence of a genuine Lie algebra some of the problems associated with quantization can be
eliminated. This applies particularly to a group-oriented approach, where the Hilbert space of the quantum theory is constructed by studying the representations of a group of observables, that often include symmetries of the classical system[1]. The standard super-Hamiltonian and super-momentum constraints are such observables, but they do not form a Lie group since their closing relations under the Poisson bracket operations do not produce a genuine Lie algebra. The presence of a genuine algebra allows the definition of a group and, hence, the use of powerful techniques from group representation theory for the construction of the appropriate Hilbert space.  

\hskip -0.25in
{\Large $\bullet$} {The Weight-$\omega$ Equation:} The issue that
arises is whether these self-commuting combinations convey any general message
about the structure of canonical general relativity[14].
An advance towards understanding their nature was made by
Markopoulou[16]. She constructed a nonlinear partial differential
equation satisfied by scalar combinations of the gravitational
constraints that
close according to the Abelian algebra (\ref{eq:GG})-(\ref{eq:HH}). 

\% The main observation was that any scalar
density ${\cal W}_{\omega}$ of arbitrary-weight can 
be written in terms of simpler combinations of the constraints and of
the spatial metric, 
\begin{equation}
{\cal W}_{\omega}(x)=g^{\omega\over 2}(x)
W_{\omega}[h(x),f(x)],
\label{eq:skata}
\end{equation}
assuming that ${\cal W}_{\omega}$ is an ultra-local function
of them.  The parameter $\omega$
denotes the weight of the corresponding scalar densities. 
The basic combinations of the constraints are
defined by 
\begin{eqnarray}
&& h(x) := g^{-{1\over 2}}(x)\ham(x) ,
\label{eq:equh}
\\
&& f(x):= g^{-1}(x) g^{ij}(x)\mom(x)\momj(x) 
\label{eq:equf}
\end{eqnarray}
and transform as scalar densities of weight zero.
Notice that both $G$ and $\Lambda_{\pm}$ can be written in the form
(\ref{eq:skata}) for weight two.

\% The requirement that the Poisson brackets of ${\cal W}_{\omega}(x)$
should vanish strongly results in a differential equation for $W_{\omega}(x)$,
\begin{equation}
{ \omega \over 2} W_{\omega}(x) W_{{\omega}f}(x) = f(x)
{W_{{\omega}f}}^2(x) - {1\over 4}{W_{{\omega}h}}^2(x) ,
\label{eq:WPDE}
\end {equation}
where the notation $W_{{\omega}h}:={{\partial W_{\omega}} \over
{\partial h}}$ and $W_{{\omega}f}:={{\partial W_{\omega}} \over
{\partial f}}$ has been used.

\% Its general solution can be found exactly, and is given[16] by
\begin{eqnarray}
W_{\omega}\bigl[h, f, B(\alpha[h,f])\bigr]
& = & \pm
\Biggl[\biggl(h - {1\over 2}B'(\alpha[h,f])
\biggr) +\sqrt
{\biggl(h - {1\over 2} B'(\alpha[h,f])
\biggr)^2-f} \ \Biggr]
^{\omega\over 2} \nonumber \\
& & \times \exp\Biggl(B(\alpha[h,f])
+{\omega\over2}
{ {1\over2} B'(\alpha[h,f]) \over 
\sqrt{\bigl(h - {1\over 2}B'(\alpha[h,f])
\bigr)^2-f}}\Biggr).
\label{eq:aasol}
\end{eqnarray}
The form of the function $\alpha[h,f]$ is 
determined by solving algebraically the equation 
\begin{equation}
\alpha=-{\omega\over 4\sqrt{(h-{1\over 2}
B'(\alpha))^2-f}}
\label{eq:asol}
\end{equation}
for a given choice of $B(\alpha)$. Complex solutions for
$W_{\omega}(x)$ can exist.  

\hskip -0.25in
{\Large $\bullet$} {A Possibility:}
Expressions (\ref{eq:aasol}) and (\ref{eq:asol}) are based on algebraic
considerations, so their physical relevance is not clear. An insight
into their origin could be gained if they were shown to be related to
systems similar to the ones discussed in [14] and [14]. 
In particular, the actions for dust and for a massless scalar field
could arise as different versions of a 
general action, parametrized by an arbitrary function of one
variable. This possibility is supported by the fact that the general solution
(\ref{eq:aasol})-(\ref{eq:asol}) has a similar dependence upon such
a function. It is also compatible with the general properties of
first-order partial differential equations[28].

\% An obstacle to this construction is that the fields
used in [13] and in [14] are unequal in number. However, it can be shown
that the relevant results $G$ and $\Lambda_{\pm}$ depend only on the
form of the action and not on the number of the canonical fields. 
An action of a single field could therefore suffice, provided that it
is parametrized by an arbitrary function of one variable and
reduces to the form of the actions in [13] and [14] for particular choices of
this function. When coupled to gravity, it could provide the general
solution of equation (\ref{eq:WPDE}) for weight two.

\% In addition, if the weight-two procedure proved to be successful it
would be extended trivially to an arbitrary weight. This follows from 
a remarkable property of equation (\ref{eq:WPDE}). If
$W_{\omega}$ is a solution of weight $\omega$ then $W_{\omega'}$, 
defined by 
\begin{equation}
W_{\omega'} := {W_{\omega}}^{{\omega'}\over{\omega}} ,
\label{eq:llllll}
\end{equation}
is a new solution of weight ${\omega}'$. Notice that both
$\omega$ and $\omega'$ must be different from zero so that the
algorithm is well-defined and invertible.
\vskip 8cm
\hskip -0.25in
{\LARGE Chapter II-2. Treating The Algebra Algebraically.}
\vskip 1cm 
\hskip -0.25in
Equation (\ref{eq:WPDE}) is related to a generalized action 
that has the properties described above. As it stands, 
the equation does not make this connection clear, so an ansatz is used to 
convert it into an appropriate form. The ansatz is parametrized by the
weight $\omega$ appearing in equation (\ref{eq:WPDE}) and expresses
$W_{\omega}$ in terms of two functions $\ltilde$ and 
$\mtilde$. Like $W_{\omega}$ they are ultra-local functions of the
basic combinations $h$ and $f$ and they transform as scalar
densities of weight zero.      
The ansatz transforms the nonlinear equation (\ref{eq:WPDE}) into a
pair of coupled quasi-linear partial differential equations for
$\ltilde$ and $\mtilde$, called the ``linear'' equation.   

\hskip -0.25in
{\Large $\bullet$} {Preliminary Remarks About The $\omega$-Equation:} 
If any of the partial derivatives of $W_{\omega}$
is trivial, equation (\ref{eq:WPDE}) implies that $W_{\omega}$ is
either a function of $f$, alone, or a constant. In both cases, 
the information concerning the Hamiltonian constraint is lost. 
These special solutions have been excluded from the following
discussion, although the reasons for excluding them will arise later,
in chapters II-4 and II-5.

\% Recall, also, that equation (\ref{eq:WPDE}) allows the existence of complex
solutions which, usually, cannot be reconciled to the
idea of a physical system. This is particularly true here, since 
these solutions are required later to be positive definite. However, a complex
combination of the  gravitational constraints is not necessarily
complex-valued. For example, the weight-one solution 
$i\sqrt{{h}^2-f}$ is positive in a region of the phase
space where $f>{h}^2$. This particular region is not
accessible to vacuum general relativity, but may be so to a coupled
system. It is therefore preferable to accept all solutions of
equation (\ref{eq:WPDE}) at this stage and make the necessary amendments
later, in chapter II-4.     

\hskip -0.25in
{\Large $\bullet$} {The $\omega$-Ansatz And The Linear Equation:}
The one-parameter family of ``ansatzes'' has the following form, 
\begin{equation}
W_{\omega}[h,f]=\ltilde^{\omega\over 2}[h,f]
\Biggl(h-\mtilde[h,f]+
\sqrt{\bigl(h-\mtilde[h,f]\bigl)^2-f}
\Biggr)^{\omega\over 2}.
\label{eq:ansatz}
\end{equation}
Each $\omega$-ansatz transforms the corresponding $\omega$-equation
(\ref{eq:WPDE}). Both signs for the square root are permitted. This is
not denoted by a $\pm$ sign in order to keep the notation simple.

\% The square root in (\ref{eq:ansatz}) is denoted by the
letter $R$, and the square-bracket notation is dropped, 
\begin{equation}
\rtilde:=\sqrt{\biggl(h-\mtilde\biggl)^2-f}. 
\label{eq:root}
\end{equation}
The partial derivatives of $W_{\omega}$ are expressed in terms of
$\lambda$, $\mu$ and $R$ according to
\begin{eqnarray}
W_{{\omega}H}&=&
{\omega\over 2}\ltilde^{\omega\over
2}(h-\mtilde+\rtilde)^{\omega\over 2}\biggl( 
{1\over\ltilde}\lambda_h-
{1\over\rtilde}\mu_h+{1\over\rtilde} \biggr), 
\nonumber\\
W_{{\omega}F}&=&
{\omega\over 2}\ltilde^{\omega\over
2}(h-\mtilde+\rtilde)^{\omega\over 2}\Bigl( 
{1\over\ltilde}\lambda_f-
{1\over\rtilde}\mu_f-{1\over{2\rtilde(h-\mtilde+\rtilde)}}\Bigr).
\label{eq:vf}
\end{eqnarray}

\% When expressions (\ref{eq:ansatz})-(\ref{eq:vf}) are used in
equation (\ref{eq:WPDE}), this becomes:  
\begin{equation}
\biggl({1\over\ltilde}\lambda_f-
{1\over\rtilde}\mu_f\biggr)
\Biggl[-f \biggl(
{1\over\ltilde}\lambda_f-
{1\over\rtilde}\mu_f\biggr)
+{h-\mtilde\over\rtilde}\Biggr]+
{1\over4}\biggl(
{1\over\ltilde}\lambda_h-
{1\over\rtilde}\mu_h\biggr)
\Biggl[\biggl(
{1\over\ltilde}\lambda_h-
{1\over\rtilde}\mu_h\Bigr)
+{2\over\rtilde}\Biggr]=0. \label{eq:square}
\end{equation}
Noticeably, the arbitrary weight $\omega$ has been eliminated.
 
\% There exist four obvious solutions of equation (\ref{eq:square}),
corresponding to four different pairs of coupled quasi-linear 
equations for $\mtilde$ and $\ltilde$:  
\begin{eqnarray}
{1\over\ltilde}\lambda_f-
{1\over\rtilde}\mu_f=0&
{\rm and}& 
{1\over\ltilde}\lambda_h-
{1\over\rtilde}\mu_h=0,
\label{eq:a}\\
{1\over\ltilde}\lambda_f-
{1\over\rtilde} \mu_f=0  &
{\rm and}&     
{1\over\ltilde}\lambda_h-
{1\over\rtilde}\mu_h=-{2\over\rtilde},
\label{eq:b}\\
{1\over\ltilde}\lambda_f-
{1\over\rtilde}\mu_f
={h-\mtilde\over\rtilde}
      &
{\rm and}&     
{1\over\ltilde}\lambda_h-
{1\over\rtilde}\mu_h=0,
\label{eq:c}\\
{1\over\ltilde}\lambda_f-
{1\over\rtilde}\mu_f
={h-\mtilde\over\rtilde}    &
{\rm and}&      
{1\over\ltilde}\lambda_h-
{1\over\rtilde}\mu_h=-{2\over\rtilde}.
\label{eq:d}
\end{eqnarray}

\% Given a $\mtilde$, any of the above pairs of
equations can be solved for the corresponding $\ltilde$, provided that the
system of the two partial equations for $\ltilde$ is not
contradictory. Then, the $\omega$-ansatz (\ref{eq:ansatz}) can be used to
produce solutions $W_{\omega}$ of the corresponding weight.
An equivalent procedure can be followed if a $\lambda$ is given initially. 

\% Furthermore, the above pairs are equivalent, in the sense that for
each weight they all lead to the same family of solutions of the
non-linear equation (\ref{eq:WPDE}). The proof can be found in Appendix A.
The most symmetric of the equivalent pairs, equation (\ref{eq:a}), is
then singled out. It is called the ``linear'' equation and is the
one related directly to the action principle. Its
solutions must 
be compared with the general solution of each $\omega$-equation
(\ref{eq:WPDE}). The surprising result is that, for all weights,
equations (\ref{eq:WPDE}) and (\ref{eq:a}) are equivalent. 

\% In particular, given a solution ${W_{\omega}}$ of the weight-$\omega$
equation, there exist unique functions  
\begin{equation}
{\bar\lambda}={-{2 {W_{\omega}}^{2\over\omega} {W_{{\omega}f}}} \over
{W_{{\omega}h}}} , 
\label{eq:C}
\end{equation}
\begin{equation}
{\bar\mu}=h + { {W_{{\omega}f}} \over 
{W_{{\omega}h}}} f + {1\over4} {W_{{\omega}h}
\over W_{{\omega}f}},
\label{eq:D}
\end{equation}
\begin{equation}
{\bar R}= { {W_{{\omega}f}} \over 
{W_{{\omega}h}}} f - {1\over4} {W_{{\omega}h}
\over W_{{\omega}f}},
\label{eq:E}
\end{equation}
that satisfy the linear equation and lead to ${W_{\omega}}$ through the
corresponding $\omega$-ansatz. The derivation of equations
(\ref{eq:C})-(\ref{eq:E}) can be found in Appendix B.  
The over-bar symbol is a reminder of the uniqueness of these
expressions. To be precise, the expression for $\bar{\lambda}$ 
is not exactly unique but holds up to an $\omega\over 2$ power of
unity. Notice that equations (\ref{eq:C})-(\ref{eq:E}) are
well-defined in general, because $W_{{\omega}h}$ and $W_{{\omega}f}$
have been required to be non-trivial functions of $h$ and $f$. Of
course, the phase space 
should be restricted to those regions where also the numerical
values of $W_{{\omega}h}$ and $W_{{\omega}f}$ are non-trivial. 
\vskip 7cm
\hskip -0.25in
{\LARGE Chapter II-3. Actions Leading To The Algebra.}
\vskip 1cm
\hskip -0.25in 
{\Large $\bullet$} The relevant action involves a scalar
field with a non-derivative coupling to gravity and, initially, two arbitrary
functions of a Lagrange multiplier. It is the simplest action that
possesses the parametrization by an arbitrary 
function of one variable and includes the actions in [13] and [14] as
sub-cases. The required parametrization arises only after the
elimination of the 
non-dynamical multiplier.
Because there is no detailed reference to an underlying
physical interpretation the following construction
should be viewed mainly as a mathematical one. 

\hskip -0.25in
{\Large $\bullet$} {The Scalar Field Action:} The action functional
$S^{\phi}$ is introduced as 
\begin{equation}
S^{\phi}[\phi, M, \gabu]=\int d^4X|\gamma|^{1\over2}\Biggl(
{1\over2}\lambda(M)\gabu\phi_{,\alpha}\phi_{,\beta}+\mu(M)\Biggr).
\label{eq:41}
\end{equation}
The dependence of the fields on the spacetime
point $X$ is not written explicitly. The functions $\lambda(M)$ and
$\mu(M)$ are continuous functions of the Lagrange multiplier $M$.
They are fixed (i.e., non-canonical) but otherwise arbitrary.
Since the scalar field has to be present in the action functional,
$\lambda(M)$ is required to be different from zero. No such
restriction is imposed on $\mu(M)$. Notice that the notation for the
two functions of the multiplier reflects the notation used in chapter II-2.

\% Keeping the conventional dimension of inverse length for the scalar field,
a consistent attribution of dimensions to the various terms
appearing in (\ref{eq:41}) is the following:
\begin{equation}
[\phi]=L^{-1},\qquad [M]=[\lambda(M)]=L^0=1,\qquad [\mu(M)]
=L^{-4}. 
\end{equation}
This means that $\mu(M(X))$ should be considered as a function of the
multiplier $M(X)$ scaled by a constant scalar function $C(X)$, 
\begin{equation}
\mu(M(X))=C(X)\rho(M(X)).
\end{equation}
The dimensions of the new functions are $[C]=L^{-4}$ and
$[\rho(M)]=L^0=1$. For simplicity, appropriate 
units can be assumed so that $C(X) = 1$, in which case $\rho$ may be
identified with $\mu$.

\hskip -0.25in
{\Large $\bullet$} {The Hamiltonian Analysis Of The Coupled System:}
The scalar field action (\ref{eq:41}) is coupled to 
the gravitational Einstein-Hilbert action $S^{gr}[\gabd]$, 
\begin{equation}
S^{gr}[\gabd]=\int d^4X|\gamma|^{1\over2} R[\gabd].
\end{equation}
The canonical analysis of the total action,
\begin{equation}
S^T:=S^{gr}+S^{\phi},
\end{equation}
results in the coupled constraints[9]
\begin{eqnarray}
{\hamtot} &:=&{\ham} +\hamphi =0,
\label{eq:hamt}\\
\momtot &:=&{\mom}+\momphi =0.
\label{eq:momt}
\end{eqnarray}
Their form is common to any theory with a non-derivative coupling to
gravity[5]. 
 
\% The gravitational parts of the constraints, ${\ham}$ and
${\mom}$, are identical to the constraints of vacuum general
relativity written out in equations (\ref{eq:ham}) and
(\ref{eq:mom}). The scalar field contributions $\hamphi$ and $\momphi$
are given by
\begin{eqnarray}
\momphi &=&\pi\phi_{,i} ,
\label{eq:mphi}\\
\hamphi &=&g^{1\over2}\Biggl(-{1\over2}{\pi^2\over g\lambda(M)}-\mu(M)
-{1\over2}{\lambda(M)\over\pi^2}g^{ij}\momphi\momjphi\Biggr).
\label{eq:hphi}
\end{eqnarray}
Notice that the kinetic energy of the scalar field has to be
positive. Equation (\ref{eq:hphi}) then implies that $\lambda(M)$ must
be negative. On the other hand, $\mu(M)$ appears as a cosmological
constant in equation (\ref{eq:hphi}) so it may take any real value. 

\hskip -0.25in
{\Large $\bullet$} The Two Equations For $M$ And $\pi$:
At this stage, the total action $S^T$ is varied with respect to the
multiplier. The latter appears only in the 
super-Hamiltonian for the scalar field, so the variation results in
the following condition 
\begin{equation}
{d\hamphi\over d M}=0.
\label{eq:papares}
\end{equation}
Equation (\ref{eq:papares}) can be written equivalently as  
\begin{equation}
{1\over2}{\pi^2\over g\lambda^2(M)}\lambda'(M)
-\mu'(M)-{1\over 2}{1\over\pi^2}\lambda'(M)g^{ij}\momphi\momjphi=0,
\label{eq:zero}
\end{equation}
where $\lambda'(M)$ and $\mu'(M)$ denote the total
derivatives of $\lambda(M)$ and $\mu(M)$ with respect to $M$.  

\% The constraints (\ref{eq:hamt}) and (\ref{eq:momt}) can now be used
to re-express equations (\ref{eq:hphi}), (\ref{eq:zero}) in terms of
the gravitational contributions to these constraints,
\begin{eqnarray}
{1\over2}{\pi^2\over g\lambda(M)}+{1\over2}{\lambda(M)\over\pi^2}
g^{ij}\mom\momj &=& h-\mu(M),
\label{eq:13}\\
{1\over2}{\pi^2\over g\lambda^2(M)}\lambda'(M)
-{1\over2}{1\over\pi^2}g^{ij}\mom\momj\lambda'(M)&=& \mu'(M).
\label{eq:14}
\end{eqnarray}
The quantity $h$ is the scalar density of weight zero defined in
equation (\ref{eq:equh}).
The aim is to solve equations (\ref{eq:13})-(\ref{eq:14}) for $\pi$ and
$M$ in terms of $\ham$ and $\mom$. Because the solution depends on the
actual form of the derivatives, some special cases must be considered
separately. 

\hskip -0.25in
{\Large $\bullet$} Solving The Two Equations For $M$ And $\pi$. The
General Case: This occurs when both the
derivatives of $\lambda$ and $\mu$ are non-trivial,
\begin{equation}
\lambda'(M)\neq 0\qquad\mu'(M)\neq 0.
\label{eq:biga}
\end{equation}
If this condition holds, equation (\ref{eq:14}) can be multiplied by
$\lambda(M)/\lambda'(M)$, resulting in the equivalent relation
\begin{equation}
{1\over2}{\pi^2\over g\lambda(M)}-{1\over2}
{\lambda(M)\over\pi^2}g^{ij}\mom\momj=
{\mu'(M)\lambda(M)\over\lambda'(M)}.
\label{eq:15}
\end{equation}

\% Equations (\ref{eq:13}) and 
(\ref{eq:15}) must be solved for $\pi$ and $M$ in terms of the gravitational
contributions to the constraints. This can be done by adding and
subtracting (\ref{eq:13}) and (\ref{eq:15}), and then cross-multiplying the
resulting equations to eliminate the field momenta. An algebraic
equation arises that determines the 
multiplier $M$ as a function of $\ham$ and $\mom$, alone,
\begin{equation}
{\mu'(M)\lambda(M)\over\lambda'(M)} =
\sqrt{\biggl(h-\mu(M)\biggr)^2-f}.
\label{eq:16}
\end{equation}
The quantity $f$ is the one defined in equation
(\ref{eq:equf}). Both signs for the square root are allowed in
(\ref{eq:16}), although this is not denoted explicitly.

\% The corresponding expression for the field momentum as a function
of $\ham$ and $\mom$, alone, is given by
\begin{eqnarray}
{1\over(g^{{1\over2}})^2}\pi^2[\ham\mom] = \lambda(M[\ham,\mom]) \times
\nonumber
\\
\times \Biggl(h-\mu(M[\ham,\mom])+
\sqrt{\Bigl(h-\mu(M[\ham,\mom])\Bigr)^2-f} \Biggr).
\label{eq:17}
\end{eqnarray}
Equations (\ref{eq:16}) and (\ref{eq:17}) are the required solutions of
the original system of equations (\ref{eq:13}) and (\ref{eq:14}).

\% The observation is that the solutions $M[\ham,\mom]$ and
$\pi^2[\ham,\mom]$ can be written solely in terms of the scalar combinations
$h$ and $f$. This follows directly from the actual form
of equations (\ref{eq:16}) and (\ref{eq:17}). 
In addition, $\lambda$ and $\mu$ can be regarded as sole 
functions of $h$ and $f$ as well, according to 
\begin{equation}
\ltilde[h,f]:=\lambda(M[\ham,\mom]),
\qquad{\rm and}\qquad
\mtilde[h,f]:=\mu(M[\ham,\mom]).
\label{eq:18}
\end{equation}
As a result, equation (\ref{eq:17}) can be put into the equivalent form 
\begin{equation}
{1\over(g^{{1\over2}})^2}\pi^2[h,f]=
{\ltilde}[h,f]
\Biggl({h}-{\mtilde}[{h},{f}]+ R[{h},
{f}] \Biggr).
\label{eq:19}
\end{equation}
Notice that the definition (\ref{eq:root}) for the function $R$ has
been used in (\ref{eq:19}).

\% If the above expression is raised to the
power of $\omega\over2$ it can be recognized 
as the $\omega$-ansatz written out in equation (\ref{eq:ansatz}).  It
satisfies the differential  equation 
(\ref{eq:WPDE}) provided that $\ltilde[h,f]$,
$\mtilde[h,f]$ and the square root in equation
(\ref{eq:19}) satisfy the common to all weights linear
equation (\ref{eq:a}). The following argument shows that this is true indeed. 
 
\% Since (\ref{eq:16}) is an algebraic equation for $M$, it
must hold identically when written in terms of an actual solution 
$M[h,f]$. Therefore, it becomes a differential equation for
$\lambda(M[h,f])\equiv\ltilde[h,f]$ 
and $\mu(M[h,f])\equiv\mtilde[h,f]$,
regardless of the particular form of the functions $\lambda(M)$ and
$\mu(M)$. Furthermore, 
equation (\ref{eq:16}) makes certain that its solution 
$M[h,f]$ satisfies the conditions
\begin{equation}
M_h\neq 0 
\qquad{\rm and}\qquad
M_f\neq 0 .
\label{eq:20}
\end{equation}

\% Equation (\ref{eq:16}) can be then multiplied by 
$M_h$ and $M_f$, resulting in the following pair of partial
differential equations for $\mu[h,f]$ and
$\lambda[h,f]$, 
\begin{equation}
{1\over\ltilde}\lambda_h-
{1\over\sqrt{(h-\mtilde)^2-f}}
\mu_h=0
\qquad{\rm and}\qquad
{1\over\ltilde}\lambda_f-
{1\over\sqrt{(h-\mtilde)^2-f}}
\mu_f=0.
\label{eq:21}
\end{equation}
This is precisely the linear equation (\ref{eq:a}). 
Notice that the gravitational phase space must be restricted to
the regions where the quantity
inside the square root as well as the whole right side of
equation (\ref{eq:19}) are positive.

\% {\Large $\bullet$} Solving The Two Equations For $M$ And $\pi$. The
Special Cases: Returning to equations (\ref{eq:13}) and (\ref{eq:14}),
the special cases are considered. There are three possibilities:  

\hskip -0.25in
(i) The Kucha\v{r}-Romano family. This occurs when both the derivatives
of $\lambda(M)$ and $\mu(M)$ are trivial,
\begin{equation}
\lambda'(M)=0\qquad{\rm and}\qquad\mu'(M)=0.
\label{eq:lzmz}
\end{equation}
This means that $\lambda(M)$ and $\mu(M)$ are constant functions 
\begin{equation} 
\mu(M)=C_1 \; \; \; \; \; \; \; \; \;  \lambda(M)=C_2 .
\end{equation}
They are required to be real and negative respectively.

\% In this case, there is no multiplier present in the total action.
Equation (\ref{eq:14}) is then satisfied trivially, both sides being equal to
zero. Accordingly, the coupled system of equations (\ref{eq:13}),
(\ref{eq:14}) for 
$M$ and $\pi^2$ reduces to the single equation (\ref{eq:13}). The
latter expresses $\pi^2$ directly as a function of $h$ and
$f$, according to
\begin{equation}
{\pi^2\over (g^{1/2})^2}=C_2\Biggl(\bigl(h-C_1\bigr)
+\sqrt{\bigl(h-C_1\bigr)^2-f}\Biggr).
\label{eq:25}
\end{equation}

\% When raised to an $\omega\over2$ power, equation (\ref{eq:25}) is
recognized as the $\omega$-ansatz. The required identification is 
\begin{eqnarray}
\ltilde[h,f]&\equiv&C_2 ,\\
\mtilde[h,f]&\equiv&C_1 .
\end{eqnarray}
The linear equation is satisfied trivially for these 
$\lambda[h,f]$ and $\mu[h,f]$,
and therefore expression (\ref{eq:25}) provides further solutions of the
differential equation (\ref{eq:WPDE}).  They are all required to be
positive. Notice that this case reduces to the Kucha\v{r}-Romano 
combination under the identification $\omega=2$, $C_1=0$ and
$C_2=-1$. 

\hskip -0.25in
(ii) The Pseudo-multiplier. This case occurs when
\begin{equation}
\lambda'(M)=0\qquad{\rm and}\qquad\mu'(M)\neq 0.
\label{eq:lzmnz}
\end{equation}
This implies that $\lambda$ is a constant function $C_2$, which is
required to be negative. 
Equation (\ref{eq:14}) then becomes 
\begin{equation}
\mu'(M)=0.
\label{eq:mz}
\end{equation}
Notice that equations (\ref{eq:lzmnz}) and (\ref{eq:mz}) 
are not 
contradictory. The first implies that $\mu(M)$ is a non-trivial function
of $M$, while the second is an algebraic equation for
determining $M$ provided that the non-trivial function $\mu(M)$ is given.

\% Any solution of equation (\ref{eq:mz}) can yield only a numerical
value for $M$. Equivalently, $M$ is not a proper 
multiplier but merely ``fixes itself a value''. However, equation
(\ref{eq:13}) can still be solved for $\pi^2$, leading to 
\begin{equation}
{\pi^2\over (g^{1/2})^2}=-C_2\Biggl(\bigl(h-C_1\bigr)
+\sqrt{\bigl(h-C_1\bigr)^2-f}\Biggr).
\end{equation} 
The constant function $C_1$ corresponds to the real numerical value of
$\mu(M)$ after the elimination of the pseudo-multiplier.  
Therefore, case (ii) is essentially identical to case (i). 

\hskip -0.25in
(iii) The Null-Vector Family. This occurs when 
\begin{equation}
\lambda'(M)\neq0\qquad{\rm and}\qquad\mu'(M)= 0.
\label{eq:lnzmz}
\end{equation}
The function $\mu(M)$ is equal to a constant function $C_1$, which is
required to be real. Equation (\ref{eq:14}) then becomes
\begin{equation}
\Biggl({\pi^2\over g\lambda(M)}-{\lambda(M)\over\pi^2}
g^{ij}\mom\momj\Biggr){\lambda'(M)\over\lambda(M)}=0.
\label{eq:30}
\end{equation}
As a result, $M$ either takes a real numerical value---thus
producing exactly the same combinations as in special cases (i) and
(ii)---or satisfies the condition 
\begin{equation}
{\pi^2\over g\lambda(M)}-{\lambda(M)\over\pi^2}g^{ij}\mom\momj=0.
\label{eq:second}
\end{equation}

\% When equation (\ref{eq:second}) is combined with equation
(\ref{eq:13}), it leads to a 
constraint between the two variables $f$ and $h$, 
\begin{equation}
f=(h-C_1)^2.
\label{eq:ctwo}
\end{equation}
This is the only case where ${\pi}^2$ drops out of its defining
equations (\ref{eq:13})-(\ref{eq:14}). However, equation (\ref{eq:ctwo}) still
leads to a self-commuting combinations  of weight zero when solved in
terms of the constant $C_1$. 
An example is when $\lambda(M)=M$ and $\mu(M)=0$. It can be
interpreted as a coordinate condition on $\gabd$ such that the
non-dynamical field $\phi$ becomes null,  
\begin{equation}
\gabu\phi_{,\alpha}\phi_{,\beta}=0.
\end{equation}
A corresponding physical system can be interpreted as null-dust.
\vskip 9cm
\hskip -0.25in
{\LARGE Chapter II-4. The Inverse Procedure.}
\vskip 1cm 
\hskip -0.25in
{\Large $\bullet$} Actions For Given Solutions: It has been shown that
for all choices of $\lambda(M)$ 
and $\mu(M)$ the action functional (\ref{eq:41}) yields solutions of 
the $\omega$-equation. The converse statement is also true. Every solution of
equation (\ref{eq:WPDE}) can be derived from an action principle of
the form (\ref{eq:41}). For the present purposes it suffices
that only a limited part of this statement is proved. 

\% In particular, it has to be shown that there exist some functions
$M[h,f]$, $\mu(M)$ and $\lambda(M)$ that satisfy the conditions 
\begin{eqnarray}
\mu\Bigl(M[h,f]\Bigr)&=&\bar\mu[h,f],\label{eq:x}\\
\lambda\Bigl(M[h,f]\Bigr)&=&{\bar\lambda}[h,f],
\label{eq:y}\\
\sqrt{\Bigl(h-\mu\bigl(M[h,f]\bigr)\Bigr)^2-
f}&=&\bar R[h,f],\label{eq:z}
\end{eqnarray}
provided that $\bar{\mu}$, $\bar{\lambda}$ and the
corresponding $W_{\omega}$ are real, negative and positive-valued
respectively. 
The over-bar symbol indicates the uniqueness of these
expressions for each given choice of $W_{\omega}[h,f]$, as
explained in chapter II-2.

\% Notice that if a solution of equations
(\ref{eq:x})-(\ref{eq:z}) exists then the problems mentioned in
chapter II-2 concerning trivial and complex combinations no
longer apply, since they are eliminated by the assumption of existence
of such a solution. This justifies the 
choices made at the beginning of chapter II-2 concerning the
subsequent treatment of these combinations. 

\% Returning to equations (\ref{eq:x})-(\ref{eq:z}), it may be observed
that the linear equation (\ref{eq:a}) makes certain that
$\bar{\lambda}[h,f]$ and $\bar{\mu}[h,f]$ can
either be functions of both $h$ and $f$ or constants. If
this is not the case, the system of the two partial differential  
equations in (\ref{eq:a}) is self-contradictory. Therefore, there are two
cases that must be considered separately:

\% (i) Constant functions:
If $\bar\mu$ and $\bar\lambda$ are negative and real constants, $C_1$
and $C_2$ respectively, then the required solution
can be found without needing to specify the form of the function
$M[h, f]$. This follows from the results 
concerning the special cases (i)-(iii) in chapter II-3.
Specifically, the required functions $\lambda(M)$ and $\mu(M)$ can be
identified as
\begin{equation}
\lambda(M)=\bar\lambda[h,f]=C_2 ,
\qquad{\rm and}\qquad
\mu(M)=\bar\mu[h,f]=C_1 .
\end{equation}
The above relations satisfy equation (\ref{eq:z}) for an 
appropriate choice of sign of the square root and, therefore, the problem of 
finding an action functional is solved.

\% (ii) Non--trivial functions:
When $\bar\mu[h,f]$   
and $\bar\lambda[h,f]$ are, respectively, real and negative non-trivial functions of $h$ and $f$ the situation is
more complicated. The observation is that the Jacobian of
$\bar\mu[h,f]$ and $\bar\lambda[h,f]$ with
respect to $h$ and $f$ is identically zero. This follows
directly from the form of the linear equation (\ref{eq:a}), and
implies that $\bar\mu[h,f]$ and
$\bar\lambda[h,f]$ are functionally dependent. As a
result, there exist at least local regions of the gravitational phase
space where $\bar\mu[h,f]$ can be solved as a 
unique and real function of $\bar\lambda[h,f]$, 
\begin{equation}
\bar{\mu}[h,f]=\kappa\bigl(\bar{\lambda}[h,f]\bigr).
\label{eq:skoto}
\end{equation} 
The reason that the above is true is because
$\bar\lambda$ and $\bar\mu$ are real-valued.    
Since $\bar\mu$ and $\bar\lambda$ depend only upon the specific
solution $W_{\omega}[h,f]$, the same is true for the
uniquely defined $\kappa$. 

\% Equations (\ref{eq:x}) and (\ref{eq:y}) then reduce to 
\begin{eqnarray}
\mu\Bigl(M[h,f]\Bigr)&=&\kappa\big(\bar\lambda[h,f],
\big)\label{eq:x2}\\
\lambda\Bigl(M[h,f]\Bigr)&=&{\bar\lambda}[h,f],
\label{eq:y2}
\end{eqnarray}
which admit the obvious solution
\begin{equation}
\mu(M)=\kappa\bigl(\lambda(M)\bigr).
\label{eq:ppp}
\end{equation}
This is the solution to the problem. 
 
\hskip -0.25in
{\Large $\bullet$} {An Application:}
As an example, consider the combination $G=h^2-f$.
Using equations (\ref{eq:D}) and (\ref{eq:C}),
the unique expressions $\bar\mu$ and $\bar\lambda$ can be found as 
\begin{equation}
\bar\mu={1\over2}\Biggl(h-{f \over h}\Biggr)\
\qquad{\rm and}\qquad
\bar\lambda=\Biggl(h-{f \over h}\Biggr).
\label{eq:halfs}
\end{equation}

\% The set of values of $h$ and $f$ for which $G$, $\bar\mu$ 
and $\bar\lambda$ are positive, real and negative respectively is given 
by the inequalities $h^2 > f$ and $h < 0$. 
For this range of values, the partial derivatives of $G$, $\bar\mu$,
$\bar\lambda$ and $\bar R$ are well defined, so the same applies to
the whole procedure in chapters II-2 and II-3. 

\% The unique real function $\kappa$ is then determined by the relation 
\begin{equation}
\kappa(\bar\lambda[h,f])={1\over2}\bar\lambda[h,f]
\end{equation}
and solves the problem.
In particular, any negative-valued, real or complex,  function
$\lambda(M)$ is fine, provided that
$\mu(M)$ satisfies the following equation,  
\begin{equation}
\mu(M)=\kappa\bigl(\lambda(M))={1 \over 2}\lambda(M). 
\end{equation}
Notice that this means that $G$ can be derived from a variety of
actions of the form (\ref{eq:41}).
\vskip 9cm 
\hskip -0.25in
{\LARGE Chapter II-5. The Algebra In Vacuum Gravity.}
\vskip 1cm 
\hskip -0.25in
{\Large $\bullet$} The above procedure is part of a
phenomenological approach towards the interpretation of  the Brown-Kucha\v{r}
algebra. The main conclusion is that the  
association of the combinations $G$ and $\Lambda_{\pm}$ with
matter is a generic property of the solutions 
(\ref{eq:aasol})-(\ref{eq:asol}). Considering the diverse origins of
the calculations in chapters 
II-2 and II-3, the coincidence of the corresponding results is remarkable, and
suggests that the role of matter in general relativity deserves to be
investigated further.      

\hskip -0.25in
{\Large $\bullet$} {The Time Evolution Generated By The Solutions:}
On the other hand, little progress has been made towards understanding the
relevance of the solutions (\ref{eq:aasol})-(\ref{eq:asol}) in 
vacuum gravity. The necessity to stay away from the constraint
surface of this theory, and possibly return to it when all
calculations have been finished, is evident throughout the previous chapters. 
This is particularly clear for the combinations $G$ and 
${\Lambda}_{\pm}$ since the time evolution they generate is,
respectively, zero and ill-defined on the constraint surface of vacuum general
relativity. The question that arises is whether this property
applies for the whole family of solutions (\ref{eq:aasol})-(\ref{eq:asol}). 

\% Consider a solution $W$ that belongs to this family. The time
evolution it generates when acting on an arbitrary local functional $F$ of the
gravitational canonical variables is the following,  
\begin{equation}
\{ F , W[h,f]' \} = \{ F , h' \} {W_h}' + \{ F , f' \} {W_f}'  . 
\label{eq:ill}
\end{equation}
The primed quantities are evaluated at the spatial point $x'$.

\% A minimum prerequisite for the use of these $W$s in the vacuum
theory is that they 
should create a constraint surface that is locally equivalent to the
usual one. In particular, under the 
replacement of $h$ by ${\cal W}$, the conditions 
$W\simeq 0$ and $f\simeq 0$ must imply the conditions
$h\simeq 0$ and $f\simeq 0$, and vice versa. Notice that
the constraint $f \simeq 0$ is equivalent to the usual
constraint $\mom \simeq 0$ due to the positivity of the spatial metric.
Equation (\ref{eq:WPDE}) then implies that $W_h$ must vanish weakly
on the constraint surface of the vacuum theory, 
\begin{equation}
W_h \simeq 0 , 
\label{eq:ill2}
\end{equation}
provided that $W_h$ and $W_f$ are both well-defined.
No restriction is imposed on the value of $W_f$ on the constraint surface.

\% When condition (\ref{eq:ill2}) is substituted in the evolution equation
(\ref{eq:ill}), the first term on the right side does not contribute because it
vanishes weakly. However, the same is true for the second term
assuming that $W_f$ is well defined. This follows from the fact 
that $f$, defined by equation (\ref{eq:equf}), is quadratic in
the super-momenta $\mom$ so that  
\begin{equation}
\{ F , f' \} \; = \; \{ F , {1 \over g'} {g^{ij}}' \} \; {\mom}' \; {\momj}'
\; + \; 2 \; \{ F , {\mom}' \} \;  {1 \over g'} {g^{ij}}' \; {\momj}'
\;  \simeq 0 
\label{eq:ill3}
\end{equation}
when the constraint $\mom \simeq 0$ is imposed. Therefore, the time evolution
associated with any well-defined solution of equation (\ref{eq:WPDE})
is trivial.

\% The inadequacy of the family of solutions
(\ref{eq:aasol}-\ref{eq:asol}) in vacuum gravity provides a further
justification of the comment made in chapter II-2 concerning the
exclusion of the trivial solutions. These solutions were 
excluded because they are not related to the action (\ref{eq:41}) and,
also, because the time evolution they generate in the vacuum theory is
trivial. 

\hskip -0.25in
{\Large $\bullet$} {Finding Vacuum Solution Of The Algebra:} 
Although the solutions
(\ref{eq:aasol})-(\ref{eq:asol}) are not regular in the sense of
Dirac[11], the possibility that they are used in a quantum theory of vacuum
gravity cannot be excluded. The combination $G$, for example, leads to
a quadratic ``Wheeler-DeWitt'' equation when imposed as a restriction
on the quantum states of the system, and no mention of its partial
derivatives is required. Despite its quadratic character, the new
equation could then result 
in an overall simplification in the quantum theory on account of its
property of generating a genuine Lie algebra.

\% This possibility should be considered in more detail. Recall that the
conditions $W_{\omega} \simeq 0$ and $h 
\simeq 0$ must imply each other when the constraint $f=0$
is imposed. However, depending on the choice of sign for the square
root in equation (\ref{eq:ansatz}), the constraint
$W_{\omega} \simeq 0$ may be satisfied identically when
$f \simeq 0$. If this is the case it cannot enforce the necessary for
equivalence Hamiltonian constraint. Conveniently, the sign-unambiguous
expression for ${\bar R}$ in equations (\ref{eq:C})-(\ref{eq:E}) makes
certain that this is not the case. This is because 
${\bar R}$ and $h-\bar{\mu}$ have the same sign when $f
\simeq 0$, which means that $W_{\omega}$ in equation (\ref{eq:ansatz}) does not
become identically trivial.

\% Regardless of whether the solutions
(\ref{eq:aasol})-(\ref{eq:asol}) can be used in a quantum
theory of vacuum gravity, the issue that matters is whether
alternative combinations that commute with themselves and lead to a
well-defined dynamical evolution can exist in the vacuum theory. 
A further insight into the origin of the new
algebra is required, and this can be attained through
finding a geometric interpretation. Significant progress is made by 
comparing the new algebra with the Dirac one. The geometric interpretation
of the latter is known, and the method for finding its physically
relevant solutions is also available. 

\% The geometry behind the Dirac algebra was recognized by
Teitelboim[19], while the procedure for
passing to its physical representations was developed by Hojman,
Kucha\v{r} and Teitelboim in the derivation of 
geometrodynamics from first principles[20]. 
The attempt to adapt their method to the requirements of the new algebra 
results in a re-examination of the equal-time formalism and
implies the need for ``classical histories''. The issue of the 
interpretation of the new algebra is set aside for the moment, and is discussed again later, in part V.

\pagebreak

\begin{center}
{\LARGE Part III:}
\vskip 9cm
{\LARGE Introducing Classical Histories}
\end{center}

\pagebreak

\% {\LARGE Chapter III-1. Motivation.}
\vskip 1cm
\hskip -0.25in
{\Large $\bullet$} Description:
The issue that is discussed in the following two parts
concerns the transition from a general canonical Hamiltonian of the
form $NH + {N^i} {H_i}$ to a specific canonical representation. The
form of the Hamiltonian is general 
enough to incorporate a variety of canonical field
theories, including general relativity. The information that
distinguishes one theory from another comes solely through the choice
of the canonical variables. The interpretation of the
functions $N$ and $N^i$ is left arbitrary.

\% Some of this has been discussed already by Hojman, Kucha\v{r} 
and Teitelboim[20] who recovered  geometrodynamics
and other canonical representations of covariant theories from an algorithm involving a few
plausible postulates.  
However, as stated by the authors themselves, the reduction of these
postulates to the minimum was not attempted and some  
redundancy was left in the system. A few redundant
requirements were pointed out at the end of their paper but, still, the exact relationship between the remaining postulates was not clarified and a further reduction seemed to be possible.

\% It is shown below that the complete set of postulates in [20] can be derived from the requirement that the canonical Hamiltonian is of the form $NH + {N^i} {H_i}$. The only other input that is needed in order that the most general canonical representation of $H$ and $H_i$ be found is the actual choice of the canonical variables. This choice involves some additional assumptions which are pointed out in a following section.

\% Besides, just as interesting as this result is the ensuing observation
that the history formalism seems to be superior to its standard (equal-time) counterpart. Whether this can be established as a general theorem, or not, depends on whether it is possible to find a {direct} link between the equal-time and the history approaches. What {\it is} established, however, is the fact that the spacetime meaning and connection of the postulates of Kucha\v{r} {\it et al} cannot be clarified in the standard formalism. The advantages of the history formalism over the standard approach are therefore genuine, at least when discussing {\it spacetime} issues. This is enough to suggest a history approach to quantum gravity. 

\% To be more precise, the original aim of Kucha\v{r} {\it et al} was to use postulates that depend exclusively on the three-dimensional hyper-surface. It is only for this reason that the relationship between their individual postulates is not clear. If the assumption of a surrounding spacetime (which is already implicit in their arguments) is used explicitly as an additional postulate then the meaning and connection of all the remaining postulates is clarified. However, the clarification of the postulates cannot be achieved in a formalism based on equal-time Poisson brackets.

\% This is because, in an equal-time formalism, Poisson brackets that involve the time derivatives of
the canonical variables cannot be defined; at least, not without the
addition of further structure. Seen from a spacetime perspective, however, these brackets ought to be treated in an equivalent way, in
which case they would give additional information about the theory's
kinematics. In the equal-time formalism the missing information is recovered precisely by the additional postulates imposed in
[20]; most notably that of the Dirac algebra.   
On the other hand, the present approach is based on a Hamiltonian 
formalism whose phase space includes the fields at general times;
i.e., is defined over the space of {\it classical histories}. The
correspondence with the spacetime picture is therefore exact
and the reduction of the postulates comes as a direct consequence.

\% The discussion in this section is organized as follows. In the remaining of chapter III-1 the existing work on the subject is reviewed, and the
main issues are emphasized. In chapter III-2 the Hamiltonian formalism
defined over the space of 
classical histories is introduced. It incorporates both constrained and
unconstrained systems and, at least for the issues of interest, is a
simpler alternative to the Dirac method. In chapter III-3 the history
formalism is employed to transform the postulated form  
$N H + N^i H_i$ of the canonical Hamiltonian to a set of kinematic
conditions on the canonical generators. These conditions define the {\it evolution} postulate which, together with the additional assumption of a surrounding spacetime, is then used in part IV to derive the canonical representations. 

\hskip -0.25in
{\Large $\bullet$} {The Dirac Algebra And The Principle Of Path Independence:}
The super-Hamiltonian and the super-momentum of vacuum general
relativity are not the only canonical generators that close according to the
Dirac algebra. The latter is satisfied by the canonical
generators of a parametrized field theory and, in a modified
form, by the generators of any field theory that is not parametrized.

\% Its universality implies that the Dirac algebra is connected
with a geometric property of spacetime that is
independent of the specific dynamics of the canonical theory. 
The fact that the Dirac algebra is a kinematic consistency 
condition was shown by Teitelboim[19] who derived it from a 
geometric argument corresponding to the integrability of
Hamilton's equations. This consistency argument---termed by Kucha\v{r}[29] ``the
principle of path independence of the dynamical evolution''---ensures
that the change in the canonical variables during the evolution 
from a given initial surface to a given final surface is independent
of the particular sequence of intermediate surfaces used in the actual
evaluation of this change. 

\% Besides the assumption of path independence---which  
applies regardless of the specific form of the canonical
Hamiltonian---Teitelboim's proof also involved explicitly the
assumption that the Hamiltonian is decomposed according to the lapse-shift
formula written in equation (\ref{eq:action}). Using these two
postulates, together, he concluded that in order
for the theory to be consistent 
the phase space should be restricted by the initial value equations 
(\ref{eq:ham}-\ref{eq:mom}) while the canonical generators
should satisfy the Dirac algebra (\ref{eq:D1}-\ref{eq:D3}).  

\% The very last statement is not completely true 
because of a mistake in the reasoning in [19] concerning the fact that the
system is constrained. Nevertheless, the correct algebra---as it
arises from the requirement of path independence---is still the Dirac
one, but is supplemented by terms $G$, $G_i$ and 
$G_{ij}$ whose first partial derivatives with respect to the
canonical variables vanish on the constraint surface: 
\begin{eqnarray}
&& \{ H (x), H (x') \} = g^{ij}(x){H}_i(x)
\delta_{,j}(x,x') + G(x,x') -\xxp ,
\label{eq:E1} \\
&& \{ H(x), {H}_i(x') \} = H(x) \delta_{,i}(x,x') + H_{,i}(x) \delta
(x,x') + G_i(x,x') ,
\label{eq:E2} \\
&& \{{H}_i(x),{H}_j (x')\} = {H}_j(x)\delta_{,i}(x,x') + G_{ij}(x,x') -\xxpp . 
\label{eq:E3}
\end {eqnarray}
The derivation of the above set of relations, which are called the ``weak
Dirac algebra'', can be found in part IV.

\hskip -0.25in
{\Large $\bullet$} {The Problem Of Deriving A Physical Theory From Just The
Canonical Algebra:}
The principle of path independence was an indication 
that a Hamiltonian theory does not have to depend exclusively on the
canonical decomposition of a given spacetime action but may have also an 
independent status. However, in any attempt to construct specific 
canonical theories via the principle of path independence,
alone, some information is found to be missing. The
weak Dirac algebra admits numerous representations whose physical relevance is
therefore doubtful.   

\% As an example, consider the case when the canonical 
variables are the spatial metric and its conjugate momentum.
The usual strong limit of the algebra (\ref{eq:E1}-\ref{eq:E3}) is
taken, where all the terms $G$, $G_i$ and $G_{ij}$ are identically zero.
The generator $H_i$ is chosen as the super-momentum of the 
gravitational field,
\begin{equation}
H_i(x) = \mom(x), 
\label{eq:mome}
\end{equation}
and the normal generator $H$ is required to transform as a scalar
density of weight one. 
Under these simplifications, the second and third Dirac relations
(\ref{eq:D2}-\ref{eq:D3}) are 
satisfied, and the Dirac algebra---which can be seen as a set of 
coupled differential equations for the canonical
generators---decouples. This leaves a single first-order 
equation for $H(x)$, equation (\ref{eq:D1}), which is
expected normally to admit an infinite number of distinct solutions.

\% In particular, it can be assumed that $H(x)$ is of the form[16]
\begin{equation}
H(x) = g^{1 \over 2} W[h,f](x).
\label{eq:myD}
\end{equation}
The weight-zero scalar densities $h$ and $f$ are the ones defined in
[16] as well as in equations (\ref{eq:equh})-(\ref{eq:equf}).
When the ansatz (\ref{eq:myD}) is used in equation (\ref{eq:D1}),
a differential equation for the function $W[h,f]$ arises,
\begin{equation}
{1 \over 2}W W_f= f {W_f}^2 - {1 \over 4} {W_h}^2 + {1 \over 4}. 
\label{eq:eksisosh}
\end{equation}
In analogy with the $\omega$-equation (\ref{eq:WPDE}), it admits a family
of solutions parametrized by an arbitrary function of one
variable. 

\% The general solution of equation (\ref{eq:eksisosh})
is obtained by solving in terms of $W$ the complete integral  
\begin{eqnarray} 
&& \Bigg( W + \sqrt{W^2 -4f + 4[h-B(a[h,f])]^2} \Bigg) \times
\nonumber\\
&& \times exp\Bigg( {{W} 
\over {W + \sqrt{W^2 -4f + 4[h-B(a[h,f])]^2}} }\Bigg) + a[h,f] = 0.
\label{eq:MYD}
\end{eqnarray}
As usual in these cases[28], the form of the function
$a[h,f]$ is determined by solving algebraically in terms of $a$ the 
equation
\begin{eqnarray} 
&& {\partial \over {\partial a}} \Bigg\{ \Bigg( W + \sqrt{W^2 -4f +
4[h-B(a)]^2} \Bigg) \times 
\nonumber\\
&& \times exp\Bigg( {{W} 
\over {W + \sqrt{W^2 -4f + 4[h-B(a)]^2}} }\Bigg) + a \Bigg\} = 0
\label{eq:MYD2}
\end{eqnarray}
for a given choice of the function $B(a)$.

\% The super-Hamiltonian of general
relativity, arising when $W[h,f]=h$, is the only one of these
solutions that is ultra-local in the field momenta. The ultra-locality is
actually related to the geometric meaning of the canonical variables
but this will be discussed in detail later, in part IV. For the moment
notice that if the weak Dirac algebra
(\ref{eq:E1}-\ref{eq:E3}) is used as the starting point of the above
calculation---which is the correct thing to be done---then a set of
differential equations arises whose precise form is not 
known and, therefore, no further progress can be made.

\hskip -0.25in
{\Large $\bullet$} {Selecting The Physical Representations Of The
Dirac Algebra:} 
Deriving geometrodynamics from plausible first principles[20],
Hojman, Kucha\v{r} and Teitelboim chose to lay the stress on the
concept of infinite dimensional groups and placed the strong Dirac algebra
at the centre of their approach. They expected that the closing
relations (\ref{eq:D1}-\ref{eq:D3}) carry enough 
information about the system to select a physical
representation uniquely but they could not extract this
information directly from them. The existence of
solutions like (\ref{eq:myD}) is the reason why. 

\% What the authors of {[20]} did, instead, was to follow an indirect
route and select the physically relevant representations by  
supplementing the strong Dirac algebra with four additional conditions. 
Specifically, they introduced the tangential and normal generators of
hyper-surface deformations, defined respectively by
\begin{eqnarray}
&& {H^D}_i(x) := {{\cal X}^{\alpha}}_i(x) {\delta \over {\delta 
{\cal X}^{\alpha}}}(x),
\label{eq:defo1}
\\
&& {H^D}(x) := {n^{\alpha}}(x) {\delta \over {\delta {\cal X}^{\alpha}}}(x),
\label{eq:defo2}
\end{eqnarray}
and acted with these on the spatial metric:
\begin{eqnarray}
&& {H^D}_k(x')g_{ij}(x) = g_{ki}(x) {\delta}_{,j}(x,x') + 
g_{kj}(x) {\delta}_{,i}(x,x') + g_{ij,k}(x) {\delta}(x,x'),
\label{eq:extra1}
\\
&& H^D(x')g_{ij}(x) = 2 n_{{\alpha};{\beta}}(x) {{\cal X}^{\alpha}}_i(x)
{{\cal X}^{\beta}}_j(x) \delta(x,x').
\label{eq:extra2}
\end{eqnarray}

\% They required that equations
(\ref{eq:extra1}-\ref{eq:extra2})---which are purely kinematic and
hold in an arbitrary Riemannian spacetime---should be satisfied
by the canonical generators, 
\begin{eqnarray}
&& \{ g_{ij}(x),H_k(x') \} = g_{ki}(x) {\delta}_{,j}(x,x') + 
g_{kj}(x) {\delta}_{,i}(x,x') + g_{ij,k}(x) {\delta}(x,x'), 
\label{eq:can1}
\\
&& \{ g_{ij}(x),H(x') \} \propto {\delta}(x,x'),
\label{eq:can2}
\end{eqnarray}
so that any dynamics in spacetime would arise as a different canonical
representation of the universal kinematics. Notice that only the
ultra-locality of the second Poisson bracket was used. 
The justification and geometric interpretation of equations
(\ref{eq:can1}) and (\ref{eq:can2}) can be found in {[20]}.

\% The strong Dirac algebra with the conditions
(\ref{eq:can1}) and (\ref{eq:can2}) 
results in a unique representation for the generators $H$ and
$H_i$. It corresponds to the super-Hamiltonian (\ref{eq:ham}) and the
super-momentum (\ref{eq:mom}) of general relativity. 
The requirement of path independence is then imposed as an
additional postulate to the algebra. It enforces the initial value constraints
(\ref{eq:ham}-\ref{eq:mom}) and, therefore, the complete set of  Einstein's 
equations is recovered. The most general scalar field Lagrangian with
a non-derivative coupling to the metric can be derived along similar
lines[29].  

\hskip -0.25in
$\bullet ${The Full List Of The Selection Postulates:} The precise assumptions used by the authors are
summarized at the end of their paper. They are written here 
in an equivalent form and, in the case of pure gravity, they are the
following:    

\% (i) The evolution postulate: The metric and its conjugate momentum are regarded as the sole canonical variables. There exists a Hamiltonian that generates the dynamical evolution of the theory. It can be casted in the lapse-shift form, equation (\ref{eq:action}), where the super-Hamiltonian and super-momentum generators are constructed entirely from the canonical variables. 

\% (ii) The representation postulate: The canonical generators must 
satisfy the closing relations (\ref{eq:D1}-\ref{eq:D3}) of the strong
Dirac algebra. 

\% (iii) Initial data re-shuffling: The Poisson bracket (\ref{eq:can1})
between the super-momentum and the configuration variable $g_{ij}$
must coincide with the kinematic relation (\ref{eq:extra1}). 

\% (iv) Ultra-locality: The Poisson bracket (\ref{eq:can2}) between the 
super-Hamiltonian and the configuration variable $g_{ij}$ must
coincide with the kinematic relation (\ref{eq:extra2}).

\% (v) Reversibility: The time-reversed spacetime must 
be generated by the same super-Hamiltonian and super-momentum as the 
original spacetime.

\% (vi) Path independence: The dynamical evolution predicted by the
theory must be such that the change in the canonical variables during
the evolution from a given initial surface to a given final one is
independent of the actual sequence of intermediate surfaces used in
the evaluation of this change.

\% Notice that there is an implicit assumption hidden in the evolution postulate. Namely, in order that the metric and the momentum be a canonical pair, the metric should be a spatial scalar and the momentum a spatial density of weight one. This is necessary since, otherwise, the $\delta$-function appearing in the basic Poisson bracket relations does not have the appropriate spatial weight. Specifically, it should be a scalar in the first argument and a density in the second. Notice, also, that in all the other postulates an assumption of a surrounding spacetime is implied. Later, this assumption will be identified explicitly, as an additional postulate concerning the choice of canonical variables. The postulates (ii)-(vi) will then be shown to be unnecessary.

\hskip -0.25in
$\bullet ${The Need For a Detailed Understanding Of The Selection
Postulates:}
The above assumptions comprise a set of plausible first principles on which the canonical formulation of a theory can be based. However, in a 
sense these principles are not completely satisfying. This is because  
they do not correspond to a minimum set and because the connection
between them is not clear. The authors of {[20]} mentioned the 
redundancy of the reversibility postulate (v) as well as the fact that the
third closing relation of the representation postulate (ii) is made
redundant by the re-shuffling requirement (iii). They stressed the
need for understanding the precise reason why some equations hold strongly
while others hold only weakly and, in particular, for clarifying the
relationship between the strong representation postulate (ii) and the
weak requirement of path independence (vi).   

\% The revised form of Teitelboim's argument makes such a
clarification a more important issue since, now, the strong
representation requirement---which is at the heart of the
approach in {[20]}---seems to be unjustified. In addition, repeating the
geometric argument used in [19] in the reverse order, it follows that
the dynamical evolution of the theory must also hold weakly. This is
in contrast to the 
strong equations used in postulates (iii) and (iv). On the other hand,
recall that any attempt to replace
these equations by weak ones will result
in a situation where the particular form of the 
differential equations that need to be solved will not be known and 
no further progress will be made.

\% Putting the issue of the weak equalities aside, the understanding of the 
exact relationship between the postulates is needed if the method
in {[20]} is to be applied to the case of a generic canonical
algebra. The reason is that---in the existing formulation of the
postulates---the overall consistency is made certain only by the fact
that the 
re-shuffling and ultra-locality assumptions (iii) and (iv) are respected by
the dynamical law of the theory (i). On the other hand, the
remaining postulates do not ensure that assumptions (iii) and (iv) are the
only ones compatible with this law. If different compatible
assumptions are used as supplementary conditions to the algebra then the
method in [20] will yield different canonical representations. 

\% However,
the dynamical law of the theory is the only assumption that enters
the derivation and geometric interpretation of the algebra besides the
principle of path independence.  
It follows that the existing formulation of the postulates will be
ambiguous if it is used as an algorithm for passing from the interpretation
of an algebra to its physical representations. 
This is of course particularly relevant to the discussion made in part II
concerning the interpretation of the new Lie algebra. 

\% Finally, there is an asymmetry in the formulation of the
postulates within which lies the main motivation 
for the discussion in this part. It concerns the identification of the
canonical generators with the generators of normal and tangential
hyper-surface deformations, that is required to hold in postulates
(iii) and (iv) for the configuration variable only. However, if  
such an identification is a fundamental principle in the canonical
theory then it should hold for both the canonical variables, in which
case additional 
information about the kinematics of the system can be extracted. 

\% In an equal-time formalism, this conjecture can neither  
be confirmed nor rejected because the action of the
deformation generators on the canonical momenta cannot be
defined. Marolf[30] used the Hamiltonian as an 
additional structure to extend the Poisson bracket from a Lie bracket
on phase space to a Lie bracket on the space of histories. Here,
instead, the equal-time formalism is put aside, and a phase
space is introduced whose Poisson bracket is defined over the space of histories from the beginning.  
\vskip 9cm
\hskip -0.25in
{\LARGE Chapter III-2. The History Phase Space.}
\vskip 1cm 
\hskip -0.25in
In this chapter, the history formulation of canonical dynamics is presented. Since most of the results are merely translated from the standard approach, the discussion is rather basic. For example, no detailed analysis is given of the Dirac method for dealing with constraints, or with the gauge transformations they generate. Only those features of the history formalism are given that are needed for the discussion that follows. The precise connection between the present section and the previous chapters will become apparent in chapter III-3, where the evolution postulate is re-formulated in terms of classical histories.
\vskip 1cm 
\hskip -0.25in
{\Large $\bullet$} {The Unconstrained Hamiltonian:}
Consider the theory described by the canonical action
\begin{eqnarray} 
&& S[q^A , p_A] = \int d^3xdt \bigg( p_A \dot{q^A} - {\cal H} \bigg),
\nonumber\\
&& {\cal H} = N H + N^i H_i.
\label{eq:UnA}
\end{eqnarray}
The functions $N$ and $N^i$ are fixed (i.e., non-canonical)  functions
of space and time. 
The generators $H$ and $H_i$ are functions of the canonical fields
$q^A$, $p_A$ and may also depend on additional fixed fields $c^K$. 
The index $A$ runs from 1 to half the total number of canonical
variables, while $K$ runs from 1 to the total number of fixed
fields.

\% The phase space can be generalized to include the canonical fields at
all times. This can be done by introducing the space of histories,
\begin{equation}
\bigg( q^A(x,t) , p_A(x,t) \bigg),
\label{eq:histoir}
\end{equation}
and defining on it the Poisson bracket 
\begin{equation}
\{ q^A(x,t) , p_B(x',t') \} = {{\delta}^A}_B {\delta}(x,x') {\delta}(t,t').
\label{eq:UnBr}
\end{equation}
The quantum analogue of the canonical fields in (\ref{eq:histoir}) is the
one-parameter family of Schr\"{o}edinger operators introduced by Isham
{\it et al} in their study of continuous time consistent
histories[3,4].

\% The Poisson bracket (\ref{eq:UnBr}) turns the space of
histories into a Poisson manifold. In terms of this bracket, the
variation of the canonical action can be written concisely in the form
\begin{eqnarray}
&& \{ S , q^A(x,t) \} \simeq 0,
\label{eq:UnV1}
\\
&& \{ S , p_A(x,t) \} \simeq 0
\label{eq:UnV2}
\end{eqnarray}
and defines a constraint surface on the space of histories. The physical
fields are defined to satisfy these relations for each value of $x$ and $t$. 
For the particular form (\ref{eq:UnA}) of the canonical action, the
weak equations (\ref{eq:UnV1}-\ref{eq:UnV2})
become\footnote{Throughout the thesis, the
functional derivative ${{\delta {\cal F}}\over{\delta q^A }}$ is
defined by
${{\delta {\cal F}}\over{\delta q^A }} =  
{{\partial {\cal F}}\over{\partial q^A }} + 
{{\partial {\cal F}}\over{\partial {q^A}_{,i} }}
{\partial}_{i} + {{\partial {\cal F}}\over{\partial {q^A}_{,ij} }}
{\partial}_{ij} + ...etc$. Sometimes ${\cal F}$ is called a functional,
although it is only a local function of the canonical variables and
a finite number of their derivatives.}   
\begin{eqnarray}
&& \dot{q^A}(x,t) \simeq \int d^3x'dt' \{ q^A(x,t) , {\cal H}(x',t') \}
\equiv \int d^3x' {{\delta {\cal H}}\over{\delta p_A }}(x',t) \delta(x,x')
\label{eq:UnH1}
\\
&& \dot{p_A}(x,t) \simeq \int d^3x'dt' \{ p_A(x,t) , {\cal H}(x',t') \}
\equiv \int d^3x' {{\delta {\cal H}}\over{\delta q^A }}(x',t) \delta(x,x'),
\label{eq:UnH2}
\end{eqnarray}
which can be recognized as Hamilton's equations in the usual
equal-time sense. This follows from the fact that the Hamiltonian in
equation (\ref{eq:UnA}) 
is by construction independent of any time derivatives, so it can be
integrated trivially over $\int dt' \delta(t,t')$. 

\% The weak equality sign is a reminder of the fact that Hamilton's
equations, and consequently the theory, are not preserved under a
general Poisson bracket. In the
equal-time formalism this presents no problem because the canonical
velocities are only defined externally but, here, they are included equally
in the phase space. For example, the Poisson bracket between a field
velocity and its conjugate momentum can be evaluated to give a
time derivative of the $\delta$-function. This is not the result that
arises when the corresponding Hamilton equation is used to replace the
field velocity before the commutation is performed. Nonetheless, since
the theory 
is about time evolution only, it is sufficient that Hamilton's
equations are preserved weakly under the Poisson bracket with the
Hamiltonian.

\% In the unconstrained theory (\ref{eq:UnA}) this follows automatically
from Hamilton's equations and the definition of the history 
Poisson bracket (\ref{eq:UnBr}) without any reference to the specific 
form of the Hamiltonian.  However, before this can be checked directly, 
the definition of the Hamiltonian has to be extended so that it can
incorporate the trivial dynamical evolution of the fixed
functions $c^K$, $N$ and $N^i$.   
This is also appropriate for the completeness of the formalism.

\hskip -0.25in
{\Large $\bullet$} {Incorporating The Fixed Functions:}
The extended unconstrained action is defined by 
\begin{eqnarray}
&& S[q^A , p_A , {\omega}_K ,\omega , {\omega}_i] = \int d^3xdt \bigg(
p_A \dot{q^A} + {\omega_K} \dot{c^K} + {\omega} \dot{N} + {\omega_i}
\dot{N^i} - {\cal H}^{ext} \bigg), 
\nonumber\\
&& {\cal H}^{ext} = N H + N^i H_i + {\omega_K} \dot{c^K} + {\omega} \dot{N}
+ {\omega_i} \dot{N^i}. 
\label{eq:ExtA}
\end{eqnarray}
The momenta ${\omega}_K$, $\omega$ and $\omega_i$ are defined
through the Poisson bracket relations
\begin{eqnarray}
&& \{ c^K(x,t) , {\omega_L}(x',t') \} = {\delta^K}_L \delta(x,x') \delta(t,t'),
\nonumber\\
&& \{ N(x,t) , {\omega}(x',t') \} = \delta(x,x') \delta(t,t'),
\nonumber\\
&& \{ N^i(x,t) , {\omega_j}(x',t') \} = {\delta^i}_j \delta(x,x')
\delta(t,t').
\label{eq:PbN}
\end{eqnarray}
The various $\delta$-functions transform in different ways depending
on the transformation properties of the corresponding canonical
variables. This is not denoted explicitly in order to keep the
notation simple. Furthermore, the above momenta are not assumed to
have any direct physical significance or interpretation. The whole
purpose of their introduction is to allow the time derivative of the
fixed functions to be calculated inside the Poisson bracket
formalism.

\% Restricting the discussion to functionals of the
canonical and the fixed variables, it follows that  
\begin{eqnarray}
&& \{ F(x,t) , \int d^3x' dt' {\cal H}^{ext}(x',t') \} =
{{\delta F} \over {\delta q^A}}(x,t) \{ q^A(x,t), \int
d^3x' dt' {\cal H}^{ext}(x',t') \}    
\nonumber\\
&& + {{\delta F} \over {\delta p_A}}(x,t) \{
p_A(x,t), \int d^3x' dt' {\cal H}^{ext}(x',t') \} + {{\delta F} \over
{\delta c^K}}(x,t) \dot{c^K}(x,t) 
\nonumber\\
&& + {{\delta F} \over
{\delta N}}(x,t) \dot{N}(x,t)  + {{\delta F} \over {\delta N^i}}(x,t)
\dot{N^i}(x,t) \simeq \dot{F}(x,t) . 
\label{eq:UnProof}
\end{eqnarray}
This implies that the extended Hamiltonian is the canonical
representation of the total time derivative operator.

\% Equivalently it may be observed that when the kinematic half
of the extended action,  
\begin{equation}
\int d^3xdt \bigg(p_A \dot{q^A} + {\omega_K} \dot{c^K} + {\omega}
\dot{N} + {\omega_i} \dot{N^i} \bigg),
\label{eq:kineterms} 
\end{equation}
is acting on the functional $F$ it produces the time derivative of $F$
in the strong sense. 
On the other hand, when the total extended action acts on any F it
yields weakly zero by definition.
It follows that the remaining
half of the action---i.e., the dynamical half corresponding to the
integral of the extended Hamiltonian---produces the total time derivative
of F in the weak sense.

\% The above  result implies that
Hamilton's equations are preserved automatically under the dynamical
evolution  of the theory. Indeed, if F is any functional of the
canonical and the fixed variables that vanishes on the constraint surface,
its total time derivative also vanishes on the same
surface. Since this derivative is weakly equal to the
commutation of F with the integral of the extended Hamiltonian, it
follows that all weakly vanishing Fs remain weakly zero under
this commutation. Choosing these Fs to be Hamilton's
equations themselves shows that the constraint surface is preserved. 
This completes the treatment of systems that are unconstrained in the usual
sense.

\hskip -0.25in
{\Large $\bullet$} {The Constrained Hamiltonian:}
The extended form of the action, equation (\ref{eq:ExtA}), arises
naturally when the functions $N$ and $N^i$ 
are either constrained canonical variables or acquire the meaning of
Lagrange multipliers. 
Both these cases are presented in their most general 
form by considering the canonical action
\begin{eqnarray}
&& S[q^A , p_A , N, \omega, N^i, {\omega}_i, {\omega}_K] = \int d^3xdt
\bigg( p_A \dot{q^A} + {\omega}_K \dot{c^K} + {\omega} \dot{N} +
{\omega_i} \dot{N^i} - {\cal H} \bigg), 
\nonumber\\
&& {\cal H} = N H + N^i H_i + {\omega}_K \dot{c^K} + {\omega} \dot{N}
+ {\omega_i} \dot{N^i}, 
\label{eq:CA}
\end{eqnarray}
which is now varied additionally with respect to $N$ and
$N^i$. The fields $c^K$ are still treated as fixed.

\% The variation of (\ref{eq:CA}) leads to the same equations as
before, namely  
\begin{eqnarray}
&& \{ S , q^A(x,t) \} \simeq 0 \Leftrightarrow \dot{q^A}(x,t)
\simeq \int d^3x' \bigg(  N {{\delta {H}}\over{\delta p_A }} +
N^i {{\delta {H_i}}\over{\delta p_A }} \bigg) (x',t) \delta(x,x'),  
\label{eq:CcV1}
\\
&& \{ S , p_A(x,t) \} \simeq 0 \Leftrightarrow \dot{p_A}(x,t)
\simeq \int d^3x' \bigg(  N {{\delta {H}}\over{\delta q^A }} +
N^i {{\delta {H_i}}\over{\delta q^A }} \bigg) (x',t) \delta(x,x'),
\label{eq:CcV2}
\\
&& \{ S , c^K(x,t) \} = 0 \Leftrightarrow \dot{c^K}(x,t) =
\dot{c^K}(x,t) \Leftrightarrow 0 = 0 ,
\label{eq:CcV3}
\\
&& \{ S , N(x,t) \} = 0 \Leftrightarrow \dot{N}(x,t) = \dot{N}(x,t)
\Leftrightarrow 0 = 0 ,
\label{eq:CcV4}
\\
&& \{ S , N^i(x,t) \} = 0 \Leftrightarrow \dot{N^i}(x,t) =
\dot{N^i}(x,t) \Leftrightarrow 0 = 0 ,
\label{eq:CcV5}
\end{eqnarray}
subject to the additional equations
\begin{eqnarray}
&& \{ S , {\omega}(x,t) \} \simeq 0 \Leftrightarrow \dot{\omega}(x,t)
\simeq \dot{\omega}(x,t) + {H}(x,t) \Leftrightarrow {H}(x,t) \simeq 0,
\label{eq:CcV6}
\\
&& \{ S , {\omega}_i(x,t) \} \simeq 0 \Leftrightarrow \dot{{\omega}_i}(x,t)
\simeq \dot{{\omega}_i}(x,t) + {H_i}(x,t) 
\Leftrightarrow {H_i}(x,t) \simeq 0
\label{eq:CcV7}
\end{eqnarray}
arising from the variation of the action with respect to $N$ and $N^i$.

\% For a functional $F[q^A,p_A,c^K,N,N^i]$ the proof of the previous
section still applies,  
\begin{equation}
\{ F(x,t) , \int d^3x' dt' {\cal H}(x',t') \} \simeq \dot{F}(x,t),
\label{eq:dra}
\end{equation}
with the weak equality referring to Hamilton's equations
(\ref{eq:CcV1}-\ref{eq:CcV2}). Again, if F is any
functional that vanishes on the surface defined by Hamilton's 
equations, its time derivative also vanishes on this surface.
Therefore, by taking Hamilton's equations
to be these Fs, it can be deduced that
equations (\ref{eq:CcV1}-\ref{eq:CcV5}) are  
preserved weakly under the dynamical evolution of the theory. On the
other hand, if F vanishes on the 
surface defined by the constraint equations
(\ref{eq:CcV6}-\ref{eq:CcV7}), its time derivative still 
vanishes on the this surface but, now, it does not follow that this is
the time derivative generated by the Hamiltonian of the theory.

\% It must be ensured also that the time derivatives of the fields
evaluated by differentiating equations (\ref{eq:CcV6}-\ref{eq:CcV7})
are compatible with the time derivatives of the same fields evaluated
from Hamilton's equations. If the constraints 
(\ref{eq:CcV6}-\ref{eq:CcV7}) do not depend on the fixed fields
$c^K$---which is the case for most of the physical theories---this 
compatibility condition results in the requirement that the algebra of 
$H$ and $H_i$ must close weakly under the history Poisson
bracket. Since $H$ and $H_i$ are by construction independent of
any time derivatives the weak closure of the algebra refers only to
the constraint equations (\ref{eq:CcV6}-\ref{eq:CcV7}). 
\vskip 9cm
\hskip -0.25in
{\LARGE Chapter III-3. The Evolution Postulate.}
\vskip 1cm 
\% {\Large $\bullet$} {The Inverse Procedure And The Evolution Postulate:}
The aim is to invert the above argument, and recover the general
canonical Hamiltonian of a theory from a set of first principles. The
need that these principles be minimal implies that 
the appropriate starting point of the derivation is the requirement
that the canonical action should have the form (\ref{eq:CA}). This
conclusion follows from the observation that equation (\ref{eq:CA}) is the only
prerequisite for the existence of a canonical algebra in the theory.

\% According to the terminology used in {[20]}, equation (\ref{eq:CA}) corresponds to the ``evolution postulate''. 
In case that this postulate turns out to be insufficient to 
determine the theory completely, the plan is that any supplementary
conditions that may be added must be such that the connection between
them remains clear throughout the derivation.

\% Initially, the most general canonical representation of the
Hamiltonian is seeked that satisfies the unconstrained version of the
postulate,  
\begin{eqnarray}
&& {\partial \over {\partial}t}{q^A}(x,t) \simeq \int d^3x'dt' \{
q^A(x,t) , (N H + N^i H_i)(x',t') \}, 
\label{eq:P1}
\\
&& {\partial \over {\partial}t}{p_A}(x,t) \simeq \int d^3x'dt' \{
p_A(x,t) , (N H + N^i H_i)(x',t') \}.
\label{eq:P2}
\end{eqnarray}
If such a Hamiltonian cannot be found, there still is the alternative
possibility of varying the action with respect to the functions $N$
and $N^i$. Equations (\ref{eq:P1}-\ref{eq:P2}) must then be
supplemented by the constraint equations   
\begin{eqnarray}
&& H(x,t) \simeq 0,
\label{eq:P3}
\\
&& {H_i}(x,t) \simeq 0,
\label{eq:P4}
\end{eqnarray}
that have to be preserved under the dynamical evolution of the
theory. This consistency requirement is included in the evolution
postulate for constrained systems, and amounts to the weak closure of
the algebra when the constraints are independent of fixed fields.

\% Notice that the time derivatives of $N$ and $N^i$ do not appear in
the equations of motion (\ref{eq:P1}-\ref{eq:P4}) which means that $N$ and
$N^i$ are allowed to have arbitrary numerical values. This is also
true, by definition, for the non-dynamical $N$ and $N^i$ in the case
of unconstrained systems. Therefore, the evolution postulate for both
constrained and unconstrained systems can be re-stated as the
requirement that the canonical action should be of the form (\ref{eq:CA})
with $N$ and $N^i$ taking arbitrary numerical values. In practice,
the lapse function is still required to be positive.    

\hskip -0.25in
{\Large $\bullet$} {The Evolution Postulate In An Equivalent Form:}
At first sight, conditions (\ref{eq:P1}-\ref{eq:P2}) do not seem to be
restrictive enough so that something definite can be drawn from them. It
seems that the canonical representations can be chosen at will, and
that any constrained theory can be created by just requiring the
closure of the resulting algebra. This view changes 
when the precise geometric meaning of the canonical fields is taken
into account. For example, in a scalar field theory, the field
$\phi(x,t)$ is not merely a spatial scalar but is also by definition the
pull-back of a spacetime scalar field. Below, the evolution postulate
is transformed to an equivalent condition on the canonical generators
that is more appropriate for the exploitation of this fact.

\% The functions $N$ and $N^i$ are taken as the lapse function
and the shift vector. In fact, this will be the case henceforth
unless stated otherwise. If the time  derivative operator in
equations (\ref{eq:P1}-\ref{eq:P2}) is decomposed according
to the lapse-shift formula, 
\begin{equation}
{\partial \over {\partial t}} = Nn^{\alpha}
{\partial \over {\partial X^{\alpha}}} + {N^i}{{\cal
X}^{\alpha}}_i {\partial \over {\partial X^{\alpha}}},
\label{eq:dxdt}
\end{equation}
and the momentum ${{\cal P}_{\alpha}}$ conjugate to
the embedding is introduced, 
\begin{equation}
\{ {{\cal X}^{\alpha}}(x,t) , {{\cal P}_{\beta}}(x',t') \} =
{{\delta}^{\alpha}}_{\beta} {\delta}(x,x') {\delta}(t,t'), 
\label{eq:Pmom}
\end{equation} 
equation (\ref{eq:P1}-\ref{eq:P2}) takes the following form,
\begin{eqnarray}
&& \{ {q^A}(x,t), H(x',t') \} \simeq \{ {q^A}[{\cal
X}](x,t) , {{\cal P}_{\beta}}(x',t') \} n^{\beta}(x',t')  ,
\label{eq:PP1}
\\
&& \{ {q^A}(x,t), {H_i}(x',t') \} \simeq \{ {q^A}[{\cal
X}](x,t) , {{\cal P}_{\beta}}(x',t') \} {{{\cal
X}^{\beta}}_i}(x',t') ,  
\label{eq:PP2}
\\ 
&& \{ {p_A}(x,t), H(x',t') \} \simeq \{ {p_A}[{\cal
X}](x,t) , {{\cal P}_{\beta}}(x',t') \} n^{\beta}(x',t') ,  
\label{eq:PP3}
\\
&& \{ {p_A}(x,t), {H_i}(x',t') \} \simeq \{ {p_A}[{\cal
X}](x,t) , {{\cal P}_{\beta}}(x',t') \} {{{\cal X}^{\beta}}_i}(x',t'). 
\label{eq:PP4}
\end{eqnarray}
Notice that the arbitrariness of $N$ and $N^i$ has been used to eliminate
the integration.

\% On the right side of the above equations the explicit dependence of
the canonical fields on the spacetime embedding is taken into account. For
the configuration fields this is just the dependence arising from
the definition of the fields as geometric objects in spacetime. For
the conjugate fields the situation is more complicated, and
equation (\ref{eq:P1}) is assumed to have been inverted to express the
momenta as 
functionals of the configuration variables, the lapse, the shift,
and the prescribed fields $c^K$. All the latter have a definite
dependence on the spacetime embedding which is then conveyed to the
conjugate canonical fields.

\% Equation (\ref{eq:P1}) is always invertible for the momenta
because, by construction, the system 
is constrained only in the quantities $N$ and $N^i$ at
the most. There is one exception to this rule when the action is not
derivable from a spacetime Lagrangian but, instead, is brought
into the form (\ref{eq:CA}) through the introduction of Lagrange
multipliers. This is the relevant case for parametrized
theories.

\% For the purposes of performing actual calculations, the evolution
postulate is to be used in the following way. Any time derivatives of the
canonical variables that arise on the right side of equations
(\ref{eq:PP1}-\ref{eq:PP4}) are replaced by the original
Hamilton's equations (\ref{eq:P1}-\ref{eq:P2}). When the theory is
unconstrained, this results in a coupled system of four functional
differential equations for $H$ and $H_i$. If a solution 
exists, it corresponds to the general canonical representation
compatible with the evolution postulate. On the other hand, when the theory is
constrained the resulting ``equations'' for $H$ and
$H_i$ are not proper differential equations, since it is sufficient
that they hold only on the constraint surface (\ref{eq:P3}-\ref{eq:P4}).

\% If the constraints (\ref{eq:P3}-\ref{eq:P4}) implied that the
canonical variables can not be treated as independent in these
``equations'' for $H$ and $H_i$, the evolution postulate for constrained
systems would not make any sense at all. However, by construction of
the canonical formalism, the constraints must be imposed only {\it
after} the Poisson  brackets have been evaluated. Therefore, even for
constrained systems, the differential equations for
$H$ and $H_i$ should be solved as if the canonical variables were
independent, and the constraints (\ref{eq:P3}-\ref{eq:P4}) should be imposed
only at the end. In practice, a term is added on each 
differential equation, the value of which is required to
vanish on the constraint surface. Exactly how this is done is shown
in part IV, in the case of general relativity.

\% Finally, notice that the replacement of the field
velocities in equations (\ref{eq:PP1}-\ref{eq:PP4}) with the original
and equivalent equations (\ref{eq:P1}-\ref{eq:P2}) does not lead to
cyclic identities as it might have been expected. The reason is that
the original equations hold in integrated form, while equations
(\ref{eq:PP1}-\ref{eq:PP4}) hold at every point in space and
time due to the arbitrariness of $N$ and $N^i$. The information
incorporated in these equations is actually so rich  
that it determines the canonical representations of the theory.  

\pagebreak
\begin{center}
{\LARGE Part IV:}
\vskip 9cm
{\LARGE Canonical Theories Derived From First Principles}
\end{center}

\pagebreak

\% {\LARGE Chapter IV-1. The Indirect Method Applied To Gravity.}
\vskip 1cm 
\hskip -0.25in
{\Large $\bullet$} {The New Set Of Postulates:} It will be shown that the only assumptions needed for the derivation of the representations of a canonical theory are (a) the evolution postulate and (b) the explicit assumption of a surrounding spacetime. That this is indeed so will be shown in an indirect way, by starting from the above two assumptions and recovering the complete set of postulates of Kucha\v{r} {\it et al}. For constrained systems, it turns out that these postulates have to be imposed weakly, which is also implied by the revised version of Teitelboim's argument.

\% The new postulates are the following:

\% (a) The evolution postulate: The configuration variable (here the metric) and its conjugate momentum are regarded as the sole canonical variables. There exists a Hamiltonian that generates the dynamical evolution of the theory. It can be casted in the lapse-shift form, equation (\ref{eq:action}), where the super-Hamiltonian and super-momentum generators are constructed entirely from the canonical variables. 

\% (b) The postulate of a surrounding spacetime: The canonical configuration variable is the pull-back of a configuration variable in spacetime through the foliation associated with the lapse function and shift vector appearing in the Hamiltonian.

\% For simplicity, both these postulates together will be called the evolution postulate in the chapters that follow.

\hskip -0.25in
{\Large $\bullet$} {Recovery Of The Re-shuffling And Ultra-locality
Postulates:} 
On the right side of equations (\ref{eq:PP1}-\ref{eq:PP2}) the
configuration fields are treated as functionals of the embedding
relative to which the decomposition of the spacetime theory has been
performed. The re-shuffling and ultra-locality postulates follow
immediately from equations (\ref{eq:PP1}-\ref{eq:PP2}) once the geometric
meaning of the configuration variable is taken into account (i.e., postulate (b) in the above notation). This is
recognized in {[20]}, although the emphasis is given on the
compatibility of the postulates with the dynamical law
(\ref{eq:P1}-\ref{eq:P2}) rather than on the fact that the
postulates are determined by this law uniquely. Referring to the
corresponding comment in chapter III-1, it is only 
because of this fact that the method in {[20]} can be used unambiguously as an algorithm for finding the physically relevant representations of a generic canonical algebra.

\% The ultra-locality and re-shuffling conditions are written down below for
the physical examples that are usually considered. The relevant
calculations can be found in Appendix C. Notice that a strong equality sign is used, with the understanding that all canonical velocities
have been eliminated through the corresponding Hamilton's equations. 
This is consistent with the general plan, according to which
an unconstrained representation of the evolution postulate is seeked
originally. If a theory is proved to be constrained the following
equations will be revised accordingly. The presence of these canonical velocities in postulates (a) and (b) is of course the reason why none of the following arguments can be applied in the standard canonical formalism.

(i) Scalar field theory: The configuration variable is the
pullback of a spacetime scalar field, 
\begin{equation}
\phi(x,t) = {\phi}[{\cal X}](x,t),
\label{eq:aygo}
\end{equation}
and, as such, is an ultra-local function of the embedding. Equations
(\ref{eq:PP1}-\ref{eq:PP2}) become 
\begin{eqnarray}
&& \{ {\phi}(x,t), H_i(x',t') \} =
{\phi}_{,{\beta}}(x,t) n^{\beta}(x,t) \delta(x,x') \delta(t,t')
\label{eq:SF1}
\\
&& \{ {\phi}(x,t), H(x',t') \} =
{\phi}_{,i}(x,t) \delta(x,x') \delta(t,t'),
\label{eq:SF2}
\end{eqnarray} 
which can be recognized as the history analogues of the re-shuffling
and ultra-locality conditions in [20]. Indeed, the 
$\delta(t,t')$ function indicates that the canonical generators are 
independent of the field velocities, the ultra-locality of the first
equation implies that the super-Hamiltonian is an ultra-local function
of the momenta, while the form of the second equation ensures that
the super-momentum just re-shuffles the data on the hyper-surface.

(ii) General relativity: The configuration variable is the
pullback of the spacetime metric, 
\begin{equation}
g_{ij}(x,t) = {\gamma}_{{\alpha}\beta}[{\cal X}](x,t) {{\cal
X}^{\alpha}}_i(x,t) {{\cal X}^{\beta}}_j(x,t).
\label{eq:aygo2}
\end{equation}
and equations (\ref{eq:PP1}-\ref{eq:PP2}) result in the 
following conditions on the canonical generators, 
\begin{eqnarray}
&& \{ g_{ij}(x,t), H_k(x',t') \} =  g_{ki}(x,t)
{\delta}_{,j}(x,x') \delta(t,t') + g_{kj}(x,t) {\delta}_{,i}(x,x')
\delta(t,t')  
\nonumber\\
&& \; \; \; \; \; \; \; \; \; \; \; \; \; \; \; \; \; \; \; \; \; \;
\; \; \; \; \; \; \; \; \; \; \; \; \; +  \; g_{ij,k}(x,t) 
{\delta}(x,x') \delta(t,t'), 
\label{eq:GR1}
\\
&& \{ g_{ij}(x,t), H(x',t') \} =
2n_{{\alpha};{\beta}}(x,t) 
{{\cal X}^{\alpha}}_i(x,t) {{\cal X}^{\beta}}_j(x,t) \delta(x,x') \delta(t,t').
\label{eq:GR2}
\end{eqnarray}
For the same reasons as in case (i) above, these can be recognized as
the history analogues of the re-shuffling and ultra-locality postulates 
(\ref{eq:can1}-\ref{eq:can2}).

(iii) Deformation and parametrized theories:
For the theory of hyper-surface deformations, the configuration
variable is the embedding itself. Equations
(\ref{eq:PP1}-\ref{eq:PP2}) become 
\begin{eqnarray}
&& \{ {\cal X}^{\alpha}(x,t), H(x',t') \} =
n^{\alpha}(x,t) \delta(x,x') \delta(t,t'),
\label{eq:HD11}
\\
&& \{ {\cal X}^{\alpha}(x,t), H_i(x',t') \} =
{{\cal X}^{\alpha}}_{i} \delta(x,x') \delta(t,t'),
\label{eq:HD22}
\end{eqnarray} 
which are the re-shuffling and ultra-locality conditions for the
deformation theory. Using the equations in (i) and in (iii)
together, the corresponding conditions for a parametrized scalar field
theory arise.  

\hskip -0.25in
{\Large $\bullet$} {The Two Jacobi Identities:}
This is the revised version of the geometric argument in [19], so many
of the following results can be found in [19] and
{[20]}. They are re-stated here only for completeness. Besides
the revision of the 
argument for constrained systems, the other difference between
this approach and the approach in [20] is that the present
discussion does not rely on the principle of path independence. The
later is also derived from the evolution postulate.

\% The discussion starts from the following two Jacobi identities, 
\begin{eqnarray}
&& \{ \{ H_j(x',t') , F(x'',t'') \} , H_i(x,t) \} + \{ \{ F(x'',t'') ,
H_i(x,t) \} , H_j(x',t') \}
\nonumber\\
&& \; \; \;  \; \; \; \; \; \; \; \; \; \;  \; \; \; \;
\; \; \; \; \; \;  + \{ \{ H_i(x,t)  , H_j(x',t') \} , F(x'',t'') \} = 0,   
\label{eq:CJ1}
\\
&& \{ \{ {H^D}_j(x',t') , F(x'',t'') \} , {H^D}_i(x,t) \} + \{ \{ F(x'',t'') ,
{H^D}_i(x,t) \} , {H^D}_j(x',t') \}
\nonumber\\
&& \;  \; \; \; \; \; \; \; \; \; \;  \; \; \; \;
\; \; \; \; \; \;  + \{ \{ {H^D}_i(x,t)  , {H^D}_j(x',t') \} ,
F(x'',t'') \} = 0, 
\label{eq:DJ1}   
\end{eqnarray}
that hold on the canonical and on the deformation history
phase space respectively. The arbitrary functional $F$ depends on both
the canonical 
variables $q^A$ and $p_A$, while the action of the deformation
generators on these variables is defined as in chapter III-3. The
notation for the normal and tangential projections of $P_{\alpha}$ is
chosen to coincide with the equal-time definitions
(\ref{eq:defo1}-\ref{eq:defo2}).

\% The only case that is considered is when the canonical Hamiltonian is
independent of the fixed fields $c^K$, which is the relevant
case for general relativity. When prescribed fields are present 
in the Hamiltonian the following derivation still applies but depends
on the actual character of these fields, and is avoided for
simplicity. An extensive account of such systems can be found in [23].

\% Having restricted $H$, $H_i$ and $F$ to be pure functionals of the
canonical variables, the first terms in the identities
(\ref{eq:CJ1}) and (\ref{eq:DJ1}) are compared. The
evolution postulate implies that  
\begin{equation}
\{ H_j(x',t') , F(x'',t'') \} = \{ {H^D}_j(x',t') , F(x'',t'') \} .
\label{eq:litsa}
\end{equation}
The use of the strong sign is due to the
replacement of the field velocities, as already explained.
Both brackets depend solely on the canonical variables because
of the restrictions imposed. Therefore, a further application of the
evolution postulate yields 
\begin{equation}
\{ \{ H_j(x',t') , F(x'',t'') \} , H_i(x,t) \} =  \{ \{ {H^D}_j(x',t')
, F(x'',t'') \} , {H^D}_i(x,t) \} , 
\label{eq:litsa2}
\end{equation}
which is valid precisely because Hamilton's equations
are preserved under the commutation with the Hamiltonian.

\% Repeating this argument when comparing the second terms in the
identities (\ref{eq:CJ1}) and (\ref{eq:DJ1}), the following equation arises,
\begin{equation}
\{ \{ F(x'',t'') , H_i(x,t) \} , H_j(x',t') \} = \{ \{ F(x'',t'') 
{H^D}_i(x,t) \} , {H^D}_j(x',t') \} .
\label{eq:parak}
\end{equation}
Equations (\ref{eq:litsa2})-(\ref{eq:parak}) implies that the
remaining terms in the identities (\ref{eq:CJ1})-(\ref{eq:DJ1}) should
be equal,  
\begin{equation}
\{ \{ H_i(x,t)  , H_j(x',t') \} , F(x'',t'') \} = \{ \{ {H^D}_i(x,t)
, {H^D}_j(x',t') \} , F(x'',t'') \} = 0 .
\label{eq:elpis}
\end{equation}

\% The Poisson bracket between the two deformation generators in
equation  (\ref{eq:elpis}) is calculated to give the history analogue
of the Dirac relation (\ref{eq:D3}), 
\begin{equation}
\{ {H^D}_i(x,t)  , {H^D}_j(x',t') \} =  {H^D}_j(x,t) {\delta}_i(x,x')
\delta(t,t') - (ix \leftrightarrow jx'). 
\label{eq:HDD3}
\end{equation}
Then the evolution postulate is used once more to give an equation 
that holds exclusively on the canonical phase space,
\begin{equation}
\{ \bigg[ \{ H_i(x,t) , H_j(x',t') \} - \bigg( H_j(x,t) {\delta}_{,i}(x,x')
\delta(t,t') - (ix \leftrightarrow jx') \bigg) \bigg] , F(x'',t'') \} = 0.
\label{eq:HCDD3}
\end{equation}
Since it holds for any choice of the functional F, the following 
relation for the super-momenta arises,  
\begin{equation}
\{ {H}_i(x,t)  , {H}_j(x',t') \} =  {H}_j(x,t) {\delta}_i(x,x')
\delta(t,t') + C_{ij}[x,t;x',t'] - (ix \leftrightarrow jx').
\label{eq:HD3}
\end{equation}
The term $C_{ij}$ is just a constant function of its arguments.

\% The same argument can be applied to the mixed Jacobi identities
\begin{eqnarray}
&& \{ \{ H_j(x',t') , F(x'',t'') \} , H(x,t) \} + \{ \{ F(x'',t'') ,
H(x,t) \} , H_j(x',t') \}
\nonumber\\
&& \; \; \;  \; \; \; \; \; \; \; \; \; \;  \; \; \; \;
\; \; \; \; \; \;  + \{ \{ H(x,t)  , H_j(x',t') \} , F(x'',t'') \} = 0,   
\label{eq:CJ2}
\\
&& \{ \{ {H^D}_j(x',t') , F(x'',t'') \} , {H^D}(x,t) \} + \{ \{ F(x'',t'') ,
{H^D}(x,t) \} , {H^D}_j(x',t') \}
\nonumber\\
&& \;  \; \; \; \; \; \; \; \; \; \;  \; \; \; \;
\; \; \; \; \; \;  + \{ \{ {H^D}(x,t)  , {H^D}_j(x',t') \} ,
F(x'',t'') \} = 0, 
\label{eq:DJ2}   
\end{eqnarray}
resulting in the relation
\begin{equation}
\{ {H}(x,t)  , {H}_i(x',t') \} =  {H}(x,t) {\delta}_i(x,x')
\delta(t,t') + {H}_{,i}(x,t) {\delta}(x,x') \delta(t,t') + C_{i}[x,t;x',t'],
\label{eq:HD2}
\end{equation}
where $C_{i}$ is constant.

\% {\Large $\bullet$} Recovery Of The Super-momentum Constraint: The
situation changes considerably when the same argument is 
applied to the identities involving the super-Hamiltonians,
\begin{eqnarray}
&& \{ \{ H(x',t') , F(x'',t'') \} , H(x,t) \} + \{ \{ F(x'',t'') ,
H(x,t) \} , H(x',t') \}
\nonumber\\
&& \; \; \;  \; \; \; \; \; \; \; \; \; \;  \; \; \; \;
\; \; \; \; \; \;  + \{ \{ H(x,t)  , H(x',t') \} , F(x'',t'') \} = 0,   
\label{eq:CJ3}
\\
&& \{ \{ {H^D}(x',t') , F(x'',t'') \} , {H^D}(x,t) \} + \{ \{ F(x'',t'') ,
{H^D}(x,t) \} , {H^D}(x',t') \}
\nonumber\\
&& \;  \; \; \; \; \; \; \; \; \; \;  \; \; \; \;
\; \; \; \; \; \;  + \{ \{ {H^D}(x,t)  , {H^D}(x',t') \} ,
F(x'',t'') \} = 0. 
\label{eq:DJ3}   
\end{eqnarray}
This leads to the relation
\begin{equation}
\{ \{ H(x,t)  , H(x',t') \} , F(x'',t'') \} = \{ \{ {H^D}(x,t)  ,
{H^D}(x',t') \} , F(x'',t'') \} ,
\label{eq:HCD1}
\end{equation}
whose left and right side is evaluated on the canonical
and on the deformation phase space, respectively.

\% Considering the Poisson bracket between the deformation 
generators, the fact arises that the Dirac algebra is not a
genuine Lie algebra but depends explicitly on the spatial
metric,
\begin{equation}
\{ {H^D}(x,t)  , {H^D}(x',t') \} = g^{ij}(x,t) {{H^D}_i}(x,t)
\delta_{,j}(x,x') \delta(t,t') - \xxp .
\label{eq:HDD1}
\end{equation}
Since the theory is by assumption independent of any
fixed fields, it follows that the metric has to be a 
canonical variable in order to appear in equation (\ref{eq:HCD1}).

\% The evolution postulate and the fact that the  
metric is a canonical variable can be used to write equation
(\ref{eq:HCD1}) exclusively in terms of variables defined on the
canonical phase space[20],   
\begin{eqnarray}
&& \{ \bigg[ \{ H(x,t), H(x',t') \} - \bigg( g^{ij}(x,t) {H_i}(x,t)
\delta_{,j}(x,x') \delta(t,t') - \xxp \bigg) \bigg], F(x'',t'') \} 
\nonumber\\
&& = - \bigg( H_i(x,t) {\delta}_{,j}(x,x') \delta(t,t') \{ g^{ij}(x,t) ,
F(x'',t'') \} - \xxp \bigg) . 
\label{eq:111}
\end{eqnarray} 
The term on the right side is the compensation
needed in order that the metric be taken inside the
Poisson brackets in the canonical phase space.

\% Because equation (\ref{eq:111}) is a linear first order equation
that is required to hold for
an arbitrary choice of functional $F$, it cannot be satisfied 
unless the super-momenta are constrained to vanish,
\begin{equation}
H_i(x,t) \simeq 0.
\label{eq:tralari} 
\end{equation}
The proof is based on the following procedure.  Both sides of
equation (\ref{eq:111}) are expanded in terms of the spatial derivatives of the
$\delta$-functions. Then, because of the linearity and the specific form of
the equation, particular choices of functionals $F$ can be found that
violate at least one of the terms in the expansion.

\% {\Large $\bullet$} Recovery Of The Weak Representation Postulate:
The constraint (\ref{eq:tralari}) leads to 
\begin{equation}
\{ \bigg[ \{ H(x,t), H(x',t') \} - \bigg( g^{ij}(x,t) {H_i}(x,t)
\delta_{,j}(x,x') \delta(t,t') - \xxp \bigg) \bigg], F(x'',t'') \} \simeq 0 ,
\label{eq:1110}
\end{equation} 
which must still hold for every choice of functional $F$.
Teitelboim argued[19] that the weak equation
(\ref{eq:1110})---which in the equal-time approach is derived from the
principle of path independence---is enough to imply  
that the expression      
\begin{equation}
\{ H(x,t), H(x',t') \} - \bigg( g^{ij}(x,t) {H_i}(x,t)
\delta_{,j}(x,x') \delta(t,t') - \xxp \bigg) 
\label{eq:lathos}
\end{equation}
vanishes strongly. 
Specifically, he argued that (\ref{eq:lathos}) must not depend on any canonical
variables because, if it did, particular choices of functionals $F$
could always be found to violate equation (\ref{eq:1110}), in a
process similar to the one described above.  The quantity
(\ref{eq:lathos}) should therefore be equal to a constant function,
which is zero[19] because of the requirement that the algebra be
closed. The requirement of closure actually implies that the constant terms
$C_{ij}$ and $C_{i}$ in equations (\ref{eq:HD3}) and (\ref{eq:HD2})
should also be zero[19] and, hence, the history analogue of the strong
Dirac algebra is derived.

\% However, this argument is not true in general, because in a
constrained system it must be ensured that all the terms in
equation (\ref{eq:1110}) are well-defined on the constraint
surface. If any of the 
first partial  derivatives of (\ref{eq:lathos}) does not vanish on the
constraint surface, the argument in [19] can be applied indeed, and
leads to the conclusion that the expression (\ref{eq:lathos}) is zero
strongly. On the other hand, if both partial derivatives of
(\ref{eq:lathos}) vanish weakly,   
well-defined choices for functionals $F$ that violate equation
(\ref{eq:1110}) cannot be found, since this would require the first
partial derivatives of any such $F$ to 
have an infinite value on the constraint surface. Consequently, the
most general expression for the algebra between the super-Hamiltonians
is the weak Dirac relation mentioned in section 2,
\begin{equation}
\{ {H}(x,t)  , {H}(x',t') \} = g^{ij}(x,t) {{H}_i}(x,t)
\delta_{,j}(x,x') \delta(t,t') + G(x,t;x',t') - \xxp  ,
\label{eq:HDDD1}
\end{equation}
where both the first derivatives of $G$ vanish on the constraint
surface (\ref{eq:tralari}). Notice that any constant terms are 
absorbed in this definition of $G$.

\% The fact that the system is constrained in $H_i$ demands for the
re-examination of the assumptions that led to equations (\ref{eq:HD3}),
(\ref{eq:HD2}) and (\ref{eq:HDDD1}). The only requirement for the
validity of the previous procedure is the preservation of any weak
equality under the commutation with the canonical generators. However,
this is included already in the definition of the evolution postulate
for constrained systems, so any strong equality signs must be replaced simply
with weak ones. 

\% This replacement results in the complete history analogue of the weak Dirac
algebra (\ref{eq:E1}-\ref{eq:E3}) as well as in the weak re-shuffling and
ultra-locality conditions  and in the rest of the
weak evolution postulate. 
Notice that, although the term ``weak'' refers currently to the
constraint surface (\ref{eq:tralari}), the arguments used
do not depend on the actual definition of the constraint
surface. Therefore, the present conclusions will remain valid in case
that the super-Hamiltonian is proved to be constrained.

\% {\Large $\bullet$} The Principle Of Path Independence: The path
independence of the dynamical evolution does not have to be assumed
separately in the present method. Instead, it is a consequence of the
evolution postulate. This can be shown directly by starting from the
evolution postulate and the derived weak Dirac algebra, and then
repeating in the reverse order the procedure used in [19]. It follows
immediately that the change in the canonical
variables during the dynamical evolution of the theory is 
independent of the path used in its actual evaluation. 
An alternative proof uses the fact that the principle of
path independence is a direct consequence of the integrability of Hamilton's
equations. The evolution postulate is just another name for these
equations and, therefore, any solution of the postulate will
lead automatically to a path-independent dynamical evolution.

\hskip -0.25in
{\Large $\bullet$} {Recovery Of The Super-Hamiltonian Constraint:}
When the representation postulate is imposed in the weak sense, the
super-Hamiltonian constraint does not follow immediately from the
closure of the Dirac algebra, as in [20], but it is also 
necessary to take into account the actual form of equations
(\ref{eq:PP1}-\ref{eq:PP4}). These
equations are considered below in the case of general relativity or,
more accurately, in the case when the configuration variable is the
pullback of the spacetime metric.

\% Referring to the corresponding comment at the end of chapter III-3, the
most general 
form of the weak evolution postulate is the following: 
\begin{eqnarray}
&& \{ {g_{ij}}(x,t), H(x',t') \} =  2 n_{{\alpha};{\beta}}(x,t) {{\cal
X}^{\alpha}}_i(x,t) {{\cal X}^{\beta}}_j(x,t) \delta(x,x')
{\delta}(t,t') 
\nonumber\\
&& \; \; \; \; \; \; \; \; \; \; \; \; \; \; \; \; \; \; \; \; \; \;
\; \; \; \; \; \; \; \; \; \; \; \; + V_{ij}(x,t;x',t'),    
\label{eq:PPGR1}
\\
&& \{ {g_{ij}}(x,t), {H_k}(x',t') \} =  g_{ki}(x,t) {\delta}_{,j}(x,x')
 {\delta}(t,t') + g_{kj}(x,t) {\delta}_{,i}(x,x') {\delta}(t,t') 
\nonumber\\
&& \; \; \; \; \; \; \; \; \; \; \; \; \; \; \; \; \; \; \; \; \; \;
\; \; \; \; \; \; \; \; \; \; \; \;   +
g_{ij,k}(x,t) {\delta}(x,x') {\delta}(t,t') + V_{ijk}(x,t;x',t'),   
\label{eq:PPGR2}
\\ 
&& \{ {p^{ij}}(x,t), H(x',t') \} = \{ {p^{ij}}[{\cal X}(x,t)] ,
{H^D}(x',t') \} + W^{ij}(x,t;x',t'), 
\label{eq:PPGR3}
\\
&& \{ {p^{ij}}(x,t), {H_k}(x',t') \} = \{ {p^{ij}}[{\cal X}(x,t)] ,
{{H^D}_k}(x',t') \}  + {{W^{ij}}_k}(x,t;x',t') .
\label{eq:PPGR4}
\end{eqnarray}

\% The tensors $V_{ij}$, $V_{ijk}$, $W^{ij}$ and ${W^{ij}}_k$ depend on
the canonical fields and are required to vanish on the constraint
surface $H_i \simeq 0$. Because of the existence of the additional terms,
the general solution of the coupled set
(\ref{eq:PPGR1}-\ref{eq:PPGR4}) cannot be found explicitly. Nevertheless, the
form of the evolution postulate allows some definite conclusions to be drawn,
a part of which can be used to prove that the Hamiltonian is
constrained.

\% The important observation{[20]} is that the conjugate momentum
$p^{ij}$ must be a tensor density of weight one in order
that the form 
$p^{ij}{\delta}g_{ij}$ that appears in the canonical action be
coordinate independent. As a result, the Poisson brackets between the 
tangential deformation generator and $p^{ij}$ depend only on the
weight of the latter, and equation (\ref{eq:PPGR4}) becomes
\begin{eqnarray}
&& \{ {p^{ij}}(x,t), {H_k}(x',t') \} = {{\delta}^j}_k {p^{im}}(x,t) 
{\delta}_{,m}(x,x') {\delta}(t,t') + {{\delta}^i}_k {p^{jm}}(x,t) 
{\delta}_{,m}(x,x') {\delta}(t,t') 
\nonumber\\
&& \; \; \; \; \; \; \; \; \; \; \; \; \; \; \; \; \; \; \; \; \; \;
\; \; \; \; \; \; \; \; \; \; \; \;  - {p^{ij}}(x,t)
{\delta}_{,k}(x,x') {\delta}(t,t') - {p^{ij}}_{,k}(x,t) {\delta} (x,x')
{\delta}(t,t') 
\nonumber\\
&& \; \; \; \; \; \; \; \; \; \; \; \; \; \; \; \; \; \; \; \; \; \;
\; \; \; \; \; \; \; \; \; \; \; \;  + {{W^{ij}}_k}(x,t;x',t') . 
\label{eq:GRmom}
\end{eqnarray}

\% Consider therefore a solution $(H, H_i)$ of the system
(\ref{eq:PPGR1}-\ref{eq:PPGR4}), taking into account equation
(\ref{eq:GRmom}). By the 
assumption of existence of such a solution, the left sides of
equations (\ref{eq:PPGR2}) 
and (\ref{eq:GRmom}) must satisfy the integrability condition
\begin{equation}
\{ \{ g_{ij}(x,t) , {H_k}(x',t') \} , {p^{mn}}(x'',t'') \} = \{ \{
{p^{mn}}(x'',t''), {H_k}(x',t') \} , g_{ij}(x,t)  \} . 
\label{eq:gnwsto}
\end{equation}    
Because the non-vanishing terms in equations (\ref{eq:PPGR2})
and (\ref{eq:GRmom}) are integrable{[20]}, the weakly vanishing terms
in the same equations should also be integrable,
\begin{equation}
\{ V_{ijk}(x,t;x',t') , {p^{mn}}(x'',t'') \} = \{
{{W^{mn}}_k}(x'',t'';x',t'), g_{ij}(x,t)  \}   .
\label{eq:neo}
\end{equation} 
This implies that functionals ${H^*}_i$ and $K_i$ can be found, satisfying   
\begin{eqnarray}
&& \{ {g_{ij}}(x,t), {{H^*}_k}(x',t') \} = g_{ki}(x,t) {\delta}_{,j}(x,x')
 {\delta}(t,t') + g_{kj}(x,t) {\delta}_{,i}(x,x') {\delta}(t,t') 
\nonumber\\
&& \; \; \; \; \; \; \; \; \; \; \; \; \; \; \; \; \; \; \; \; \; \;
\; \; \; \; \; \; \; \; \; \; \; \;   +
g_{ij,k}(x,t) {\delta}(x,x') {\delta}(t,t')  ,
\label{eq:eid1}
\\
&& \{ {p^{ij}}(x,t), {{H^*}_k}(x',t') \} = {{\delta}^j}_k {p^{im}}(x,t) 
{\delta}_{,m}(x,x') {\delta}(t,t') + {{\delta}^i}_k {p^{jm}}(x,t) 
{\delta}_{,m}(x,x') {\delta}(t,t') 
\nonumber\\
&& \; \; \; \; \; \; \; \; \; \; \; \; \; \; \; \; \; \; \; \; \; \;
\; \; \; \; \; \; \; \; \; \; \; \;  - {p^{ij}}(x,t)
{\delta}_{,k}(x,x') {\delta}(t,t') - {p^{ij}}_k(x,t) {\delta} (x,x')
{\delta}(t,t') ,
\label{eq:eid2}
\\
&& \{ {g_{ij}}(x,t), {{K}_k}(x',t') \} = V_{ijk}(x,t;x',t') ,
\label{eq:eid3}
\\
&& \{ {p^{ij}}(x,t), {{K}_k}(x',t') \} = {{W^{ij}}_k}(x,t;x',t') .
\label{eq:eid4}
\end{eqnarray}

\% It follows from equations (\ref{eq:eid1}-\ref{eq:eid4}) 
that every solution $H_i$ of the weak evolution postulate can be
written as the sum of two terms,  
\begin{equation}
H_i  = {H^*}_i + K_i .
\label{eq:sum}
\end{equation} 
Furthermore, the form of ${H^*}_i$ is uniquely fixed by equations
(\ref{eq:eid1}) and (\ref{eq:eid2}), and corresponds to the
super-momentum of general relativity, 
\begin{equation}
{H^*}_i = \mom,
\label{eq:grisstar}
\end{equation} 
written out in equation (\ref{eq:mom}).

\% It can now be shown that the super-Hamiltonian of the theory is
constrained. As in [20], this follows from the preservation of the
super-momentum constraint under the dynamical evolution, resulting in the
condition    
\begin{equation}
\{ H(x,t) , H_i(x',t') \} \simeq 0.
\label{eq:fanfara}
\end{equation}    
Using equations (\ref{eq:sum}) and (\ref{eq:grisstar}), this
condition can be written as 
\begin{equation}
\{ H(x,t) , \bigg[ \mom(x',t') + K_i(x',t') \bigg]\} \simeq  0 
\label{eq:weaklymal}
\end{equation}
or, equivalently, as 
\begin{equation} 
\{ H(x,t) , \mom(x',t') \} \simeq 0 .
\label{eq:weaklymal2}
\end{equation}
This follows from equations
(\ref{eq:eid3}-\ref{eq:eid4}) and from the fact that 
${{W^{ij}}_k}$ and $V_{ijk}$ vanish on the constraint surface
(\ref{eq:tralari}).

\% The evolution postulate can be used once more to rewrite equation
(\ref{eq:weaklymal2}) as follows,
\begin{equation}
\{ H(x,t) , \mom(x',t') \} \simeq \{ H(x,t) , {H^{D}}_i(x',t')
\} \simeq 0 . 
\label{eq:persil}
\end{equation}
Notice that this is a special application of the evolution postulate
where the arbitrary test functional $F$ has been replaced by the
super-Hamiltonian.  
The latter must transform necessarily as a scalar
density of weight one[20],
\begin{equation}
\{ H(x,t) , {H^{D}}_i(x',t') \} \simeq  H(x,t) {\delta}_{,i}(x,x')
{\delta}(t,t') + H_{,i}(x,t) {\delta}(x,x') {\delta}(t,t') ,
\label{eq:psada}
\end{equation} 
so by combining equations (\ref{eq:persil}) and (\ref{eq:psada}) the
constraint $H \simeq 0$ arises.  
Recall that the actual definition of the constraint surface
does not affect the validity of any of the above arguments, and hence
the procedure just described remains consistent under the
additional constraint. 

\hskip -0.25in
{\Large $\bullet$} {The Representations Of The Weak Principle:} 
Although the understanding of the relationship between the strong and the weak equations is no longer an issue (in the revised algorithm no strong equations are used) there is still need to clarify the relation between the ``strong'' and ``weak'' representations of the evolution postulate. 
In particular, there is need to understand exclusively in terms of the evolution postulate how the standard representation of general
relativity arises and, also, to find out if the new
representations are physically equivalent to the standard one. ``Physically
equivalent'' means that they must generate weakly the same equations of motion
and lead to the same constraint surface.

\% A preliminary examination of this problem has already been carried out when proving that the super-Hamiltonian of the theory is constrained. 
Indeed, equation (\ref{eq:sum}) shows that the standard
representation of the super-momentum is recovered from the evolution
postulate as the special case $K_i = 0$. Also, equations
(\ref{eq:PPGR2}) and (\ref{eq:GRmom}) imply  that 
all solutions $H_i$ generate weakly the same equations of motion.
Finally, it follows from equation (\ref{eq:sum}) and from the fact that both partial derivatives of $K_i$ vanish on the constraint surface $H_i = 0$ that the constraints $H_i$ and $\mom$ imply each
other. The representations $H_i$ and $\mom$ are therefore physically
equivalent, and the privileged position occupied by $\mom$ is
merely because the standard description of the system is minimal.

\% On the other hand, whether the same is true for the representations of $H$ cannot be said without further examination. The basic complication arises because equation (\ref{eq:P1}) can only be inverted {\it implicitly} in order that the momenta be defined as functionals of the embedding. In addition, when the field velocities are replaced on the right side of equations (\ref{eq:PPGR1}) and (\ref{eq:PPGR3}), the resulting expressions are not the same for all representations. 
This spoils the method used when deriving the representations for the super-momentum.
 
\% The issue concerning the physical equivalence of the ``weak''
representations is therefore still unclear. It 
would be certainly interesting if     
representations could be found that are not equivalent to
the standard super-Hamiltonian, but this possibility is rather remote
considering the restrictions imposed on the spacetime character of any such
representations by Lovelock's theorem[31]. 

\% The question arises whether an additional postulate is missing, that could uniquely select the ``strong'' representation of general relativity. This is not difficult to be found, and simply corresponds to requiring all the weakly vanishing terms $V_{ij}$, $V_{ijk}$, $W^{ij}$ and ${W^{ij}}_k$ in equations (174)-(177) to be identically zero.
This requirement indeed leads to the standard canonical representation derived by Kucha\v{r} {\it et al} in [20].
However, it cannot be justified by any {\it physical} principle because, at the very least, it is a general property of constrained systems to allow many different sets of equivalent constraints. Therefore, the best that can be achieved is an actual proof that all the ``weak'' representations are equivalent. 

\% The conclusion that postulates (a) and (b) are the only assumptions needed in the derivation of the most general canonical representation is more transparent in the following chapter. There, the history algorithm is applied directly to a simple unconstrained system, namely the scalar field theory on a given metric background.

\vskip 8cm
\hskip -0.25in
{\LARGE Chapter IV-2. The Direct Method.}
\vskip 1cm 
\hskip -0.25in
Up to now, the intricate nature of general relativity has not allowed
any details of the history formalism to be revealed. For this reason,
a sufficiently simpler system is considered below. It is derived directly
from the evolution postulate, thus providing a clear illustration of the
new formalism.

\% {\Large $\bullet$} {Preliminaries:} 
A scalar field theory with a non-derivative coupling to the metric is
considered. Because of the restriction on the coupling, the Lagrangian
of the theory can be written in the following form,
\begin{equation}
{\cal L}(x,t)= {\cal L} \bigg[ {\f},{\fdot},N,g,N^{i}{\f}_{,i},g^{kj}
{\f}_{,k} {\f}_{,j} \bigg](x,t).
\label{eq:scfL}
\end{equation}
To be precise, the Lagrangian may depend on the additional combinations
\begin{eqnarray}
g^{ij}N_{,i}N_{,j} \; , 
\nonumber 
\\
g^{ij}{\f}_{,i}N_{,j} \; ,
\nonumber
\\
N^{i}N_{,i} \; ,
\nonumber
\\
g_{ij}N^{i}N^{j} \;  
\end{eqnarray}
which are also compatible with the assumption of the coupling. 
However, using similar arguments to the ones that follow, it can
be shown that this dependence is trivial. For
simplicity, it is taken to be trivial from the beginning.  

\% The evolution postulate states that the following conditions should
be satisfied, 
\begin{eqnarray}
&& \{ {\phi}(x,t), H(x',t') \} \simeq \{ {\phi}[{\cal
X}](x,t) , {{\cal P}_{\beta}}(x',t') \} n^{\beta}(x',t')  ,
\label{eq:SFEP1}
\\
&& \{ {\phi}(x,t), {H_i}(x',t') \} \simeq \{ {\phi}[{\cal
X}](x,t) , {{\cal P}_{\beta}}(x',t') \} {{{\cal
X}^{\beta}}_i}(x',t') ,  
\label{eq:SFEP2}
\\ 
&& \{ {\pi}(x,t), H(x',t') \} \simeq \{ {\pi}[{\cal X}](x,t) , {{\cal
P}_{\beta}}(x',t') \} n^{\beta}(x',t') ,   
\label{eq:SFEP3}
\\
&& \{ {\pi}(x,t), {H_i}(x',t') \} \simeq \{ {\pi}[{\cal X}](x,t) , {{\cal
P}_{\beta}}(x',t') \} {{{\cal X}^{\beta}}_i}(x',t'),  
\label{eq:SFEP4}
\end{eqnarray}
The weak sign refers to the still unknown Hamilton's equations.
Initially, the most general canonical representation for $H$ and
$H_i$ is seeked that satisfies the unconstrained version of the
evolution  postulate.

\% Recall that, according to the plan described in chapter III-3, any time
derivatives of the canonical variables in equations
(\ref{eq:SFEP1})-(\ref{eq:SFEP4}) should be replaced with the
corresponding expressions arising from Hamilton's equations,   
\begin{eqnarray}
&& {\partial \over {\partial}t}{\phi}(x,t) \simeq \int d^3x'dt' \{
\phi(x,t) , (N H + N^i H_i)(x',t') \}, 
\label{eq:Pipi1}
\\
&& {\partial \over {\partial}t}{\pi}(x,t) \simeq \int d^3x'dt' \{
\pi(x,t) , (N H + N^i H_i)(x',t') \}.
\label{eq:Pipi2}
\end{eqnarray}
This replacement results in a system of coupled
differential equations for the generators $H$ and $H_i$, the general
solution of which corresponds to the general canonical representation
compatible with the postulate.

\% The above procedure is rather formal, so it is replaced below
with one that is more suitable for the present purposes. Specifically,
the super-Hamiltonian is not considered to be the ``unknown''
of the problem but, instead, the Legendre relation between the field
velocity and the momentum takes its place. Because the
information incorporated in the latter is not equally rich, the need
that the Lagrangian be treated as a ``new'' unknown will arise at some
stage. 

\% {\Large $\bullet$} {The Unknown Legendre:} According to these
modifications, the momentum is expressed as a functional of the
embedding according to   
\begin{equation}
{\pi}[{\cal X}](x,t) := {\Pi}\bigg[ {\f},{\fdot},N,g,N^{i}{\f}_{,i},g^{kj}
{\f}_{,k} {\f}_{,j} \bigg](x,t) .
\label{eq:SFLEGENDRE}
\end{equation}
The function $\Pi$ is the required Legendre relation. Its form is the
most general possible, assuming that the Lagrangian of the
theory is given by equation (\ref{eq:scfL}). Notice that the variables
treated as arguments of $\Pi$ are independent of each other, since the
system is unconstrained by assumption. 

\% To proceed further, the following history Poisson brackets are needed:
\begin{eqnarray}
&& \{ {\phi}, {{{\cal P}_{\beta}}'} \} = {\phi}_{,\beta} \delta \delta ,
\label{eq:QW1}
\\
&& \{ \dot{\phi}, {{{\cal P}_{\beta}}'} \} = {\phi}_{,\beta} \delta
\dot{\delta} + {\phi}_{,\beta\alpha} {\dot{\cal X}}^{\alpha} \delta \delta , 
\label{eq:QW2}
\\
&& \{ N, {{\cal P}_{\beta}}'  \} = -n_{\beta} \delta
\dot{\delta} +n_{\beta} N^m {\delta}_{,m} \delta - {1 \over 2} N 
{{\gamma}_{\mu{\nu},\beta}} n^{\mu} n^{\nu}  \delta \delta ,
\label{eq:QW3}
\\
&& \{ g, {{\cal P}_{\beta}}'  \} = 2g {{\cal X}_{\beta}}^m
{\delta}_{,m} \delta + g {{\gamma}_{\mu{\nu},\beta}} {{\cal X}^{\mu}}_m
{{\cal X}^{\nu}}^m \delta \delta ,
\label{eq:QW4}
\\
&& \{ N^i {\phi}_{,i}, {{\cal P}_{\beta}}'  \} = {{\cal X}_{\beta}}^i
{\phi}_{,i} 
\delta \dot{\delta} + N n_{\beta} g^{im} {\phi}_{,i} {\delta}_{,m}
\delta - N^m
{{\cal X}_{\beta}}^i {\phi}_{,i} {\delta}_{,m} \delta + 
\nonumber\\
&& + N {{\gamma}_{\mu{\nu},\beta}} n^{\mu} {{\cal X}^{\nu}}^i {\phi}_{,i}
\delta \delta + {\phi}_{,\beta} N^i 
{\delta}_{,i} \delta + {\phi}_{,\beta\alpha} N^i {{\cal
X}^{\alpha}}_{,i} \delta \delta ,  
\label{eq:QW5}
\\
&& \{ g^{ij}{\phi}_{,i}{\phi}_{,j}, {{\cal P}_{\beta}}'  \} = -{{\cal
X}_{\beta}}^i g^{jm} {\phi}_{,i}{\phi}_{,j} {\delta}_{,m} \delta
-{{\cal X}_{\beta}}^j g^{im} {\phi}_{,i}{\phi}_{,j} {\delta}_{,m}
\delta +
\nonumber\\
&& + {\phi}_{,\beta} g^{ij} {\phi}_{,j}  
{\delta}_{,i} \delta + {\phi}_{,\beta\alpha} g^{ij} {\phi}_{,j} {{\cal
X}^{\alpha}}_{,i} \delta \delta - {{\gamma}_{\mu{\nu},\beta}} {{\cal
X}^{\mu}}^i {{\cal X}^{\nu}}^j {\phi}_{,i}{\phi}_{,j} \delta \delta . 
\label{eq:QW6}
\end{eqnarray}
In the above equations, $\delta(x,x') \delta(t,t')$, ${\partial \over
{\partial x^i}} \delta(x,x') \delta(t,t')$ and $\delta(x,x') {\partial
\over {\partial t}} \delta(t,t')$ are denoted, respectively, by $\delta
\delta$,  ${\delta}_{,i}$ $\delta$ and $\delta$ $\dot{\delta}$.
If an expression is evaluated at $(x',t')$ it is primed.

\% Starting from equations (\ref{eq:SFEP3})-(\ref{eq:SFEP4}), 
the terms that are proportional to the time derivative of
the delta function are considered, 
\begin{equation}
\bigg\{ {\p}[{\cal X}](x,t) , {{{\cal P}_{\beta}}'}(x',t') \bigg\} \simeq
\bigg[ {{\partial {\Pi}} \over {\partial \fdot}} {\f}_{,\beta} - {{\partial {\Pi}}
\over {\partial N}} n_{\beta} + {{\partial {\Pi}} \over 
{\partial [N^m{\f}_{,m} ]}} {\f }_{,k}  {{\cal X}_{\beta}}^k \bigg] \delta 
\dot{\delta} + \; \; rest \; \;  of \; \; terms.
\label{eq:DOT}
\end{equation}
Since the left side of the above equation is independent of
any time derivatives by construction, the terms multiplying the
derivative of the delta function must vanish weakly. However, the
system has been assumed to be unconstrained, so the only way
that these terms can vanish is that they vanish strongly.

\% Taking the normal and tangential projections of equation (\ref{eq:DOT}), two
differential equations are obtained, 
\begin{eqnarray}
&& {{\partial \Pi} \over {\partial \fdot}} N^{-1} [\fdot - N^m{\f}_{,m}
] + {{\partial \Pi} \over {\partial N}} = 0  \nonumber\\
&& {{\partial \Pi} \over {\partial [N^m{\f}_{,m} ]}} +  {{\partial \Pi}
\over {\partial \fdot}} = 0.
\label{eq:twoequ}
\end{eqnarray}
The most general Legendre relation $\Pi$ that solves the above
equations has the form
\begin{equation}
{\p}[{\cal X}](x,t) = {\Pi}\bigg[ N^{-1}[{\fdot}-N^{i}{\f}_{,i}],{\f}, g^{kj}
{\f}_{,k} {\f}_{,j}, g  \bigg](x,t). 
\label{eq:step1}
\end{equation}

\% Next, equations (\ref{eq:SFEP2}) and (\ref{eq:SFEP4}) are
compared. Because the field momentum must transform as a scalar 
density of weight one (appendix F), these equations fix the form of the
super-momentum uniquely. It is given by 
\begin{equation}
{H_i}(x,t) = \p {\f}_{,i} (x,t).
\label{eq:supermcc}
\end{equation}
To be precise, a function $c_i(x,t)$ of space and time should also be added
on the right side of the above equation, but it is set to zero because 
the Lagrangian is of the form (\ref{eq:scfL}).  

\% Recall that the requirement concerning the weight of $\pi$ has been the only
assumption used in the derivation of equation (\ref{eq:supermcc}). 
By imposing this requirement on the function $\Pi$ as well, and by
making use of the Poisson bracket relations
(\ref{eq:QW1})-(\ref{eq:QW6}), the following differential equation arises,
\begin{equation}
\Pi = [g^{1 \over 2}] {{\partial \Pi} \over {\partial {[g^{1 \over 2}]} }}.
\label{eq:trans}
\end{equation}
It admits the solution
\begin{equation}
{\p}[{\cal X}](x,t) = [g^{1 \over 2}] {\it \Pi}\bigg[
N^{-1}[{\fdot}-N^{i}{\f}_{,i}], g^{kj} {\f}_{,k} {\f}_{,j} , {\f} \bigg]
[{\cal X}(x,t)], 
\label{eq:step2}
\end{equation}
where ${\it \Pi}$ transforms as a scalar density of weight zero. It can 
be shown that ${\it \Pi}$ drops out of the remaining evolution
postulate, which means that all the information incorporated in ${\it
\Pi}$ has been used. The final representation for the 
latter is given by equation (\ref{eq:step2}).

\% {\Large $\bullet$} The Unknown Lagrangian: Returning to equations
(\ref{eq:SFEP3})-(\ref{eq:SFEP4}), the remaining terms are considered. The
relevant Poisson brackets are 
\begin{eqnarray}
&& \bigg\{ \bigg[ N^{-1}[{\fdot}-N^{i}{\f}_{,i}] \bigg], {\cal
P}_{\beta}  \bigg\} n^{\beta} = g^{km} {\f}_{,m} {\delta}_{,k} \delta
\; \; + \; \; ultralocal \; \; terms , 
\nonumber\\
&& \bigg\{ g^{1 \over 2}, {{\cal P}_{\beta}}'  \bigg\} n^{\beta}  = 0 \;
\; + \; \; ultralocal \; \; terms ,  
\nonumber\\
&& \bigg\{ [g^{mj} {\f}_{,m} {\f}_{,j}] , {{\cal P}_{\beta}}'  \bigg\}
n^{\beta}   = 2 \bigg[ N^{-1}[{\fdot}-N^{i}{\f}_{,i}] \bigg] g^{km}
{\f}_{,m} {\delta}_{,k} \delta \; \; + \; \; ultralocal \; \; terms ,   
\nonumber\\
&& \bigg\{ \f , {{\cal P}_{\beta}}'  \bigg\} n^{\beta} = 0 \;
\; + \; \; ultralocal \; \; terms,
\label{eq:help2}
\end{eqnarray}
and lead to the following equation,
\begin{equation}
\bigg\{ \p , H'  \bigg\} = g^{1 \over 2}
\bigg[ {{\partial {\Pi}} \over {\partial \bigg[
N^{-1}[{\fdot}-N^{i}{\f}_{,i}] \bigg]}} + 2
N^{-1}[{\fdot}-N^{i}{\f}_{,i}]  {{\partial {\Pi}} \over {\partial
{[g^{kj} {\f}_{,k} {\f}_{,j}] } }}  \bigg] g^{km} {\f}_{,m}
{\delta}_{,k} \delta. 
\label{eq:trexa}
\end{equation}

\% The Lagrangian (\ref{eq:scfL}) is now treated as the new
unknown of the problem. The super-Hamiltonian on the left side of equation
(\ref{eq:trexa}) is expressed as a function of 
$\pi$ and ${\cal L}$ according to the usual Legendre definition
\begin{equation}
H := {1 \over N} \Bigg[ \bigg(\pi \dot{\phi} - {\cal L} \bigg) - N^i
H_i \Bigg] 
\label{eq:gamoto}
\end{equation}
and, similarly, the function $\Pi$ appearing on the right side of
(\ref{eq:trexa}) is replaced by the definition
\begin{equation}
\Pi := {\partial \over {\partial \dot{\phi}}}{\cal L}.
\label{eq:gamoto2}
\end{equation}

\% Putting the definitions (\ref{eq:gamoto})-(\ref{eq:gamoto2}) in
equation (\ref{eq:trexa}), the quantities $H_i$, $\dot{\phi}$, $N$ and
$N^i$ drop out identically and a differential equation arises,
\begin{equation}
{L}_{RR} + 2 R {L}_{RS} + 2 {L}_{S} = 0.
\label{eq:skato}
\end{equation}
As before, the subscripts indicate partial differentiation.

\% The function $L$ is defined by
\begin{equation}
{\cal L} := N g^{1 \over 2} L,
\label{eq:skato0}
\end{equation}
while the quantities $R$ and $S$ are defined by 
\begin{equation}
R := N^{-1}[{\fdot}-N^{i}{\f}_{,i}]
\label{eq:skato1}
\end{equation}
and
\begin{equation}
S := g^{kj} {\f}_{,k} {\f}_{,j}.
\label{eq:skato2}
\end{equation}

\% The general solution of (\ref{eq:skato}) fixes the final form of
${\cal L}$,  
\begin{equation}
{\cal L} = N g^{1 \over 2} L[ (R^2-S), \f ],
\label{eq:skato6}
\end{equation}
which can be recognized as the most general Lagrangian for a field
theory with a non-derivative coupling to the metric[19]. In order that
the derivation be complete, the ultra-local terms in equations
(\ref{eq:SFEP3})-(\ref{eq:SFEP4}) must be compared. These terms are cancelled
identically and, therefore, overall consistency is reached.

\pagebreak

\begin{center}
{\LARGE PART V:}
\vskip 9cm
{\LARGE The Geometric Interpretation Of The New Algebra.}
\end{center}

\pagebreak

\% {\LARGE Chapter V-1. The Generators Of Deformations.}
\vskip 1cm 
\hskip -0.25in
{\Large $\bullet$} {Preliminaries:} The method developed in [20] is used
below as a means of finding the physical solutions of the genuine Lie
algebra. As explained in chapters III-1 and IV-1, this is possible
precisely because the ultra-locality and re-shuffling assumptions
follow uniquely from the evolution postulate. The argument that leads
to the required representations is based on the following
observation. 

\% If the canonical generators $K$, $K_i$ are pure functionals of the canonical
variables, they satisfy exactly the same algebra that is satisfied by the
corresponding generators of hyper-surface deformations. It is only when
prescribed fields are present in the theory that the canonical algebra may 
differ, according to the discussion in chapter IV-1. A representation
for the deformation generators $K^D$, $K^D_i$ arises when the
deformation vector field is decomposed in an appropriate basis
$({\mu}^{\alpha}, {{\mu}^{\alpha}}_i)$,  
\begin{eqnarray}
&& K := {\mu}^{\alpha}[{\cal X}]P_{\alpha},
\label{eq:mlkke}
\\ 
&& K_i := {{\mu}^{\alpha}}_i[{\cal X}]P_{\alpha},   
\label{eq:mlkkel}
\end{eqnarray}
so that the genuine Lie algebra is satisfied,
\begin{eqnarray}
\{K(x,t),K(x',t')\}&=&0 , \,\label{eq:BK1} \\
\{K(x,t), K_i(x',t')\}&=&K_{,i}(x,t)\delta(x,x') \delta(t,t') + K(x,t)
\delta_{,i}(x,x') \delta(t,t') , \, 
\label{eq:BK2}\\
\{ K_i (x,t), K_j (x',t')\}&=&K_j(x,t)\delta_{,i}(x,x') \delta(t,t')
-\xxpp . \label{eq:BK3}  
\end {eqnarray}

\% If the basis $({\mu}^{\alpha}, {{\mu}^{\alpha}}_i)$ is used
in the decomposition of the spacetime Lagrangian the
canonical generators are guaranteed to satisfy equations
(\ref{eq:BK1})-(\ref{eq:BK3}), provided that no ``remnants'' of
${\mu}^{\alpha}$ or ${{\mu}^{\alpha}}_i$ have been left in the theory
in the form of prescribed functions. This is always the case 
for an unconstrained system, since any such remnants can
be eliminated through parametrization. On the other hand, for a system that is
constrained, the situation is more complicated and additional
assumptions have to be made. These are discussed in part VI.

\% {\Large $\bullet$} The Representation For The Deformation
Generators: The most general representation for $K^D$ and $K^D_i$ is
required, treating the embedding ${\cal X}^{\alpha}$ and its conjugate
momentum $P_{\alpha}$ as the sole variables. Here, there are no
ultra-locality or re-shuffling requirements because the generators
$K^D$ and $K^D_i$ are unknown themselves. However, there is
still sufficient information to select the representation uniquely. The
requirement that plays the role of the additional selection 
criteria is evidently the linearity in the momentum conjugate to the
embedding.

\% The problem should not be made unnecessarily complicated, so the
basis ${{\mu}^{\alpha}}_i$ is identified with the usual
coordinate basis ${{\cal X}^{\alpha}}_i$. Accordingly, the
identification of $K^D_i$ with the generator of hyper-surface
deformations $H^D_i$ is implied. In this case, equation (\ref{eq:BK3})
is satisfied identically while the second relation (\ref{eq:BK2})
depends only on the transformation properties of $K^D$ and, for the
moment, can be ignored. 

\% Using equations (\ref{eq:mlkke}) and (\ref{eq:mlkkel}), the Poisson
bracket in equation (\ref{eq:BK1}) is expanded according to  
\begin{equation}
P_{\alpha}(x,t) \{ {\mu}^{\alpha}(x,t) , P_{\beta}(x',t') \}
{\mu}^{\beta}(x',t') - (x \leftrightarrow x') (t \leftrightarrow t') = 0 
\label{eq:edrft}
\end{equation}
or, equivalently, according to 
\begin{equation}
P_{\alpha}(x,t) {\mu}^{\beta}(x',t') { {\delta {\mu}^{\alpha}} \over
{\delta {\cal X}^{\beta}} }(x,t) \delta(x,x') \delta(t,t') - (x
\leftrightarrow x') (t \leftrightarrow t') = 0 . 
\label{eq:functionK}
\end{equation}
The way in which the functional derivative of ${\mu}^{\alpha}$ is
expressed reflects the
assumption that ${\mu}^{\alpha}$ is a local functional of the
embedding; that is, a function of ${\cal X}$ and of a finite number of
its spatial and time derivatives. Under this assumption, the
functional derivative can be expanded according to
\begin{equation}
{ {\delta {\mu}^{\alpha}} \over {\delta {\cal X}^{\beta}} } := 
{ {\partial {\mu}^{\alpha}} \over {\partial {\cal X}^{\beta}} } + 
{ {\partial {\mu}^{\alpha}} \over {\partial {\cal X}^{\beta}_i} }
{\partial}_i +
{ {\partial {\mu}^{\alpha}} \over {\partial {\dot{\cal X}}^{\beta}} }
{\dot{\partial}} +
{ {\partial {\mu}^{\alpha}} \over {\partial {\dot{\cal X}}^{\beta}_i} }
{\dot{\partial}_i} + \; . \; . \; . \; ,
\label{eq:expans}
\end{equation}
where the series has finite terms. 

\% Next, the various terms in (\ref{eq:functionK}) are evaluated at $(x,t)$
by using the series of identities
\begin{eqnarray}
A(x',t') \delta \delta &=& A \delta \delta   ,
\nonumber
\\
A(x',t') \delta {\dot \delta} &=& A \delta {\dot \delta} +
\dot{A} \delta \delta  ,
\nonumber
\\
A(x',t') \delta {\ddot \delta} &=& A \delta {\ddot \delta} +    
2 {\dot A} \delta {\dot \delta} +  {\ddot A} \delta \delta ,
\nonumber
\\
&{e.t.c. ,}&
\nonumber
\\
A(x',t') {\delta}_{,i} \delta &=& A {\delta}_{,i} \delta
+ A_{,i} \delta \delta  ,
\nonumber
\\
A(x',t') {\delta}_{,ij} \delta &=& A {\delta}_{,ij} \delta 
+  A_{,i} {\delta}_{,j} \delta +  A_{,j} {\delta}_{,i} \delta  +
A_{,ij} \delta  \delta,
\nonumber
\\
&{e.t.c. ,}&
\nonumber
\\
A(x',t') {\delta}_{,i} \dot{\delta} &=&  A(x',t) {\delta}_{,i}
\dot{\delta} +  \dot{A}(x',t) {\delta}_{,i} \delta ,
\nonumber
\\
A(x',t') {\delta}_{,i} \dot{\delta} &=& A(x,t') {\delta}_{,i}
\dot{\delta} + A_{,i}(x,t') \delta \dot{\delta}   ,
\nonumber
\\
A(x',t') {\delta}_{,i} \dot{\delta} &=&  A {\delta}_{,i} \dot{\delta}
+ \dot{A} {\delta}_{,i} \delta  + A_{,i} \delta \dot{\delta}   
+ \dot{A_{,i}} \delta \delta ,
\nonumber
\\
&{e.t.c. ,}&
\label{eq:rhyuikw}
\end{eqnarray}
and are then collected according to the corresponding derivatives of
the $\delta$-functions. The notation for the
$\delta$-functions is the same as the one used before.

\% Because of equation (\ref{eq:expans}), each
particular term has to vanish. This implies that 
the partial derivatives of ${\mu}^{\alpha}$ with respect to any spatial
or time derivatives of the embedding must be zero. Specifically, an iteration arises for the partial derivatives of 
${\mu}^{\alpha}$, which then reduces to zero because the partial
derivatives with respect to the highest-order spatial and time
derivatives of the embedding have to be trivial. Crucial to the proof
is the assumption that ${\mu}^{\alpha}$ is a local functional of the
embedding, so that highest-order spatial and time derivatives exist indeed.

\% Furthermore, because of the symmetry under the interchange of
$K^D(x,t)$ with $K^D(x',t')$, the ultra-local terms in equation
(\ref{eq:expans}) vanish identically. As a result, no condition is
imposed on the partial derivative of ${\mu}^{\alpha}$ with respect to
the embedding. The general solution for $K^D$ then takes the form 
\begin{equation}
K^D(x,t) = {\mu}^{\alpha}({\cal X})(x,t)P_{\alpha}(x,t) ,
\label{eq:hkjipye}
\end{equation}
where ${\mu}^{\alpha}({\cal X})$ is an ultra-local function of the embedding.
This combination transforms as a scalar density of weight-one in equation
(\ref{eq:BK2}) and provides the final solution to the representation
problem.

\% The fact that ${\mu}^{\alpha}$ is an ultra-local function of the
embedding implies that ${\mu}^{\alpha}({\cal X})$ corresponds to the
pull-back of a spacetime vector field. This is the 
required geometric interpretation of the new Lie algebra. 
\vskip 9cm
\hskip -0.25in
{\LARGE Chapter V-2. The Canonical Generators.}
\vskip 1cm
\hskip -0.25in 
{\Large $\bullet$} {The Decomposition Of The Spacetime Theory:}
Having fixed the interpretation of the
corresponding deformation generators, the evolution postulate can be
used in exactly the same way as in parts III and IV. It determines the
corresponding ultra-locality and re-shuffling conditions, the form of
the canonical algebra and, finally, the actual form of the canonical
representations. Equivalently, the spacetime action of the theory can
be decomposed in terms of the new basis $({\mu}^{\alpha}, {{\cal
X}^{\alpha}}_i)$. This is a much simpler procedure, so it is the one
that is followed below.

\% Suppose that the vector field ${\mu}^{\alpha}$ is decomposed in the usual
$({n}^{\alpha}, {{\cal X}^{\alpha}}_i)$ basis according to
\begin{equation}
{\mu}^{\alpha} := A {n}^{\alpha} + B^i {{\cal X}^{\alpha}}_i .
\label{eq:sdfrghj}
\end{equation}
The functions $A$ and $B^i$ are identified as the normal
and tangential projections of the vector field,
\begin{eqnarray}
&& A = - {n}_{\alpha} {\mu}^{\alpha} ,
\label{eq:qerty}
\\
&& B^i ={{\cal X}_{\alpha}}^i {\mu}^{\alpha}  .
\label{eq:perty}
\end{eqnarray}
They transform as a spatial scalar and as a spatial vector respectively. 
Also suppose that the deformation vector of the foliation can be
expanded in terms of the basis $({\mu}^{\alpha}, {{\cal X}^{\alpha}}_i)$
according to 
\begin{equation}
{\dot{\cal X}}^{\alpha} = M {\mu}^{\alpha} +  M^i {{\cal X}^{\alpha}}_i .
\label{eq:lasdqkmbf}
\end{equation}
The functions $M$ and $M^i$ can be viewed as the ``new lapse'' and the
``new shift''.

\% The unconstrained canonical action (\ref{eq:UnA}) is then written in
terms of $A$, $B^i$, $M$ and $M^i$ as follows,
\begin{eqnarray} 
&& S[q^A , p_A] = \int d^3xdt \bigg( p_A \dot{q^A} - {\cal H} \bigg),
\nonumber\\
&& {\cal H} = M K + M^i K_i ,
\nonumber\\
&& K = A H + B^i H_i 
\nonumber\\
&& K_i = H_i .
\label{eq:UnA2}
\end{eqnarray}
The generators $H$ and $H_i$ are the usual generators arising from the 
decomposition of the action with respect to the lapse function and
shift vector. The functions  
$A$ and $B^i$ are precisely the ``remnants'' of the vector field
${\mu}^{\alpha}$. This means that the generators
$K$ and $H_i$ may not close according to the genuine algebra
(\ref{eq:BK1})-(\ref{eq:BK1}), and indeed they do not. As a result,
the theory has to be parametrized.

\hskip -0.25in
{\Large $\bullet$} An Application:
As an example, the action for a massless scalar field is
considered. It is decomposed in terms 
of the new basis. The corresponding canonical generators are given by
\begin{eqnarray}
K^{\phi} &=& { \pi^2 \over {2 g^{1 \over 2}} } + {g^{1 \over 2} \over
2} g^{ij} {\phi}_{,i} {\phi}_{,j}  , 
\label{eq:hphi2}\\
\momphi &=&\pi\phi_{,i} .
\label{eq:mphi2}
\end{eqnarray}

\% When the theory is parametrized, the non-physical
degrees of freedom $({\cal X}^{\alpha}, P_{\alpha})$ imply the constraints
\begin{eqnarray}
&& K^T := K^{\phi} + {\mu}^{\alpha} P_{\alpha} \simeq 0,
\label{eq:qws}
\\
&& {H^T}_i := \momphi + {{\cal X}^{\alpha}}_i  P_{\alpha} \simeq 0.
\label{eq:qwsa}
\end{eqnarray}
Notice that the constraints have been projected along the
appropriate basis 
$({\mu}^{\alpha}, {{\cal X}^{\alpha}}_i)$.

\% Using the definitions (\ref{eq:qerty})-(\ref{eq:perty}), and the
history Poisson brackets appearing in Appendix C, the required result
arises:
\begin{equation}
\{ K^T(x,t) , K^T(x',t') \} = 0 .
\end{equation}
In addition, $K$ transforms as
a scalar density of weight one in equation (\ref{eq:BK2}), so the
complete genuine Lie algebra is generated.

\pagebreak

\begin{center}
{\LARGE Part VI:}
\vskip 9cm
{\LARGE Discussion And Acknowledgments}
\end{center}

\pagebreak

\% {\LARGE Chapter VI-1. The Results, Their Relation, Their Extent And Limitations.}
\vskip 1cm
\hskip -0.25in
{\Large $\bullet$} {The Genuine Algebra:} 
A genuine Lie algebra was discovered by Brown and Kucha\v{r} in the context of gravity coupled to matter fields. A differential equation was constructed by Markopoulou that is satisfied by any scalar combinations of the gravitational constraints that close according to this algebra. Despite that, little was known about the significance of this equation either in gravity coupled to matter or in vacuum gravity. An insight into the origin of the algebra in the coupled theory was gained by constructing an action whose variation leads precisely to the general solution of Markopoulou's equation. 

\% However, this was not achieved via a canonical reduction, the physical meaning of which is transparent. Instead, the required algebraic manipulations were performed ``outside'' the canonical method, thus failing to connect the new algebra with the physical relevance of the theory. In particular, it remains unclear whether the new algebra can be maintained after the elimination of the scalar field momentum from the coupled action. Such an elimination is usually followed by a parametrization of the theory in terms of a privileged ``time'' associated with the (non-canonical) scalar field. If it can be shown that the new algebra is still present in the reduced action, a clear interpretation of the algebra in terms of ``matter-time'' will be achieved. This possibility was not explored here, and remains as a project for the future.

\% Priority was given instead to understanding the importance of the new algebra in vacuum relativity. It was shown that all the gravitational combinations derived from the action functional generate a time evolution that is either zero or ill-defined on the constraint surface of the {\it vacuum} theory. As a result, alternative combinations were seeked which not only satisfy the new algebra but also generate the appropriate dynamical evolution of Einstein's theory.

\% {\Large $\bullet$} {The Use Of Classical Histories In The Thesis:} For this purpose, an algorithm was generalized, originally used by Kucha\v{r} {\it et al} in the derivation of geometrodynamics from first principles. The need to adapt this algorithm to the requirements of the new algebra eventually led to the concept of classical histories. By using a
Hamiltonian formalism defined over the space of histories, it was
shown that the canonical representations of any theory can be derived from a minimal set of postulates. These depend only on the foliation through which the original spacetime theory is decomposed; therefore, they have a clear geometric interpretation.

\% This clarification---which is essential for the application of the algorithm to the new algebra---can only be achieved in the history formalism. This is simply because the new set of postulates contain derivatives of the canonical variables, and these cannot be defined in the standard approach. The implications for the quantization of general relativity are straightforward, and provide additional support for the consistent histories approach to quantum gravity[7,8,9,10].

\% A mistake in the original algorithm was also corrected. For
unconstrained systems, both the old and new algorithms give identical results. For systems subject to constraints, the revised algorithm implies that certain strong equations (in the sense of Dirac) have to be replaced by weak ones. As a result, new (weak) canonical representations of the evolution postulate arise.

\% The issue of the interpretation of the new algebra was finally
discussed. The interpretation amounts to the decomposition of the
deformation vector in terms of a particular foliation (other than the standard lapse-shift) so that the projections of the embedding momenta on the spacetime basis associated with the foliation close according to the given algebra. For the particular case of the new algebra the
appropriate decomposition was shown to involve the projection on a
spacetime vector field which is not normalized.

\% Having fixed the issue of the interpretation of the new algebra
the generators of any canonical field theory are then made compatible with it. This is achieved by decomposing the spacetime Lagrangian with respect to the new spacetime basis and then parametrizing the result. The 
parametrization is essential in order that the canonical generators
be pure functionals of the canonical variables. If this is not
the case---i.e., if remnants of the spacetime basis are left in the
theory---the procedure may not be effective.

\hskip -0.25in
{\Large $\bullet$} Self-Commuting Combinations In The Vacuum Theory:
Unfortunately, general relativity is a constrained system, so the
remnants arising from the vector field cannot be parametrized 
without changing the physical content of the theory.  In the usual
A.D.M. decomposition the corresponding remnants drop out of the
canonical action because of the normalization conditions imposed on
the normal vector field. Such a normalization cannot be used in the
case of a spacetime vector field, which must be by definition independent of the embedding.

\% However, having located the root of the problem, it may be possible that this final difficulty can be overcome, possibly through the use of a covariant normalization condition on the new spacetime basis. 
If such a normalization can be attained the resulting gravitational constraints would be of particular importance, considering that the geometric interpretation of the new algebra is unique.

\hskip -0.25in
{\Large $\bullet$} {Histories And Explicit Spacetime Invariance:}
An objection raised against canonical quantization of field theories in general, and Einstein's theory of spacetime in particular, is that the canonical formulation depends on a foliation and necessarily destroys the spacetime picture since it fails to keep track of the spacetime invariances of the action. In canonical quantum field theory, divergences are typically encountered when posing questions about probabilities at a single instant of time (a spacelike hypersurface). Such problems prompted previous attempts to establish the canonical structure in spacetime setting; either by using the Peirels bracket[32], or by imposing symplectic structure and inner product on the space of solutions[30], or by working in the multi-symplectic formalism[33].

\% The history symplectic structure is rather complicated for spacetime fields. Even for a scalar field on a given background the momentum conjugate to the spacetime field should be a spacetime scalar density, while the standard momentum depends on an auxiliary structure. 
The history symplectic structure that was used in the thesis depends on such a foliation, since the canonical fields are functions of the hypersurface coordinates $(x^i,t)$ rather than of the original spacetime coordinates $X^{\alpha}$. In order that the two pairs of coordinates be related, a foliation $X^{\alpha}={\cal X}^{\alpha}(x^i,t)$ of spacetime into hypersurfaces is required. The possibility of constructing a {\it spacetime} history canonical formalism is currently investigated in collaboration with K. Kucha\v{r}[21]

\hskip -0.25in
{\Large $\bullet$} {The Consistent Histories Approach:}
It is remarkable that most of the results described here could not
have been achieved in the equal-time formulation without adding
further structure. Whether the superiority of the history
formalism can be established as a mathematical theorem depends on whether it is possible to construct a direct link between the equal-time and the history approaches. The clarification of this matter will certainly affect the way in which the consistent histories program is viewed. At the very least, there is the option of reformulating the complete canonical framework in terms of classical histories and searching for the classical analogues of the structures used in the constistent histories aprroach. At the very least, a classical insight into the idea of coarse-graining and into the related issue of probability is expected to be gained. 

\pagebreak      

\hskip -0.25in
{\LARGE Acknowledgements.}
\vskip 1cm       
\hskip -0.25in
I am indebted to the Alexander S. Onassis Public Benefit
Foundation for the financial support they have provided. I am also  
grateful to Karel Kucha\v{r} for a careful reading of the preprints
and for suggestions. The thesis is dedicated to my supervisor
Chris Isham, to my family
$A$$\varrho$$\tau$$\varepsilon$$\mu$$\eta$,
$\Gamma$$\iota$$\omega$$\varrho$$\gamma$$o$,
$\Theta$$o$$\delta$$\omega$$\varrho$$o$ and to my friends
$K$$\omega$$\sigma$$\tau$$\alpha$, $N$$\iota$$\kappa$$o$, 
$N$$\tau$$\iota$$\nu$$\alpha$, $X$$\alpha$$\varrho$$\eta$. 

\pagebreak

\% {\LARGE Appendix A.}
\vskip 1cm
\hskip -0.25in
The equivalence between cases 
(\ref{eq:a}) and (\ref{eq:b}) is demonstrated; the same argument applies when
comparing any two cases from the set (\ref{eq:a})--(\ref{eq:d}). The sets of functions satisfying equations (\ref{eq:a}) and
(\ref{eq:b}) are denoted respectively by $(\mtilde_1,\ltilde_1,R_1)$ and
$(\mtilde_2,\ltilde_2,R_2)$. For the two cases to be equivalent
the following three conditions must be
satisfied: 
\begin{enumerate}
\item
$\mtilde_1$, $\ltilde_1$, $R_1$ should satisfy (\ref{eq:a})---the
functions $\mtilde_1$,
$\ltilde_1$ regarded as known---
\item
$\mtilde_2$, $\ltilde_2$, $R_2$ should satisfy (\ref{eq:b}) ~~~~~~~~~~~~~~~~~~and
\item
$W[h,f]={\ltilde_1}^{\omega\over2}(h-\mtilde_1+\rtilde_1)^{\omega\over2}=
{\ltilde_2}^{\omega\over2}(h-\mtilde_2+\rtilde_2)^{\omega\over2}$.
\end{enumerate}

\% To avoid comparing directly the differential equations arising from conditions 1 and 2 the following trick is performed. 
Equations (\ref{eq:a}) and (\ref{eq:b}) are inserted into
equations  (\ref{eq:vf}) that determine the partial
derivatives of $W[h,f]$. By condition 3, these equations must be the
same for both cases. As a result, the differential equations 
(\ref{eq:a}) and (\ref{eq:b}) are trasformed into the pair of algebraic equations
\begin{equation}
W_h={{\ltilde_1}^{\omega\over2}(h-\mtilde_1+\rtilde_1)^{\omega\over2}\over\rtilde_1}=-
{({\ltilde_2})^{\omega\over2}(h-\mtilde_2+\rtilde_2)^{\omega\over2}\over\rtilde_2}
\label{eq:iiia}
\end{equation}
\begin{equation}
W_f={{\ltilde_1}^{\omega\over2}(h-\mtilde_1+\rtilde_1)^{{\omega-2}\over2}\over\rtilde_1}={{\ltilde_2}^{\omega\over2}(h-\mtilde_2+\rtilde_2)^{{\omega-2}\over2}\over\rtilde_2}.
\label{eq:iiib}
\end{equation}
By  comparing condition 3 with equations (\ref{eq:iiia}) and
(\ref{eq:iiib}) the following {\it consistent} solution arises (notice that the system is
over--determined):  
\begin{equation}
\mtilde_2=2h-\mtilde_1\qquad{\ltilde_2}^{\omega\over2}=({-\ltilde_1})^{\omega\over2}\qquad
\rtilde_2=-\rtilde_1.
\end{equation}
This proves equivalence.

\% For completeness, the corresponding results arising from comparing
cases (\ref{eq:a}) with (\ref{eq:c}) and (\ref{eq:a}) with
(\ref{eq:d}) are written down:
\begin{equation}
\mtilde_3=2h-\mtilde_1,\qquad
({\ltilde_3})^{\omega\over2}={({-\ltilde_1})^{\omega\over2}(h-\mtilde_1+\rtilde_1)\over
(h-\mtilde_1-\rtilde_1)},\qquad
\rtilde_3=\rtilde_1,
\end{equation}
\begin{equation}
\mtilde_4=\mtilde_1,\qquad
({\ltilde_4})^{\omega\over2}={({-\ltilde_1})^{\omega\over2}(h-\mtilde_1+\rtilde_1)\over
(h-\mtilde_1-\rtilde_1)},\qquad
\rtilde_4=-\rtilde_1,
\end{equation}
where $(\mtilde_3,\ltilde_3)$ and $(\mtilde_4,\ltilde_4)$
satisfy respectively equations (\ref{eq:c}) and (\ref{eq:d}).

\vskip 9cm
\% {\LARGE Appendix B.}
\vskip 1cm
\% It is shown that the linear and the weight-$\omega$ equations are
equivalent. The ansatz relation (\ref{eq:ansatz}) can be considered 
as an one-parameter family of maps, $a_{\omega}$, from  the set of functions
$(\mtilde,\ltilde, R)$ 
satisfying equation (\ref{eq:a}) to the set of solutions $W_{\omega}$ of
the corresponding 
weight-$\omega$ differential equation, on the assumption that $W_{\omega}$,
$W_{{\omega}H}$ and $W_{{\omega}F}$ are not identically zero (see the
remark at the beginning of this chapter).
The reason that $R$ is included in the set of functions
$(\mu,\lambda,R)$ is that 
although $R$ is a function of $\mu$, defined by equation
(\ref{eq:root}), it is not fully
specified by $\mu$ due to the sign ambiguity.

\% One would like to know whether the map $a_{\omega}$ is
one-to-one and, most importantly, whether it is onto. To check the
latter,  one
supposes that $W$ is any solution of the $\omega$--equation
(\ref{eq:WPDE}), where---for simplicity---the subscript 
$\omega$ of $W$ is omitted. A set of functions  $(\mtilde$,
$\ltilde$, $R)$ obeying the linear equation 
(\ref{eq:a}) is therefore required, with the property of 
producing through the $a_{\omega}$ map the
given solution $W$. This is similar to the procedure followed
in Appendix A in order to show that equations
(\ref{eq:a})--(\ref{eq:d}) are equivalent;  
the difference is that the requirement that  
 at least one of the cases (\ref{eq:a})--(\ref{eq:d}) leads to 
$W$ is lifted---$W$  is now only confined to obey
equation (\ref{eq:WPDE}) for some weight $\omega$.

\% The three conditions that  must be  satisfied  are:
\begin{enumerate}
\item The original differential equation
\begin{equation}
{\omega\over2}W W_f=f W_f^2-{1\over 4}
W_h^2,\qquad W\neq 0\quad 
W_f\neq 0\quad W_h\neq 0.
\end{equation}
\item  The linear equation
\begin{equation}
{1\over\ltilde}\lambda_f
-{1\over\rtilde}\mu_f=0
\quad{\rm and}\quad
{1\over\ltilde}\lambda_h
-{1\over\rtilde}\mu_h=0;
\qquad
R=\sqrt{\bigl(h-\mu\bigr)^2-f}.
\end{equation}
\item The ansatz relation
\begin{equation}
W=\ltilde^{\omega\over2}\biggl(h-\mtilde+
\sqrt{(h-\mtilde)^2-f} 
\biggr)^{\omega\over2}.\label{eq:36}
\end{equation}
\end{enumerate}

\% The third condition can be written as 
\begin{equation}
W^{2\over{\omega}}=\ltilde\biggl(h-\mtilde+
\sqrt{(h-\mtilde)^2-f} 
\biggr),
\label{eq:366}
\end{equation}
where this relation is valid up to a $2\over{\omega}$ power of
unity.   
Equation (\ref{eq:366}) can now be solved for $\mtilde$,  resulting in
\begin{equation}
\mtilde=h-{1\over2}\Bigl({W^{2\over{\omega}}\over\ltilde}+
{\ltilde\over W^{2\over{\omega}}}f\Bigr).
\label{eq:3a}
\end{equation}

\% Differentiating $\mtilde$  with respect to both $h$ and
$f$ gives
\begin{eqnarray}
\mtilde_h&=&1-{1\over\omega}
\Biggl      ({{\mit
v}^{{2-\omega}\over\omega}\over\ltilde}-{\ltilde\over {\it
v}^{{2+\omega}\over\omega}}f\Biggr)  
W_h+{1\over2}
\Biggl({W^{2\over\omega}\over\ltilde^2}-{1\over {\it
v}^{2\over\omega}}f\Biggr) 
\ltilde_h \qquad{\rm and}\nonumber\\  
\mtilde_f&=&-{1\over2}{\ltilde\over W^{2\over\omega}}
-{1\over\omega}
\Biggl({W^{{2-\omega}\over\omega}\over\ltilde}-{\ltilde\over
W^{{2+\omega}\over\omega}}f\Biggr) 
W_f+{1\over2}
\Biggl({W^{2\over\omega}\over\ltilde^2}-{1\over {\it
v}^{2\over\omega}}f\Biggr) 
\ltilde_f.\label{eq:3c}
\end{eqnarray}

\% Conditions 2 and 3---being now in the same form---are compared:
In particular, substituting equation  
(\ref{eq:3a}) into  the expression for $R$---used in condition 2---one
finds that  
\begin{equation}
\rtilde={1\over2}
\Biggl({W^{2\over\omega}\over\ltilde}-{\ltilde\over {\it
v}^{2\over\omega}}f\Biggr), 
\end{equation}
and hence  condition  2 becomes
\begin{equation}
\mu_h={1\over2}
\Biggl({W^{2\over\omega}\over\ltilde^2}-{1\over {\it
v}^{2\over\omega}}f\Biggr) 
\lambda_h
\qquad{\rm and}\qquad
\mu_f={1\over2}
\Biggl({W^{2\over\omega}\over\ltilde^2}-{1\over {\it
v}^{2\over\omega}}f\Biggr)  
\lambda_f.
\label{eq:2c}
\end{equation}

\% When equations (\ref{eq:2c})---derived from the second
condition---are compared with  
equations (\ref{eq:3c})---derived from the third
condition---they lead to the following pair of equations for $\ltilde$:
\begin{equation}
\omega-\Biggl({{\it
v}^{{2-\omega}\over\omega}\over\ltilde}-{\ltilde\over {\it
v}^{{2+\omega}\over\omega}}f\Biggr) 
W_h=0
\qquad{\rm and}\qquad
{\ltilde\over W^{2\over\omega}}+{2\over\omega}
\Biggl({W^{{2-\omega}\over\omega}\over\ltilde}-{\ltilde\over
W^{{2+\omega}\over\omega}}f\Biggr) 
W_f=0.
\label{eq:B}
\end{equation}

\% The above set of equations admits a common solution for $\ltilde$,
 we call it ${\bar\lambda}$,  given by 
\begin{equation}
{\bar\lambda}=-{2 W^{2\over\omega} W_f\over W_h}.
\label{eq:C1}
\end{equation}
Equations (\ref{eq:B}) and (\ref{eq:C}) are all well defined since---by
assumption---$W$, $W_h$, and $W_f$ are not 
identically zero, but for
the two equations in (\ref{eq:B}) to be consistent, they must lead either
to an identity or at least to a valid equation when
${\bar\lambda}$ is substituted into them. Indeed, by doing so, they
both reduce to 
\begin{equation}
{\omega\over2}W W_f=f W_f^2-{1\over 4}W_h^2,
\end{equation}
which is of course true by virtue of condition 1. This proves that the
$a_{\omega}$ map
(\ref{eq:ansatz}) (from the set of functions $(\mtilde,\ltilde, R)$ to
the set of solutions of the corresponding $\omega$--equation
(\ref{eq:WPDE})) is {\it onto}.

\% The expression for  ${\bar\lambda}$ is now substituted back into equation 
(\ref{eq:3a}) to give the relevant expression  for $\mtilde$, 
\begin{equation}
{\bar {\mu}}=h+{W_f\over 
W_h}f+{1\over4}{W_h
\over W_f},
\label{eq:DDDD1}
\end{equation}
and the one for $\rtilde$,
\begin{equation}
{\bar R}= {W_f \over W_h} f
-{1\over 4}{W_h \over W_f}
\label{eq:EEEEE1}.
\end{equation}
Note that the expression for ${\bar R}$ is now sign-unambiguous.

\% To check if the map $a_{\omega}$ is one--to--one, two sets of 
functions $(\mtilde_1,\ltilde_1,R_1)$ and $(\mtilde_2,\ltilde_2,R_2)$
are considered. They are required 
 to satisfy the linear pair of equations (\ref{eq:a}) and
provide---through the $a_{\omega}$ map---the same solution $W$
of the original 
$\omega$--equation. 
The problem is already solved in Appendix A and leads to the
following three conditions,
\begin{eqnarray}
\ltilde_1^{\omega\over2}(h-\mtilde_1+\rtilde_1)^{\omega\over2}&=&\ltilde_
2^{\omega\over2}
(h-\mtilde_2+\rtilde_2)^{\omega\over2},
\nonumber\\
{\ltilde_1^{\omega\over2}\over\rtilde_1}(h-\mtilde_1+\rtilde_1)^{\omega
\over 2}&=&
{\ltilde_2^{\omega\over2}\over\rtilde_2}
(h-\mtilde_2+\rtilde_2)^{\omega\over2},
\nonumber\\
{\ltilde_1^{\omega\over2}\over\rtilde_1}(h-\mtilde_1+\rtilde_1)^{{\omega-
2}\over2}      &=&
{\ltilde_2^{\omega\over2}\over\rtilde_2}(h-\mtilde_2+\rtilde_2)^{{\omega-
2}\over2},
\end{eqnarray}
which admit the almost trivial solution
\begin{equation}
\mtilde_1=\mtilde_2\qquad\ltilde_1^{\omega\over
2}=\ltilde_2^{\omega\over 2}\qquad R_1=R_2.
\end{equation}

\% The word ``almost'' is used because of the ambiguity in the expression
for the  $\lambda$'s. However, if the equivalence class of $\lambda$ is
defined as the set of functions that differ
from $\lambda$ by an $\omega\over2$ power of unity (it can be easily shown
that this defines an equivalence relation) then each anzatz--map
(\ref{eq:ansatz}) 
becomes {\it one--to--one}.
Hence $\bar\mu$, $\bar R$ and the equivalence
class of $(\bar\lambda)$---given respectively by equations
(\ref{eq:DDDD1}), (\ref{eq:E1}) and (\ref{eq:C1})---are unique.
This completes the proof.

\vskip 9cm
\% {\LARGE Appendix C.}
\vskip 1cm
\% In the following we denote $\delta(x,x') \delta(t,t')$ by $\delta
\delta$,  ${\partial \over {\partial x^i}} \delta(x,x') \delta(t,t')$
by ${\delta}_{,i}$ $\delta$ and ${\partial \over {\partial t}}
\delta(x,x') \delta(t,t')$
by $\delta$ $\dot{\delta}$.
If some expressions are calculated at $(x',t')$ they will be simply
primed.

\begin{eqnarray}
&& \{ {\cal X}^{\alpha} , {\cal P}_{\beta}  \} =
{{\delta}^{\alpha}}_{\beta}   \delta \delta
\nonumber\\
&& \{ {\gamma}_{\alpha{\epsilon}}, {\cal P}_{\beta}  \} =
{\gamma}_{\alpha{\epsilon},\beta} \delta \delta
\nonumber\\
&& \{ {\gamma}^{\alpha{\epsilon}}, {\cal P}_{\beta}  \} =
{{\gamma}^{\alpha{\epsilon}}}_{,\beta} \delta \delta 
\nonumber\\
&& \{ {{\delta}^{\alpha}}_{\epsilon}, {\cal P}_{\beta}  \} = 0
\nonumber\\
&& \{ {{\cal X}^{\alpha}}_i, {\cal P}_{\beta}  \} =
{{\delta}^{\alpha}}_{\beta} {\delta}_{,i} \delta
\nonumber\\
&& \{ {{\cal X}_{\alpha}}_i, {\cal P}_{\beta}  \} =
{\gamma}_{\alpha{\beta}} {\delta}_{,i} \delta +
{\gamma}_{\alpha{\mu},\beta} {{\cal X}^{\mu}}_i \delta \delta
\nonumber\\
&& \{ {\dot{\cal X}}^{\alpha} , {\cal P}_{\beta}  \} =
{{\delta}^{\alpha}}_{\beta} \delta \dot{\delta}
\nonumber\\
&& \{ n^{\alpha}, {\cal P}_{\beta}  \} = -n_{\beta} {\cal
X}^{{\alpha}m} {\delta}_{,m} \delta -{1 \over
2}{{\gamma}_{\mu{\nu},\beta}} n^{\mu} n^{\nu} n^{\alpha} \delta \delta
-{{\gamma}_{\mu{\nu},\beta}} n^{\mu} {\gamma}^{{\alpha}\nu} \delta\delta 
\nonumber\\ 
&& \{ n_{\alpha}, {\cal P}_{\beta}  \} = -n_{\beta} {{\cal X}_{\alpha}}^m {\delta}_{,m} \delta -{1 \over2}{{\gamma}_{\mu{\nu},\beta}} n^{\mu} n^{\nu} n_{\alpha} \delta \delta
\nonumber\\
&& \{ g_{ij}, {\cal P}_{\beta}  \} = {{\cal X}_{\beta}}_i
{\delta}_{,j} \delta + {{\cal X}_{\beta}}_j {\delta}_{,i} \delta +
{{\gamma}_{\mu{\nu},\beta}} {{\cal X}^{\mu}}_i {{\cal X}^{\nu}}_j
\delta \delta
\nonumber\\
&& \{ g^{ij}, {\cal P}_{\beta}  \} = -{{\cal X}_{\beta}}^i
g^{jm} {\delta}_{,m} \delta -{{\cal X}_{\beta}}^j
g^{im} {\delta}_{,m} \delta - {{\gamma}_{\mu{\nu},\beta}} {{\cal X}^{\mu}}^i {{\cal X}^{\nu}}^j
\delta \delta
\nonumber\\
&& \{ {{\delta}^i}_j, {\cal P}_{\beta}  \} = 0
\nonumber\\
&& \{ {{\cal X}^{\alpha}}^i, {\cal P}_{\beta}  \} =
-n^{\alpha} n_{\beta} g^{im} {\delta}_{,m} \delta -{{\cal
X}^{\alpha}}^{m} {{\cal X}_{\beta}}^{i} {\delta}_{,m} \delta -
{{\gamma}_{\mu{\nu},\beta}} {{\cal X}^{\alpha}}_m {{\cal X}^{\nu}}^m
{{\cal X}^{\mu}}^i \delta \delta
\nonumber\\
&& \{ {{\cal X}_{\alpha}}^i, {\cal P}_{\beta}  \} = -n_{\alpha} n_{\beta} g^{im} {\delta}_{,m} \delta -{{\cal
X}_{\alpha}}^{m} {{\cal X}_{\beta}}^{i} {\delta}_{,m} \delta -
{{\gamma}_{\mu{\nu},\beta}} n_{\alpha} n^{\nu}
{{\cal X}^{\mu}}^i \delta \delta
\nonumber\\
&& \{ g, {\cal P}_{\beta}  \} = 2g {{\cal X}_{\beta}}^m
{\delta}_{,m} \delta + g {{\gamma}_{\mu{\nu},\beta}} {{\cal X}^{\mu}}_m
{{\cal X}^{\nu}}^m \delta \delta
\nonumber\\
&& \{ N, {\cal P}_{\beta}  \} = -n_{\beta} \delta
\dot{\delta} +n_{\beta} N^m {\delta}_{,m} \delta - {1 \over 2} N 
{{\gamma}_{\mu{\nu},\beta}} n^{\mu} n^{\nu}  \delta \delta 
\nonumber\\
&& \{ N^i, {\cal P}_{\beta}  \} = {{\cal X}_{\beta}}^i
\delta \dot{\delta} + N n_{\beta} g^{im} {\delta}_{,m} \delta - N^m
{{\cal X}_{\beta}}^i {\delta}_{,m} \delta + N
{{\gamma}_{\mu{\nu},\beta}} n^{\mu} {{\cal X}^{\nu}}^i \delta \delta 
\nonumber\\
\label{eq:PB1}
\end{eqnarray}

\pagebreak
                          
\% {\LARGE Bibliography}

\vskip 1cm

\% [1] C.J. Isham, in {\it the 19th International Colloquium on Group
Theoretical Methods in Physics}, Salamanca, Spain, 1992.

\% [2] K.V. Kucha\v{r}, in {\it Proceedings of General Relativity and
Relativistic Astrophysics}, Winnipeg, 1991.

\% [3] C.J. Isham and N. Linden, {\it J. Math. Phys.} {\bf 36},
5392 (1995). 

\% [4] C. J. Isham, N. Linden, K. Savvidou and S. Schreckenberg: J. Math. Phys. {\bf 39}, 1818 (1998).

\% [5] K. Savvidou: {\sl The action operator in continuous time histories}, J. Math. Phys. {\bf 40}, 5657 (1999). 

\% [6] K. Savvidou and C. Anastopoulos: {\sl Histories quantization of parametrized systems and the problem of time}, gr-qc/9912077.

\% [7] R. B. Griffiths: J. Stat. Phys. {\bf 36}, 219 (1984).

\% [8] R. Omn\`{e}s: J. Stat. Phys. {\bf 53}, 893 (1988).
 
\% [9] M. Gell-Mann and J. Hartle: in Complexity, Entropy and the Physics of Information, SFI Studies in the Science of Complexity, Vol. VIII, 425 (Addison-Wesley, Reading) (1990)

\% [10] C. J. Isham: J. Math. Phys. {\bf 35}, 2157 (1994).

\% [11] P.A.M. Dirac, {\it  Lectures on Quantum Mechanics}
(Yeshiva University, New York, 1964).

\% [12] C.J. Isham, in {the Les Houches Summer School on Relativity,
Groups and Topology}, Les Houches, France, 1983.

\% [13] J.D. Brown and K.V. Kucha\v{r}, {\it Phys. Rev.\/} D
{\bf 51}, 5600 (1995).

\% [14] K.V. Kucha\v{r} and J.D. Romano, {\it Phys. Rev.\/} D
{\bf 51}, 5579 (1995). 

\% [15] J. D. Brown and D. Marolf: Phys. Rev. {\bf D53}, 1835 (1996).

\% [16] F.G. Markopoulou, {\it Class. Quant. Grav.} {\bf 13}, 2577 (1996).

\% [17] I. Kouletsis, {\it Action functionals of scalar fields and gravitational constraints that generate a genuine Lie albebra}, Class. Quant. Grav. {\bf 13}, 3085 (1996).

\% [18] I. Kouletsis, {\it A classical history theory: geometrodynamics regained}, gr-qc/9801019 (updated version to be submitted to J. Math. Phys.). 

\% [19] C. Teitelboim, {\it Ann. Phys.} {\bf 79}, 542 (1973).

\% [20] S.A. Hojman, K.V. Kucha\v{r} and C. Teitelboim, {\it
Ann. Phys.} {\bf 96}, 88 (1976).

\% [21] I. Kouletsis and K. V. Kucha\v{r}, {\it Spacetime Hamiltonian formalism} (in preparation, title is provisional)

\% [22] R. Arnowitt, S. Deser, and C.W. Misner, in
{\it Gravitation: An Introduction to Current Research,} edited
by L. Witten (Wiley, New York,1962). 

\% [23] K.V. Kucha\v{r}, {\it J. Math. Phys.} {\bf 17}, 777-820
(1976) and {\it J. Math. Phys.} {\bf 18}, 1589 (1977).

\% [24] K.V. Kucha\v{r}, in {\it Proceedings of Quantum Gravity
2}, Oxford, 1980.

\% [25] K.V. Kucha\v{r}, {\it J. Math. Phys.} {\bf 13}, 768 (1972).

\% [26] C.J. Isham and K.V. Kucha\v{r}, {\it Annals Phys.} {\bf 164},
288 and 316, 1985.

\% [27] K.V. Kucha\v{r} and C. Torre, {\it Phys. Rev.\/} D {\bf 43},
419, 1991.

\% [28] P.R. Garabedian, {\it Partial differential equations},  1964

\% [29]	K.V. Kucha\v{r}, {\it J. Math. Phys.} {\bf 15}, 708 ,1973.

\% [30]	D. Marolf, {\it Ann. Phys.} {\bf 236}, 374 (1994).

\% [31]	D. Lovelock, {\it J. Math. Phys.} {\bf 12},
498 (1971) 

\% [32] R. E. Peierls: Proc. Roy. Soc. (London) {\bf 214}, 143 (1952).

\% [33] J. Marsden {\it et al}: physics/9801019.

\end{document}